\pdfoutput=1
\documentclass[preprint,12pt,3p,authoryear]{elsarticle}

\biboptions{square,semicolon,sort&compress, sort}

\graphicspath{{figures/}{../figures/}}
\usepackage[space]{grffile}  

\usepackage[T1]{fontenc}

\usepackage{mathptmx} 

\usepackage{latexsym}
\usepackage{textcomp}
\usepackage{longtable}
\usepackage{multirow,booktabs}
\usepackage{amsfonts,amsmath,amssymb}
\usepackage[utf8]{inputenc}
\usepackage[english]{babel}
\addto\captionsenglish{}

\usepackage{todonotes}
\usepackage{soul}
\makeatletter
\if@todonotes@disabled
\else

\fi
\makeatother

\usepackage{siunitx}
\usepackage[version=4]{mhchem}
\usepackage{subfiles}
\usepackage{pdflscape}
\usepackage[labelfont=bf]{caption}

\usepackage[inline]{enumitem}
\newlist{mylist}{enumerate*}{1}
\setlist[mylist]{label=(\arabic*)}

\usepackage[section]{placeins}
\makeatletter
\AtBeginDocument{%
  \expandafter\renewcommand\expandafter\subsection\expandafter{%
    \expandafter\@fb@secFB\subsection
  }%
}
\makeatother

\usepackage{hyperref}
\hypersetup{colorlinks=true,allcolors={blue},pdfborder={0 0 0}}

\newcommand{\Ls}{\(\mathrm{L}_\mathrm{s}\)}
\newcommand{\degree}{\(^\circ\)}

\journal{Elsevier Journal}

\begin{document}

\begin{frontmatter}



\title{Planet Four: Probing Springtime Winds on Mars by Mapping the Southern Polar CO$_2$ Jet Deposits}

\cortext[cor1]{Corresponding author}
\author[lasp]{K.-Michael Aye\corref{cor1}}
\ead{michael.aye@lasp.colorado.edu}
\address[lasp]{Laboratory for Atmospheric and Space Physics, University of Colorado at Boulder, Boulder, CO 80303, USA}

\author[gemini,asiaa,YCAA,Yale]{Megan E. Schwamb}
\address[gemini]{Gemini Observatory, Northern Operations Center, 670 North A’ohoku Place, Hilo, HI 96720, USA}
\address[asiaa]{Institute for Astronomy and Astrophysics, Academia Sinica; 11F AS/NTU, National Taiwan University, 1 Roosevelt Rd., Sec. 4, Taipei 10617, Taiwan}
\address[YCAA]{Yale Center for Astronomy and Astrophysics, Yale University,P.O. Box 208121, New Haven, CT 06520, USA}
\address[Yale]{Department of Physics, Yale University, New Haven, CT 06511, USA}

\author[lasp]{Ganna Portyankina}

\author[psi]{Candice J. Hansen}
\address[psi]{Planetary Science Institute, 1700 E. Fort Lowell, Suite 106, Tucson, AZ 85719, USA}

\author[oxford]{Adam McMaster}
\author[oxford]{Grant R.M. Miller}
\author[adler]{Brian Carstensen}
\author[adler]{Christopher Snyder}
\author[adler]{Michael Parrish}
\author[adler]{Stuart Lynn}
\author[asiaa,asu]{Chuhong Mai}
\address[asu]{School of Earth and Space Exploration, Arizona State University, Tempe, AZ 85287, USA}
\author[adler]{David Miller}
\author[oxford]{Robert J. Simpson}
\author[adler,stsci]{Arfon M. Smith}

\address[oxford]{Oxford Astrophysics, Denys Wilkinson Building, Keble Road, Oxford OX1 3RH, UK}
\address[adler]{Adler Planetarium, 1300 S. Lake Shore Drive, Chicago, IL 60605, USA}
\address[stsci]{Space Telescope Science Institute, 3700 San Martin Drive, Baltimore, MD 21218, USA}
%

\begin{abstract}
    The springtime sublimation process of Mars' southern seasonal polar \ce{CO2} ice cap features dark fan-shaped deposits appearing on the top of the thawing ice sheet.
    The fan material likely originates from the surface below the ice sheet, brought up via \ce{CO2} jets breaking through the seasonal ice cap.
    Once the dust and dirt is released into the atmosphere, the material may be blown by the surface winds into the dark streaks visible from orbit.
    The location, size and direction of these fans record a number of parameters important to quantifying seasonal winds and sublimation activity, the most important agent of geological change extant on Mars.
    We present results of a systematic mapping of these south polar seasonal fans with the Planet Four online citizen science project.
    Planet Four enlists the general public to map the shapes, directions, and sizes of the seasonal fans visible in orbital images.
    Over 80,000 volunteers have contributed to the Planet Four project, reviewing 221 images, from \emph{Mars Reconnaissance Orbiter's} HiRISE (High Resolution Imaging Science Experiment) camera, taken in southern spring during Mars Years 29 and 30.
    We provide an overview of Planet Four and detail the processes of combining multiple volunteer assessments together to generate a high fidelity catalog of \num{\sim{} 400000} south polar seasonal fans.
    We present the results from analyzing the wind directions at several locations monitored by HiRISE over two Mars years, providing new insights into polar surface winds.
\end{abstract}%


\end{frontmatter}


\section{Introduction}%
\label{sec:intro}

Mars has a predominantly \ce{CO2} atmosphere with pressure levels buffered by seasonal CO2 polar caps \citep{leighton1966}.
In the winter atmospheric \ce{CO2} falls as snow or condenses directly onto the surface, forming a seasonal ice layer with a thickness of up to \SI{1}{\m}, depending on the latitude.
In the spring the south polar region of Mars exhibits a host of exotic phenomena associated with sublimation of the seasonal \ce{CO2} polar cap, and sublimation winds \citep{smith2001} contribute to atmospheric circulation.

\begin{figure}
\centering
\includegraphics[width=0.77\columnwidth]{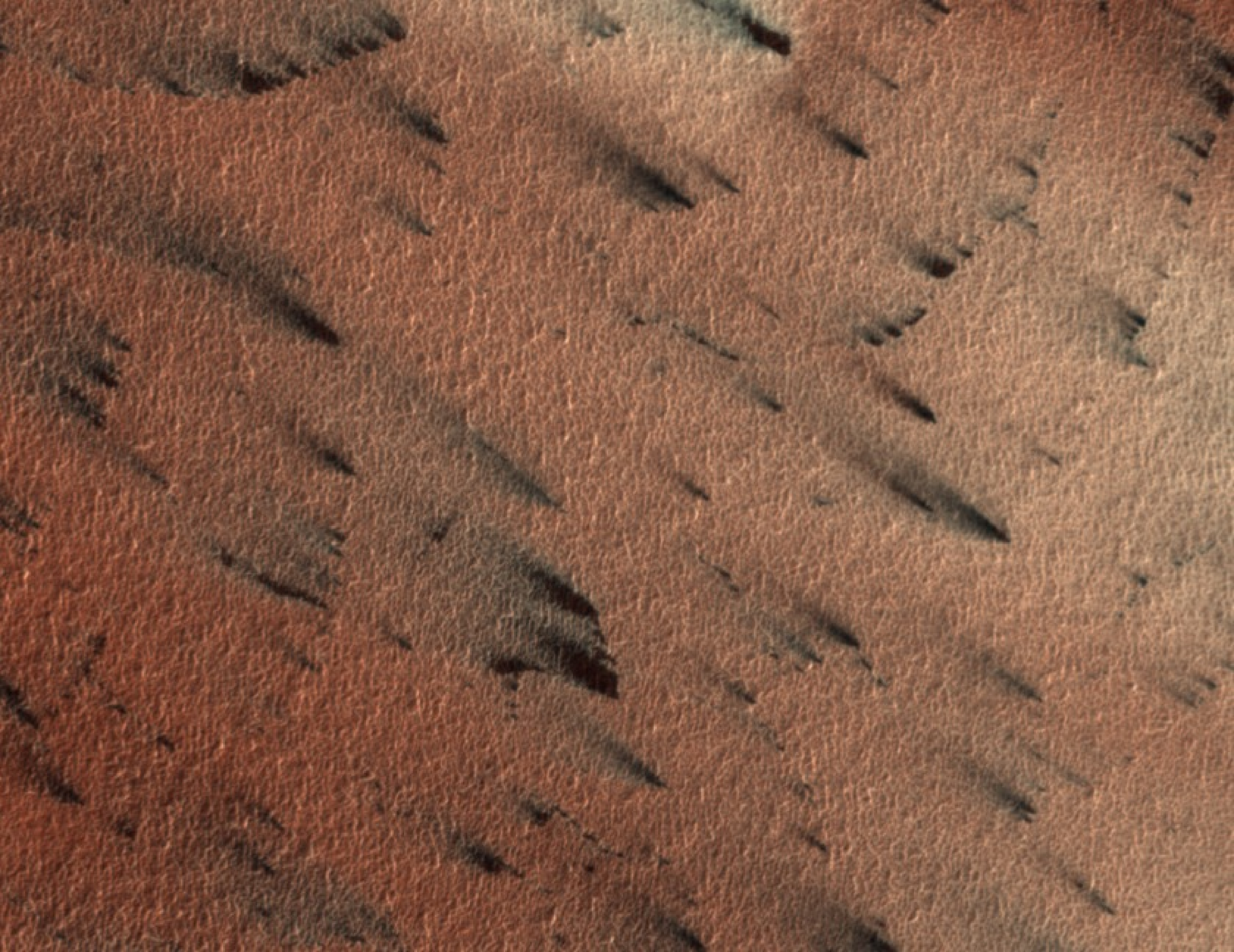}
\caption{\label{fig:fans} Subsection of HiRISE image \nolinkurl{ESP_011960_0925}, taken at (LAT, LON) \SIlist{-87.303;167.970}{\degree}; \Ls{} \SI{209.1}{\degree}.
The image is approximately \SI{321.4}{m} long and \SI{416.6}{m} wide%
}
\end{figure}

In the south polar region images from the \emph{Mars Reconnaissance Orbiter} (MRO) High Resolution Imaging Science Experiment (HiRISE, \citet{mcewen2007}) document activity best described by the ``Kieffer'' model \citep{kieffer2007,piqueux2003,hansen2010}:
\begin{enumerate}
    \item Over the winter \ce{CO2} anneals to form a translucent slab of impermeable ice.
Penetration of sunlight through the \ce{CO2} ice, which warms the ground below, results in basal sublimation of the ice.
\item The laboratory measurements done by \citet{hansen2005a} show that up to \SI{70}{\%} of the solar energy that reaches the top surface of a \SI{1}{\meter} thick slab layer can be transmitted through it.
Recent laboratory experiments by \citet{kaufmann2016} were able to trigger dust eruptions from a layer of dust inside a \ce{CO2} ice slab under Martian conditions, lending further credence to the proposed \ce{CO2} jet and fan production model.
\item Trapped gas escapes through ruptures in the ice, eroding and entraining material from the surface below \citep{devilliers2012}.
\item When this dust-laden gas is expelled into the atmosphere the dust settles in fan-shaped deposits on the top of the ice in directions oriented by the ambient wind, as shown in Figure~\ref{fig:fans} \citep{thomas2010,thomas2011}.
\item When the layer of seasonal ice sublimates in summer, the fans fade, as the material mostly blends back into the surface \citep{hansen2010}.
\item The compressed \ce{CO2} gas streams of the jets are believed to erode the surface, carving uniquely Martian spidery channels originally identified in images from the Mars Orbiter Camera \citep{piqueux2003a}, now referred to as araneiforms \citep{hansen2010}.
\end{enumerate}

The number, time history, area covered and changes in direction of the fans provide a wealth of information on the spring sublimation process and spring winds.
Apart from few wind direction estimations from remotely observed dunes \citep{Ewing2010-qn} and surface rover wind measurements \citep{Newman2017-ir,Greeley2006-lj}, no wide spread wind measurements exist for Mars.
The science goals enabled by cataloging fan measurements fall into two categories:

\begin{enumerate}
\item  Enhance our understanding of spring winds and provide constraints for global and mesoscale circulation models.
The length, width, and direction of these fans are snapshots in time of the local wind direction.
Changes in the orientation of the fans over time records changes in wind direction.
These markers can be compared to predictions from global and mesoscale circulation models (e.g.\ \citet{smith2015}) to improve our understanding of Mars' weather in the polar regions.
Dust injected into the atmosphere can be estimated.

\item  Extend our understanding of the sublimation process and its efficacy as an agent of change on the Martian surface.
The number of fans as a function of time record sublimation activity while the overlying ice thickness and insolation change during the season.
The areal coverage of the fans allows us (with reasonable assumptions about particle size) to estimate the amount of material eroded from the surface on seasonal timescales.
Inter-annual variability and the relationship of timing of seasonal activity to global dust storms can be quantified with this data-set (These are topics of future papers).

\end{enumerate}

Although the value of this data-set is clear, the sheer number of fans (on the order of hundreds of thousands) present in HiRISE images from multiple locations and times observed over many Mars years has proven to be a daunting data-set to catalog.
Attempts at developing automated detection algorithms have been unsuccessful at identifying the locations and shapes of these seasonal fans in images from orbit in a reliable fashion \citep{aye2010}.
However, there is an increasing interest to use the outcomes of Citizen Science projects as training data for neural networks (e.g.\ \citet{Banerji2010-ut,Bird2018-zx,Bowley2018-bi,Nguyen2018-tj,Peng2018-rm,Alger2018-xb}), hence we believe that these two lines of research will become strongly complimentary in the near future.

The task of mapping the dark fans is simply pattern recognition, and the human brain is ideally suited for this task, easily capable of spotting and outlining these features.
With the advent of the Internet, tens of thousands of people across the globe can be enlisted to assist scientists with tasks that are impossible to automate.
This citizen science or crowd-sourcing approach, where independent assessments from multiple non-expert classifiers are combined, has become an established technique as the data volumes have continued to grow.
This method has been applied to nearly all areas in astronomy and planetary science \citep{marshall2014} (see reference therein) including galaxy morphology \citep{lintott2008,willett2013}, identification of planet transits \citep{fischer2012, schwamb2012}, crater counting~\citep{bugiolacchi2016,robbins2014} and to a sister project of the here presented efforts, Planet Four: Terrains~\citep{schwamb2017}.
In collaboration with the Zooniverse\footnote{\href{http://www.zooniverse.org}{http://www.zooniverse.org}} \citep{lintott2011,fortson2012}, the largest collection of online citizen science projects, we have developed Planet Four\footnote{\href{http://www.planetfour.org}{http://www.planetfour.org}}, a web portal to enlist the general public to identify and map the seasonal fans in HiRISE images of Mars' polar regions.

In this paper we present the first results from the Planet Four project, a catalog of seasonal fans from two Mars years, MY 29 and 30, of HiRISE monitoring of the Martian South Polar region.
In Section~\ref{sec:HiRISE}, we provide an overview of the HiRISE South Pole Seasonal Processes Monitoring Campaign and the specific HiRISE observations used in this study.
In Section~\ref{sec:planet_four_intro}, we present the Planet Four project and the online classification interface.
Section~\ref{sec:data_reduction} details the process for assessing and combining the volunteer classifications to create a catalog of seasonal features.
In Section~\ref{sec:data_validation} we examine our catalog's validity by comparing results between volunteers and science team members.
Section~\ref{sec:results} presents general statistical results of the catalog, and finally, we use the catalog for an initial probing into regional winds in Section~\ref{sec:regional}.
We summarize our conclusions in Section~\ref{sec:conclusions}.
All place names referred to in this paper are informal and not approved by the International Astronomical Union.
Full machine-readable versions of the catalogs and tables presented in this paper are also available from
\href{https://www.planetfour.org/results}{https://www.planetfour.org/results}.

\section{HiRISE Instrument and Seasonal Processes Monitoring Campaign}%
\label{sec:HiRISE}

The \emph{Mars Reconnaissance Orbiter (MRO)} has the ability to turn off nadir to target a specific location.
In its inclined orbit there are numerous opportunities to achieve repeat coverage in the polar region.
In order to study seasonal processes the HiRISE team selected a limited number of regions of interest (ROIs) in the Martian south polar region to image throughout the spring season.
Time is defined on Mars by the orbital longitude \Ls{}, where southern spring begins at \Ls{}=\SI{180}{\degree}.

Originally, the HiRISE monitoring campaigns were numbered by their ordinal number of seasons the MRO mission had been observing Mars.
This work focuses on the observations from seasons 2 and 3 which have more regular repeat HiRISE imaging of ROIs over multiple years, compared to season 1 HiRISE monitoring campaign.
To be able to compare with other missions and modeling, we also identify our data using the convention of Martian years, established by \citet{clancy2000} and \citet{piqueux2015}, where Mars Years 29 and 30, also written as MY29 and MY30, correspond to HiRISE seasons 2 and 3.
Every day, citizen scientists are making more fan measurements for later Mars years and the catalog continues to grow.
The longer timespan covered by the catalog will be discussed in future paper(s).

\begin{figure}[p]
\centering
\vspace*{-2cm}
\includegraphics[width=0.75\columnwidth]{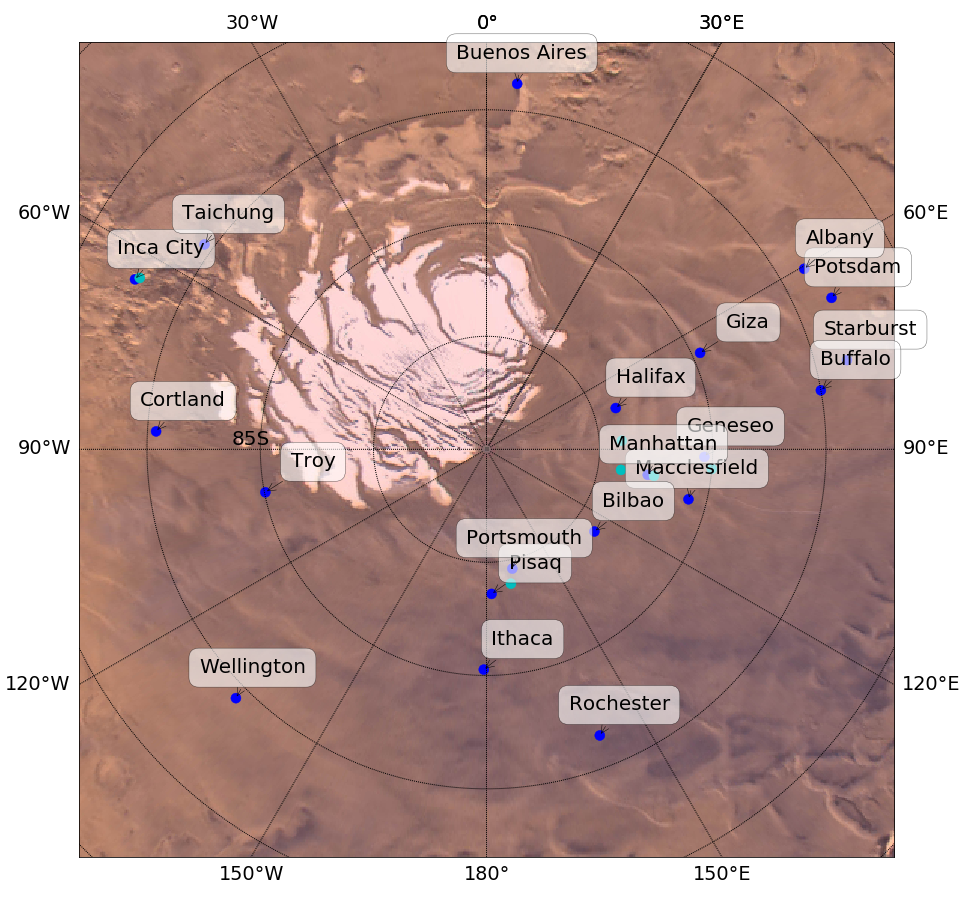}
\caption{\label{fig:monitoring_overview} Map overview of the regions of interest for the seasonal monitoring campaign of HiRISE.
For readability, the following regions are shown as cyan-colored unlabeled dots: Inca City Ridges, Schenectady, Troy, Manhattan Cracks, Manhattan Classic, Atka, Halifax, Oswego edge.%
}

\includegraphics[width=0.8\columnwidth]{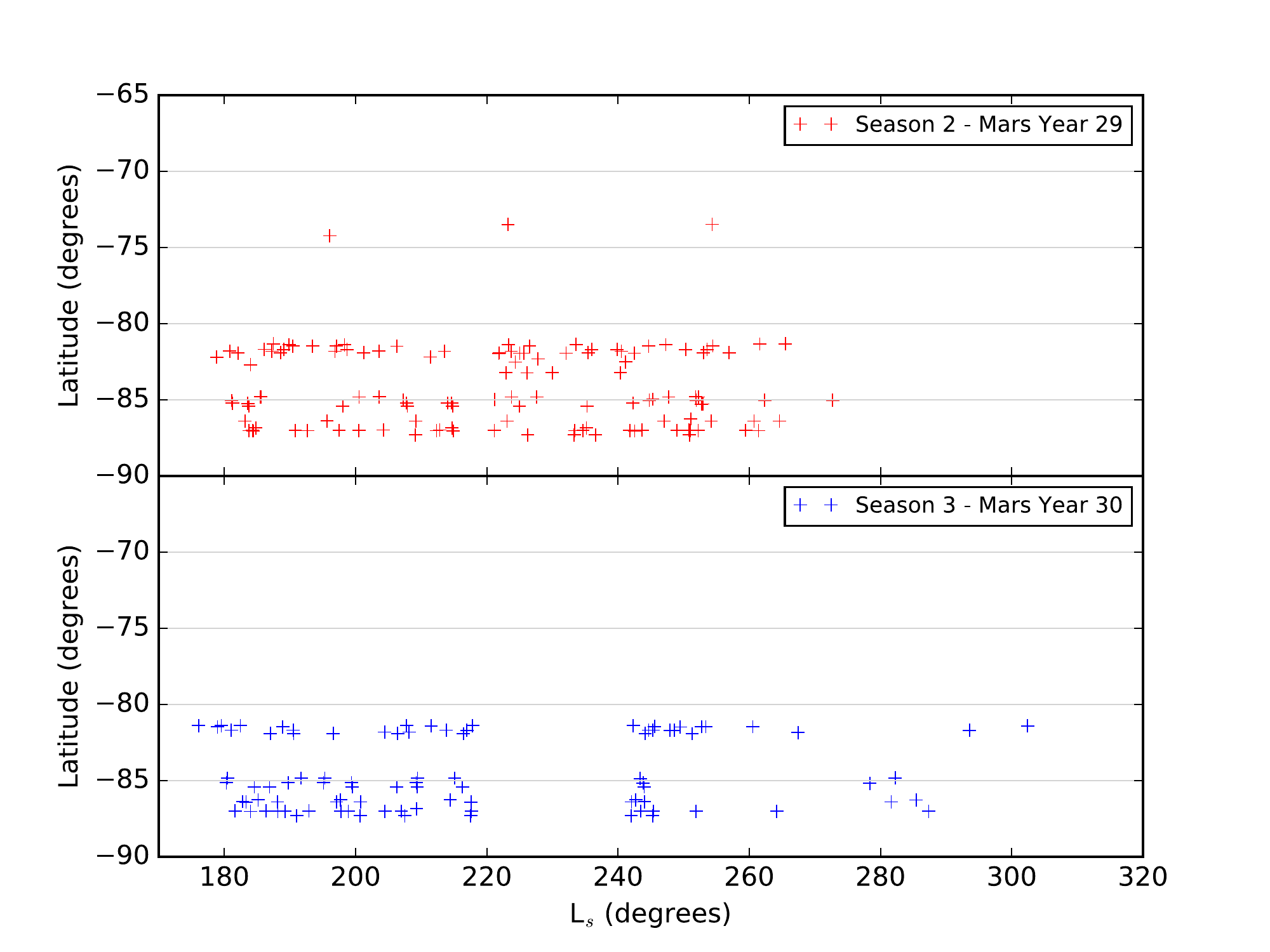}
\caption{\label{fig:temporal_coverage} Temporal and latitude coverage for the season 2 and season 3 HiRISE monitoring campaign observations reviewed on Planet Four.%
}
\end{figure}

Figures~\ref{fig:monitoring_overview} and \ref{fig:temporal_coverage} provide an overview of the observed locations and times in solar longitudes of the HiRISE data used in this work.
Table~\ref{tab:regions} lists the ROIs selected for analysis using Planet Four.
221 high quality images from southern spring season 2 and 3 (i.e.\@ MY 29 and 30) were selected for analysis on Planet Four (see Table~\ref{tab:obsids}).
The reduced HiRISE products were obtained from the National Aeronautics and Space Administration's (NASA) Planetary Data System (PDS) HiRISE PDS Data Node\footnote{\href{http://hirise-pds.lpl.arizona.edu/PDS/}{http://hirise-pds.lpl.arizona.edu/PDS/}}.

\begin{table}
\begin{tabular}{lllcc}
\toprule
\textbf{Latitude} & \textbf{Longitude} & \textbf{Informal Name} & \textbf{ \# of Images} & \textbf{\# of Images} \\
\textbf{(degrees)} & \textbf{(degrees East)} & &  \textbf{MY 29} & \textbf{MY 30} \\
\midrule
-73.53& 339.5 & Binghamton & 2 & 0 \\
-74.22 & 168.5 &  Caterpillar & 1 & 0 \\
-81.38 & 295.8   & Inca City       & 7 & 7 \\
-81.46 & 296.3 & Inca City Ridges & 7 & 8 \\
-81.68 &  66.3 & Potsdam           &  7 & 9 \\
-81.80 & 76.1  & Starburst          &  7 & 3 \\
-81.93 & 60.4 & Albany &  5 & 0 \\
-81.9  &   4.8 &  Buenos Aires      &  7 & 7 \\
-82.2 & 225.2 & Wellington &  2 & 0 \\
-82.3     & 306 & Taichung & 1 & 0 \\
-82.5     & 80.0 & Buffalo & 2 & 0 \\
-82.69 & 273.1 & Cortland & 1 & 0 \\
-83.2 & 158.4 & Rochester &  4 & 0 \\
-84.82 &  65.7 & Giza              & 11 & 7\\
-85.0 & 95.0 & Schenectady & 1 & 0 \\
-85.02 & 259.0 & Troy & 1 & 0  \\
-85.13 & 180.7 & Ithaca            & 10 & 6 \\
-85.18 & 92.0 &  Geneseo &  0 & 1 \\
-85.4  & 103.9 &  Macclesfield     &  7 & 7 \\
-86.25 &  99.0 & Manhattan Cracks  &  1 & 5 \\
-86.39 &  99.0 & Manhattan Classic &  8 & 9 \\
-86.8  & 178.0 & P\'{i}saq    &  3 & 1 \\
-86.98 & 169.7 & Atka & 3 & 0 \\
-86.99 & 99.1 & Manhattan Frontinella &  5 & 3 \\
-87.0 & 72.3 &  Halifax  & 3 & 0 \\
-87.0  &  86.4 & Oswego edge       &  6 & 10 \\
-87.0  & 127.3 & Bilbao      &  7 & 3 \\
-87.3  & 167.8 & Portsmouth          &  5 & 6 \\
\bottomrule
\end{tabular}
 \caption{\label{tab:regions} Regions of interest studied with Planet Four that were monitored during both seasons 2 (Mars Year 29) and 3 (Mars Year 30) HiRISE Southern Seasonal Processes Campaign.
A full list of the images is available as supplemental data in the file \nolinkurl{P4_catalog_v1.0_metadata.csv}
The Latitude and Longitude values are the mean value over the center latitudes and longitudes of the respective HiRISE observations.
All informal names are internal designations used by the Planet Four team and not approved by the International Astronomical Union.}
\end{table}

HiRISE is a pushbroom imager.
It has ten \num{2048}-pixel detectors in the cross-track direction, which covers \SI{\sim{}6}{km} at the spacecraft altitude of \SI{300}{km} (MRO is in an elliptical \SI{255}{km} by \SI{320}{km} orbit).
An image is built up in the along-track dimension as the spacecraft travels in its orbit, with a ground velocity of \SI{\sim{}3} {\kilo\meter\per\second}.
A typical size image has \num{\sim{}60000} pixels along-track, thus covers a \SI[product-units = brackets]{6 x 18}{\kilo\meter\squared} area.
Color is available in the center \SI{20}{\%} of the image.
A full description of the camera is found in~\citet{mcewen2007}.

\begin{figure}[p]
\centering
\includegraphics[width=0.95\columnwidth]{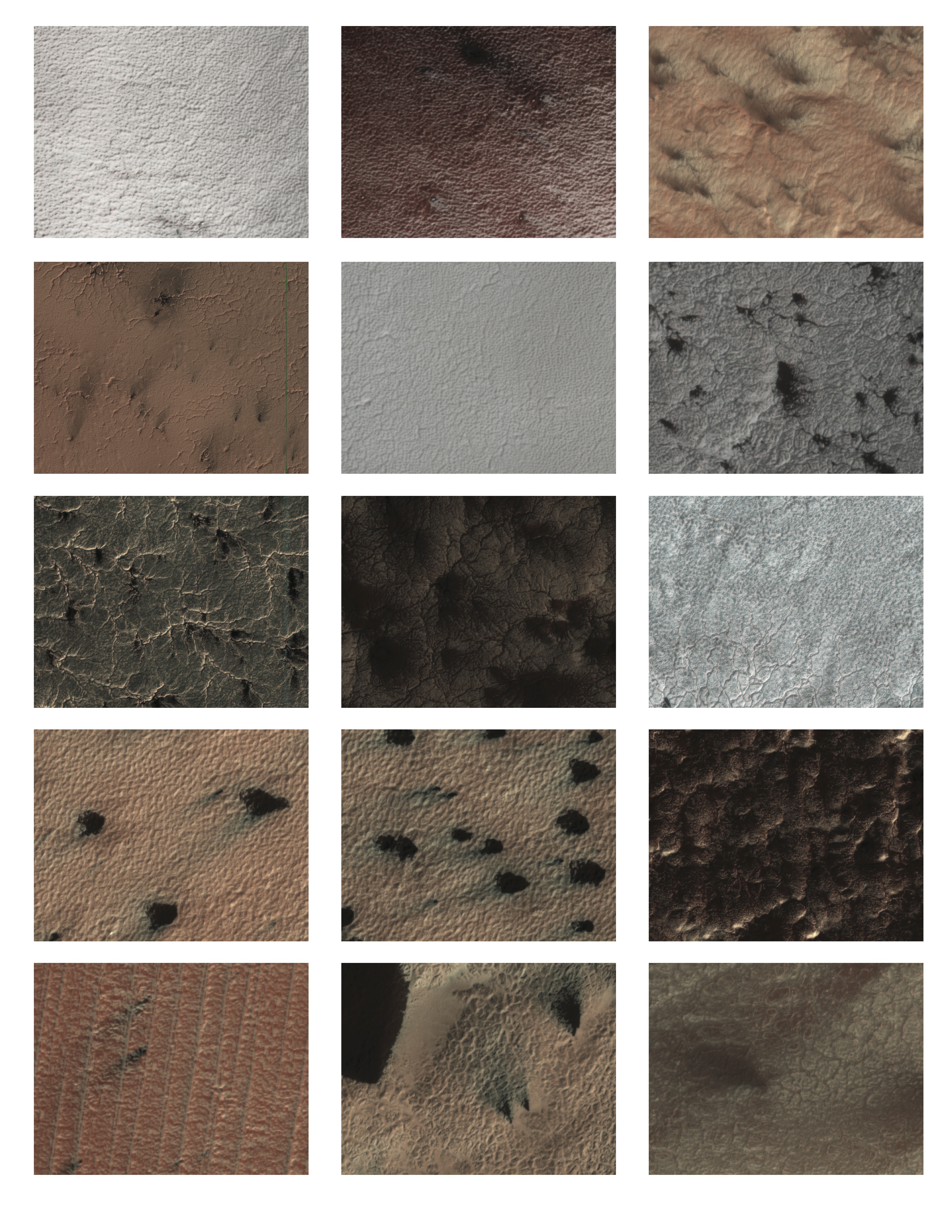}
\caption{\label{fig:samplecutouts1} Randomly selected sample of Planet Four tiles characteristic of the season 2 and season 3 HiRISE monitoring campaign.
Each tile has \num{840 x 648} pixels, but its ground resolution varies with HiRISE binning modes.
This is reflected in the \nolinkurl{map_scale} column of the Planet Four catalog files.%
}
\end{figure}

\begin{figure}[p]
\centering
\includegraphics[width=0.95\columnwidth]{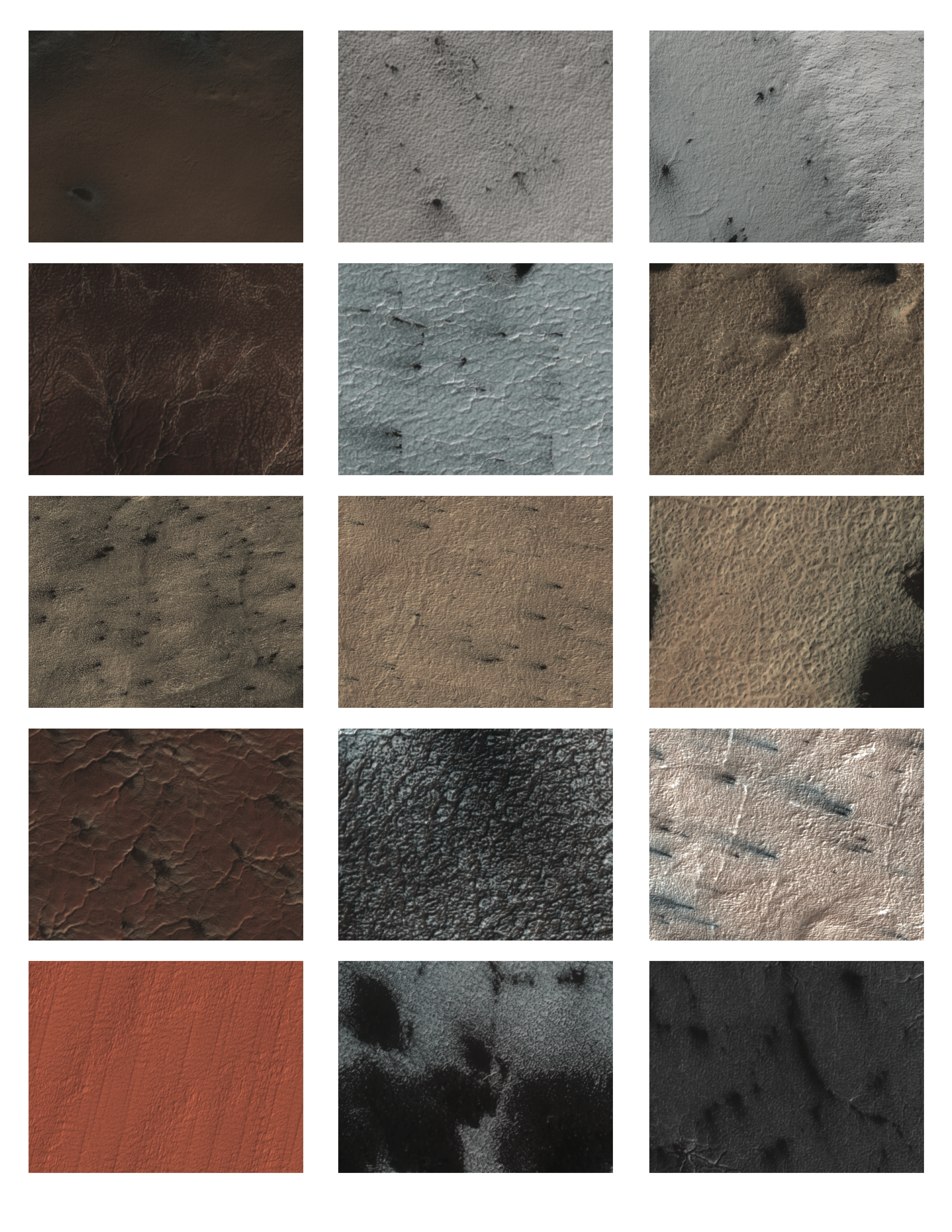}
\caption{\label{fig:samplecutouts2} Randomly selected sample of Planet Four tiles characteristic of the season 2 and season 3 HiRISE monitoring campaign.
Each tile has \num{840 x 648} pixels, but its ground resolution varies with HiRISE binning modes.
This is reflected in the \nolinkurl{map_scale} column of the Planet Four catalog files.%
}
\end{figure}

It is generally easier to identify the fans in the color portion of the image, so only the \SI{\sim{}1}{km} wide color (RGB) sub-image was used for the Planet Four image set.
A visitor to the Planet Four website is presented with a sub-image from a RGB non-mapped projected HiRISE image.
Each HiRISE frame (typically several hundred megabytes in size) is divided into \num{840 x 648} pixel sub-images that we will refer to as ``tiles''.
To avoid edge effects, the tiles are generated such that there is a 100-pixel overlap with the neighboring tiles.
We avoid showing volunteers tiles where part or most of the tile is blank.
Due to the variable length and width of HiRISE images, there is typically a small region on the right and bottom edges of the non-map projected HiRISE image that cannot be made into a full-sized tile and thus is not searched for seasonal features with Planet Four.
Pixel sampling scales per tile are typically 24.7 cm/pixel when HiRISE is in \num{1 x 1} binning mode, and the seasons 2 and 3 observations span binning resolutions of \num{1 x 1} to \num{4 x 4}.
For the seasons 2 and 3 monitoring campaign, a HiRISE image is associated with 36 to 635 tiles (see Table~\ref{tab:obsids}).
\begin{table}
\begin{tabular}{ccccccc}
\toprule
\textbf{Observation ID} & \textbf{Latitude} & \textbf{Longitude} & \textbf{L\(_s\)} & \textbf{Start Time} & \textbf{North} & \textbf{\# of} \\
& \textbf{[deg]} & \textbf{[deg east]} & \textbf{[deg]} & & \textbf{Azimuth} & \textbf{Tiles} \\
\midrule
ESP\_011296\_0975 & -82.197 & 225.253 & 178.8 & 2008-12-23  & 110.6 & 91 \\
ESP\_011341\_0980 & -81.797 & 76.13 & 180.8 & 2008-12-27    & 110.2 & 126 \\
ESP\_011348\_0950 & -85.043 & 259.094 & 181.1 & 2008-12-27  & 123.6 & 91 \\
ESP\_011350\_0945 & -85.216 & 181.415 & 181.2 & 2008-12-27  & 99.7  & 126 \\
ESP\_011351\_0945 & -85.216 & 181.548 & 181.2 & 2008-12-27  & 128.0 & 91 \\
ESP\_011370\_0980 & -81.925 & 4.813 & 182.1 & 2008-12-29    & 110.6 & 126 \\
ESP\_011394\_0935 & -86.392 & 99.068 & 183.1 & 2008-12-31   & 139.4 & 72 \\
ESP\_011403\_0945 & -85.239 & 181.038 & 183.5 & 2009-01-01  & 106.5 & 164 \\
ESP\_011404\_0945 & -85.236 & 181.105 & 183.6 & 2009-01-01  & 134.1 & 91 \\
ESP\_011406\_0945 & -85.409 & 103.924 & 183.7 & 2009-01-01  & 111.3 & 126 \\
ESP\_011407\_0945 & -85.407 & 103.983 & 183.7 & 2009-01-01  & 138.8 & 91 \\
ESP\_011408\_0930 & -87.019 & 86.559 & 183.8 & 2009-01-01   & 148.9 & 59 \\
ESP\_011413\_0970 & -82.699 & 273.129 & 184.0 & 2009-01-01  & 112.8 & 108 \\
ESP\_011420\_0930 & -87.009 & 127.317 & 184.3 & 2009-01-02  & 157.3 & 54 \\
ESP\_011422\_0930 & -87.041 & 72.356 & 184.4 & 2009-01-02   & 157.0 & 54 \\
ESP\_011431\_0930 & -86.842 & 178.244 & 184.8 & 2009-01-03  & 148.6 & 54 \\
ESP\_011447\_0950 & -84.805 & 65.713 & 185.5 & 2009-01-04   & 113.0 & 218 \\
ESP\_011448\_0950 & -84.806 & 65.772 & 185.6 & 2009-01-04   & 138.8 & 59 \\
\bottomrule
\end{tabular}
\caption{\label{tab:obsids} Partial table of used HiRISE observations to indicate spatial and temporal coverage.
Full table published in the online version. The center coordinates for all HiRISE pointings used in this study.
Latitudes are planeto-centric and the given north azimuth angle is for the non-map-projected data that went into the Planet Four system.}
\end{table}
For the analysis presented here  23,723 tiles derived from 129 full frame HiRISE season 2 monitoring images and 19,181 tiles derived from 92 season 3 HiRISE images were reviewed by Planet Four volunteers.
A characteristic sample of Planet Four tiles is presented in Figures~\ref{fig:samplecutouts1} and~\ref{fig:samplecutouts2}.

\section{Planet Four}%
\label{sec:planet_four_intro}
Here we describe the Planet Four classification interface and the information generated by volunteers visiting the Planet Four website.

\subsection{Classification Web Interface}
Planet Four volunteers are asked to identify and outline fans in the presented tiles. Sometimes the fan has an indeterminate direction, in which case we call them ``blotches''.
Although less useful for wind regime studies the blotches are sites where the ice has ruptured and released material, so they are important to studying the sublimation process of the polar \ce{CO2} ice sheet.
Thus, volunteers are asked to identify and mark blotches as well. Positions, orientations, and sizes of fans and blotches are obtained via a web interface (see Figure~\ref{fig:tutorial}) built upon the Zooniverse's Application Programming Interface (API), which communicates with their custom built Ouroboros web platform (described in \ref{sec:ouroboros}).
Each tile is assessed by approximately 30--100 independent reviewers.
To ensure reviewers have no prior information that may influence their judgment, tiles are randomly served to the classifier, and no identifying information about the parent HiRISE image is presented in the Planet Four web interface.
The volunteer is blind to the location on the South Pole, time of season the observation was taken, and responses from other classifiers while reviewing a given tile.
Planet Four was launched originally in English; later on the websites, classification interface, and help material have also been  translated into several languages , including traditional and simplified character Chinese, German, and Magyar (Hungarian).
For the analyses presented here, all Planet Four classifications are treated the same, regardless of what language the volunteer was using in the classification web interface.

\subsubsection{Tutorial}%
\label{sec:tutorial}
First time visitors to the Plant Four website are presented with a short inline interactive tutorial that explains the task and guides the classifier on how to use the marking tools.
Additional training material is also available elsewhere on the site.
The tutorial is shown only once for those classifiers using the Planet Four web interface logged-in with a registered Zooniverse account.
Volunteers using the site in the non-logged-in mode, are presented with the tutorial each time they visit the Planet Four website.
Other than the frequency of the tutorial appearing, the user experience on Planet Four, including the tutorial content, are exactly the same for logged-in or non-logged in volunteers.

\subsubsection{Marking Tools}
\begin{figure}[p]
\centering
\includegraphics[width=0.95\columnwidth]{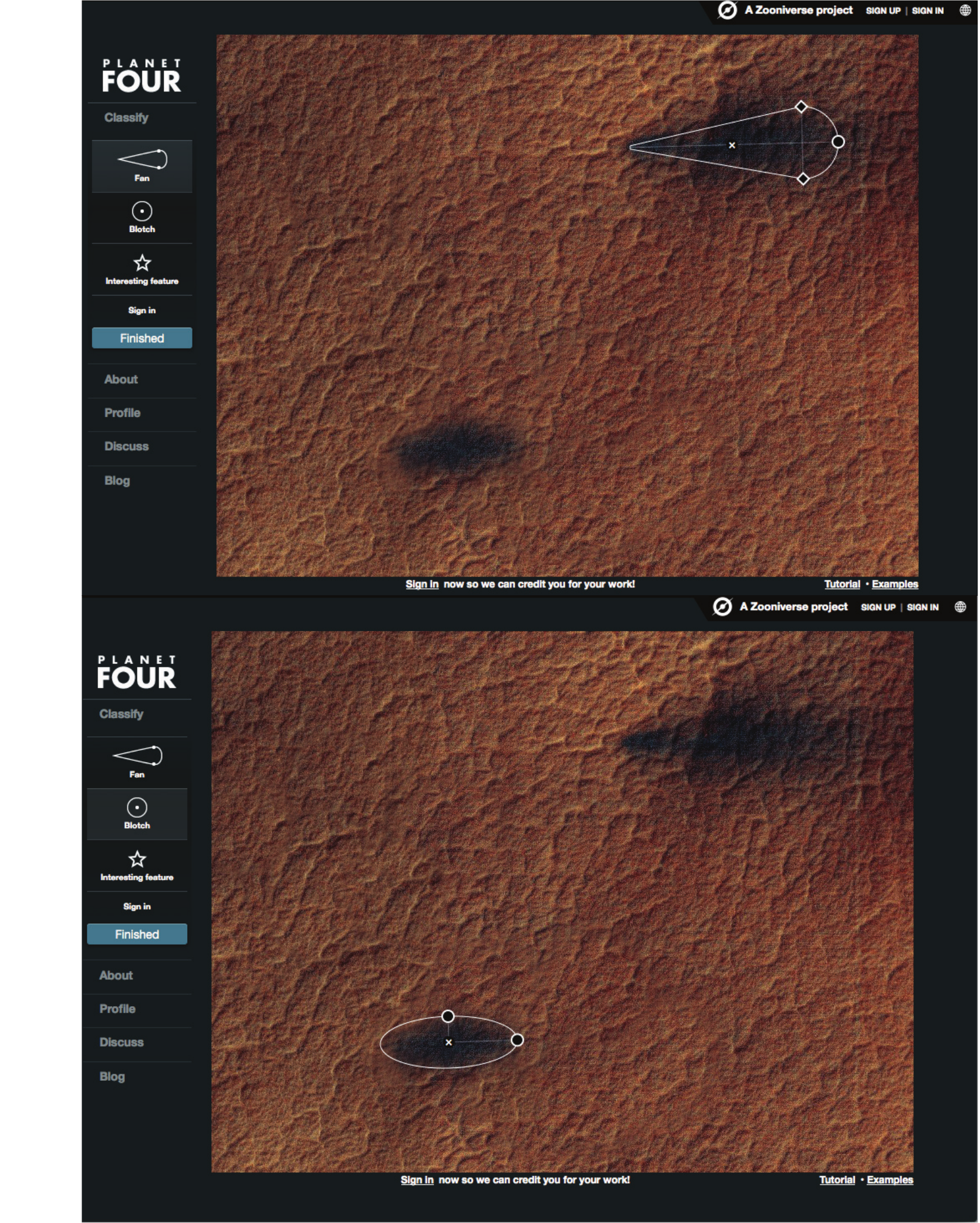}
\caption{\label{fig:tutorial} The fan (above) and blotch (below) marker on the Planet Four tutorial image.
Black circles and diamonds are the marker handles that can be used to adjust the shape and orientation in the web classification interface.
The ``x'' is used to delete the marker.%
}
\end{figure}

Fans and blotches are drawn by selecting the appropriate tool in the classification interface (see Figure~\ref{fig:tutorial}), clicking on the tile displayed, and dragging to resize the marker to the appropriate shape and orientation.
The fan tool generates a triangle with a rounded base with the user controlling the endpoint of the fan.
The default opening angle for the fan marker is set to \ang{5}.
The blotch tool simply produces an ellipse with the user controlling the size and orientation of the major axis.
For blotches, the default length of the minor axis is 0.75 times the pixel length of the major axis drawn.
Once a blotch or fan marking has been made, a classifier can edit the initial parameters by manipulating handles on the marker.
For blotches, the length of the major and minor axes and rotation can be adjusted.
For fans, the opening angle, orientation, and length can be modified.
If only a single mouse click is made on the interface, than the minimum sized fan or blotch marker is produced: a fan with a length of 10 pixels and an opening angle of \ang{1} or an ellipse with both axes equal to 10 pixels.
Additionally, there is an `Interesting Feature' tool available for volunteers to highlight the position of anything that they deem worth review by the Planet Four Science Team.
The Interesting Feature marker is not resizable.
All markers drawn in the web interface can be repositioned or removed by the classifier.

\subsection{Classification Database}%
\label{sec:classification_database}
Once the volunteer is done making markings, if any, and hits the `Finished' button, the classification (which we define as the sum total of all the markings or lack of markings made by the volunteer) is submitted to the Ouroboros API to be saved to a database.
At this point, the classifier can move on to view the next tile by hitting the `Next' button or can choose instead to enter the Planet Four discussion tool (discussed in further detail in Section~\ref{sec:Talk}).
Once the classification has been submitted, it cannot be revised.
For blotches, the center position, rotation angle, and pixel lengths of the major and minor axes of the ellipse are recorded.
For fans, the starting position, distance in pixels from the starting point to the end of the fan, opening angle, and rotation angle are saved to the database.
For interesting features, only the pixel location is stored.
If no features are marked, the database records the classification as a non-marking.
A tile identifier and timestamp for each classification is also stored in the database.

If the volunteer is logged in with a registered Zooniverse account, the classifications are tracked in the database via the associated username.
For non-logged-in classifications, a unique session id is generated and used to link the classifications  completed by a given IP address and web browser.
The non-logged-in identifier does not exactly correspond one-to-one to a unique individual.
If a person classifiers non-logged-in and changes their IP address, their new classifications would be stored under a different identifier.
Additionally, if a volunteer initially participates as a non-logged-in classifier on Planet Four and then registers for a Zooniverse account, the previous classifications stored in the database are not linked to the Zooniverse username and remain associated with the unique non-logged-in session identifier.

We note there are occasional spurious or duplicate entries stored in the classification database, typically due to a glitch in the classifiers' browser or a minor bug in the Ourborous framework.
These entries compose a very small percentage of the total volunteer classifications.
They are easily identified and removed from the analysis presented here. Further details are provided in~\ref{sec:dupes_and_spurious}.
Additionally the Planet Four classification interface  originally recorded a different angle than the intended spread angle from the fan marking tool.
This was identified and subsequently fixed in the software.
The true spread angle of the fan marker drawn by the volunteers is recoverable from the values stored recorded in the database, and we have adjusted the classifications effected.

\subsection{Talk Discussion Tool}%
\label{sec:Talk}
Associated with the Planet Four classification interface is a dedicated object-orientated discussion tool known as ``Talk''\footnote{\href{http://talk.planetfour.org}{http://talk.planetfour.org}}.
Each Planet Four tile assessed on the main classification interface has a dedicated page on the Planet Four Talk website.
Volunteers can access these pages directly through the classification interface after submitting their classification.
With Talk, volunteers can write comments, add searchable Twitter-like hash tags, create longer side discussions, and group similar tiles together in collections.
For the analysis presented here, we focus strictly on the volunteer markings from the main user interface, and do not include a complete analysis of the data from the Talk tool.

\subsection{Site History}
Planet Four was publicly launched on 2013 Jan 8 as part of the British Broadcasting Corporation's (BBC) Stargazing Live, three nights of live astronomy programing (2013 Jan 8--10) on BBC Two in the United Kingdom.
Review of Season 2 and 3 tiles span from January 2013 to March 2015 with 9,809,637 classifications produced in total.
The majority of classifications for Seasons 2 and 3 were obtained during the BBC Stargazing period, but subsequently data from HiRISE's other seasonal monitoring campaigns were mixed with the Season 2 and Season 3 classifications.
The results from data outside season 2 and 3 which are still in the process of being reviewed on the Planet Four website will be the topic of subsequent publications.
\begin{figure}
\centering
\includegraphics[width=0.95\columnwidth]{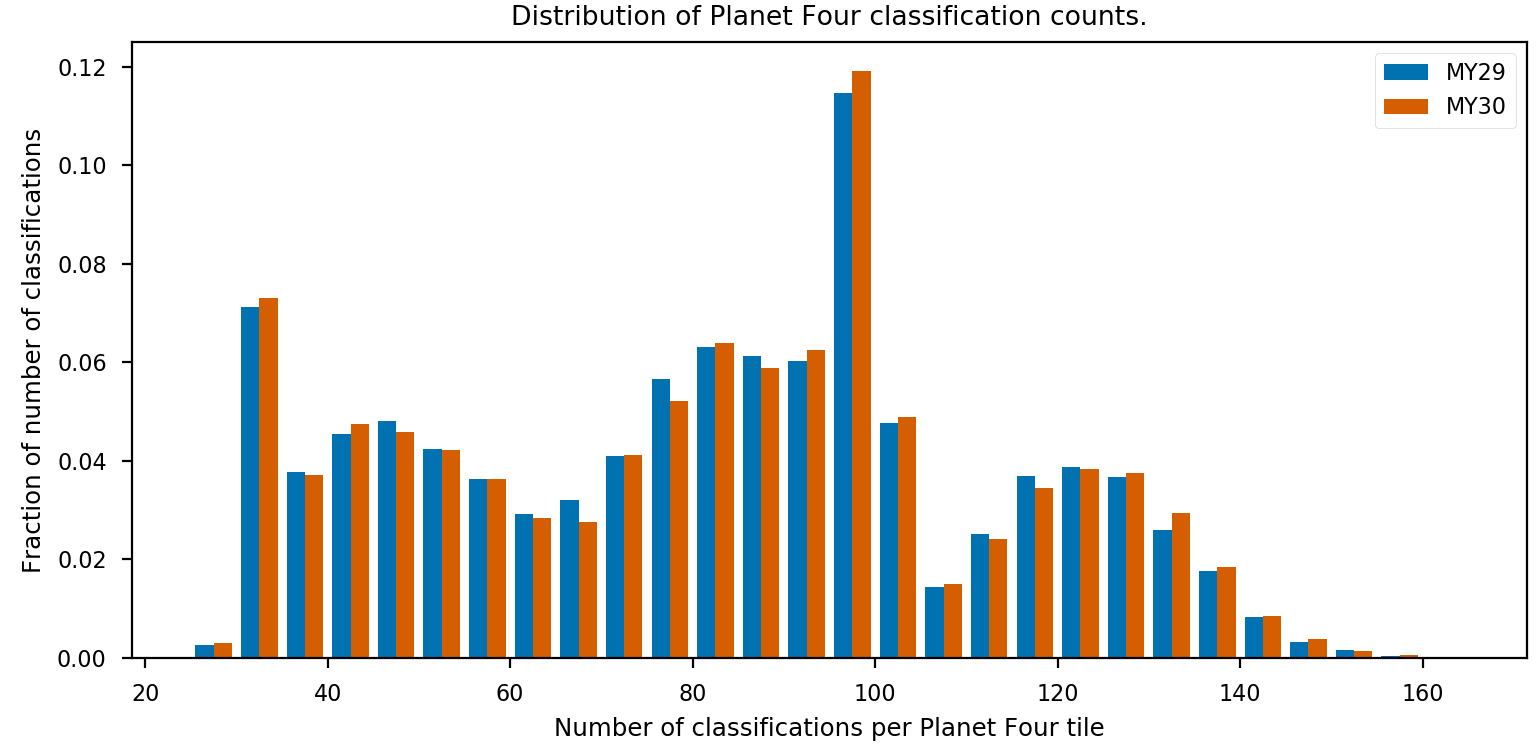}
\caption{\label{fig:cutout_count} Distribution of the number of Planet Four classifications for Season 2 (MY29) and Season 3 (MY30) tiles with a bin size of 5.
The distribution peaks at the two different retirement values of 100 and 30.
Due to performance issues in the webserver's queueing system, the retirement values were at times not enforced, leading to the spread-out distributions at values higher than the retirement values.%
}
\end{figure}
Figure~\ref{fig:cutout_count} plots the distribution of classifications per tile for Seasons 2 and 3.
Due to the high classification rate at launch, tiles were set to retire from rotation in the web interface after 100 independent assessments (counting duplicates) to ensure that the project would continue to serve data over the Stargazing period.
Over time the classification rate dropped significantly from launch, and on 2013 Dec 9 the retirement threshold for a tile was lowered to a more reasonable --- and statistically acceptable --- value of 30 to better accommodate the actual work rate on Planet Four.
This value is similar to the image retirement threshold that was used by the Zooniverse's Milky Way Project \citep{simpson2012}, which enlists the general public in a similar task, drawing circles on space-based infrared images to identify the shape and size of star formation bubbles.

\subsection{User Statistics}
\begin{figure}
\centering
\includegraphics[width=0.77\columnwidth]{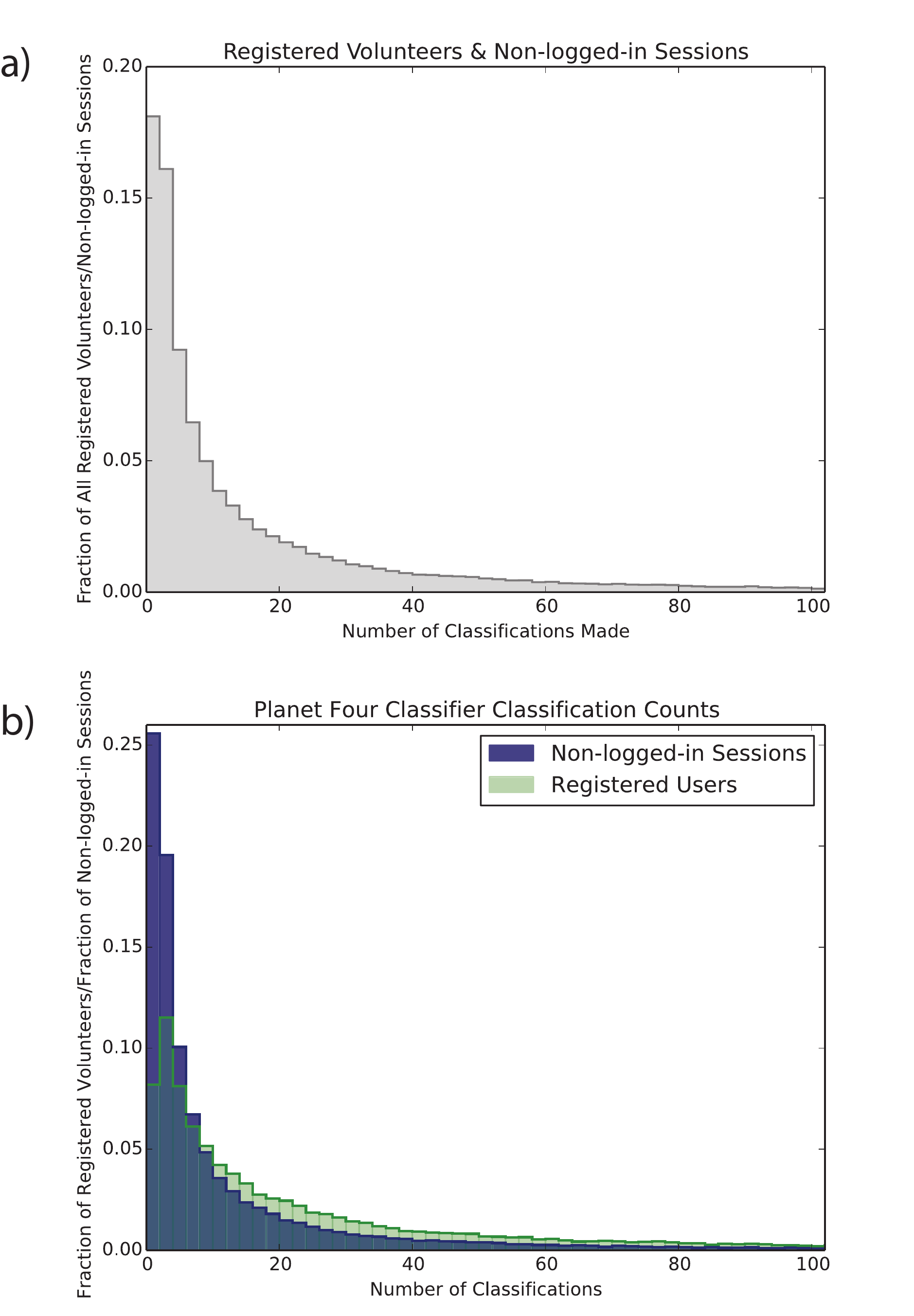}
\caption{\label{fig:classificationhist} Distribution of volunteer classifications.
Figure a shows the combined distribution tallied together for both logged-in and non-logged in sessions.
Figure b shows the volunteer classification count individually for registered and non-logged volunteers.
Both histograms use a bin size of 2.%
}
\end{figure}

\num{36433} registered volunteers and \num{48094} non-logged-in sessions have classified at least one tile in our MY29/30 data-set.
Volunteers made in total \num{9461062} classifications with a median of 7 and average of 41 classifications per registered volunteer/non-logged-in session.
The highest number of different classifications (i.e.\ submitted Planet Four tiles) by the same volunteer was \num{31808}.
After clean-up, Planet Four volunteers drew a combined \num{3460056} blotches, \num{2694415} fans, and \num{805903} interesting features.
Figure~\ref{fig:classificationhist} shows the distribution of volunteer classifications for Seasons 2 and 3 tiles combined.
Individual registered volunteers (median of 14 and average of 69 classifications per user) tend to contribute slightly more classifications than a individual non-logged in session (median of 4 and average of 21 classifications per session).
A given volunteer/session reviews only a small percentage of the entire sample of HiRISE tiles.
Only \SI{15}{\%} of classifiers (\num{12483} registered volunteers and non-logged-in sessions) have contributed more than 50 classifications.
Most volunteers contribute a few classifications of Planet Four tiles before leaving the site.
This is a typical response for web-based projects \citep{crowston2008,zachte2012} and is similar to the volunteer behavior found on other Zooniverse projects \citep{sauermann2015}.

\section{Data reduction}%
\label{sec:data_reduction}

\begin{figure}
\centering
\includegraphics[width=0.77\columnwidth]{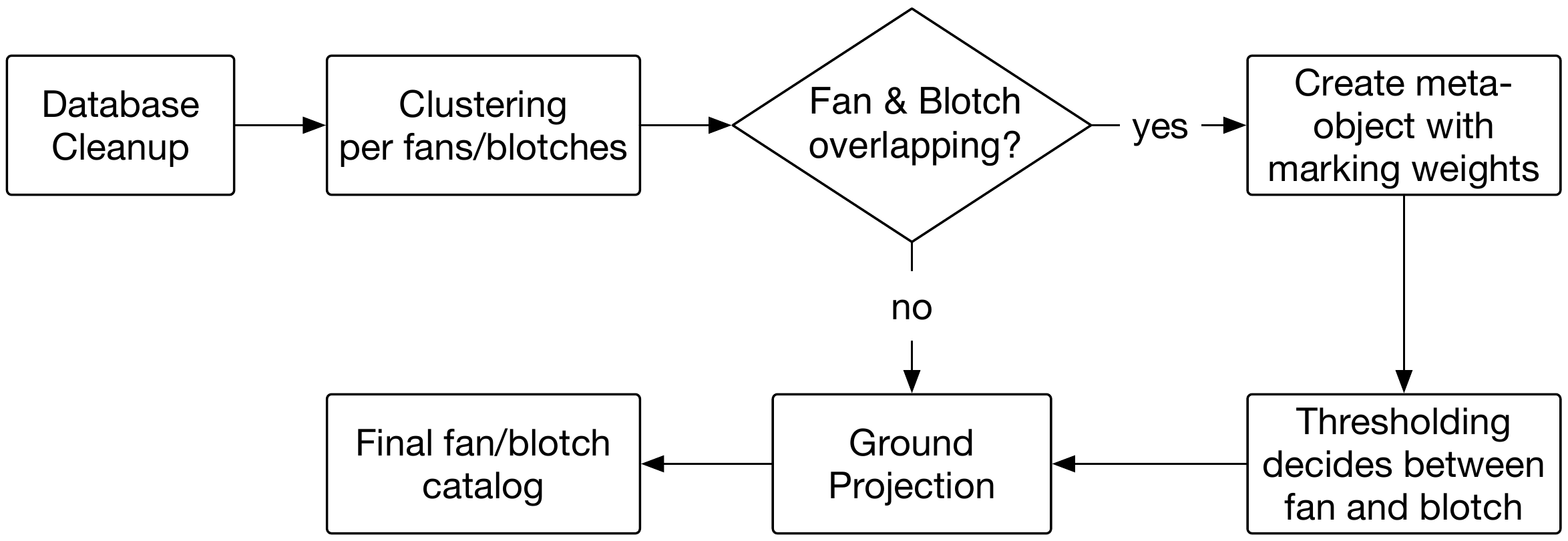}
\caption{\label{fig:data_reduction_intro} Overview of conceptual steps of the Planet Four data reduction pipeline.%
}
\end{figure}

In order to create fan and blotch object catalogs from the Planet Four markings, a reduction pipeline was implemented, for which the code is open source and made available\footnote{The pipeline is located at \href{https://github.com/michaelaye/planet4}{https://github.com/michaelaye/planet4}.
}.
The pipeline is based on the Python programming language, interfacing also to the US Geological Survey's (USGS) Integrated Software for Imagers and Spectrometers (ISIS) \citep{anderson2004,becker2007a}, and making use of the ``scikit-learn'' package for machine-learning related tasks \citep{pedregosa2011}.
This data reduction pipeline has five main conceptual stages (see Fig.~\ref{fig:data_reduction_intro}):
\emph{Cleanup}, where the Planet Four classification data is cleaned, normalized and converted to a binary database (Section~\ref{sec:cleanup}),
\emph{Clustering}, where the markings of the many different volunteers are being combined into, ideally, one resulting average object (Section~\ref{sec:clustering}),
\emph{Combination}, where we combine fans and blotches markings that seem to address the same visible object in the image into a meta-object for further processing during the next stage (Section~\ref{sec:fnotching}),
\emph{Thresholding}, where a cut on the required number of volunteers that voted for either fan or blotch will decide if the previously created meta-object should be considered a fan or a blotch (Section~\ref{sec:thresholding}),
and finally \emph{Ground Projection}, where we project the HiRISE image pixel coordinates of the resulting fan and blotch markings into latitude and longitude coordinates on Mars (Section~\ref{sec:ground_projection}).

\subsection{Database Cleanup}%
\label{sec:cleanup}
After the removal of the tutorial data (see \ref{sec:tutorial}), and a first cleaning for spurious, incomplete and duplicate classification database entries (see Section~\ref{sec:dupes_and_spurious}), we normalize all angles from the Planet Four classification interface, and finally produce a binary database in the format of HDF5 (Hierarchical Data Format, version 5) for the remainder of the data processing.
Normalizing of angles is required because the Planet Four system records blotches with an angular range from -180 to 180 while ellipses possess a degree-2 rotational symmetry.
This means only the range of 0 to 180 degrees is required to fully describe blotches, once the radii are sorted in a consistent way (semi-major axis first).
Volunteers randomly start to draw the ellipses required to mark blotches either from the semi-minor axis or the semi-major axis, making it error-prone to cluster on these parameters without normalization.
The cleaned raw Planet Four classifications as used by this work's analyses are provided as supplemental data to this work in the file \nolinkurl{P4_catalog_v1.0_raw_classifications.csv}.
Further details about the format of the raw classifications are described in \ref{sec:raw_classification}.

\subsection{Clustering}%
\begin{figure}
\centering
\includegraphics[width=1\columnwidth]{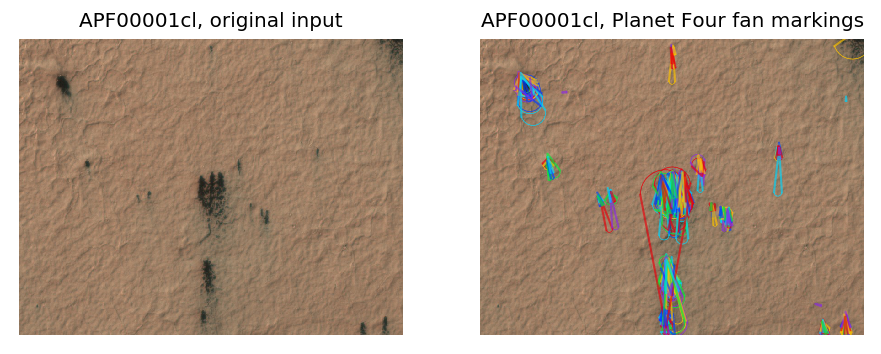}
\caption{\label{fig:fan_markings}
Fan markings for Planet Four tile \texttt{APF00001cl} of HiRISE image \nolinkurl{ESP_012322_0985}.
\textbf{Left}: The cut-out tile that is shown to the Planet Four volunteers.
\textbf{Right}: 51 different users have classified this image.
The colors cycle through randomly for the markings of different users.
With such a large number of different volunteers classifying, the ``sensitivity'' for detection is increased, as notable by a few markings that outline even the smallest potential dark deposit candidates.
However, when the ``crowd'' does not agree with these, i.e.\ if the potential cluster does not reach the \nolinkurl{min_samples} number of required members, the clustering pipeline discards these entries, as shown in Fig.~\ref{fig:fans_clustered}.%
}
\end{figure}

\label{sec:clustering}
We identify fans and blotches by combining together the multiple volunteer assessments from each Planet Four tile.
To identify and precisely locate the marked features from the multiple classifications performed by many (between 30--100, see \ref{sec:ouroboros}) volunteers per Planet Four tile, we perform a clustering analysis on the data.
Figure~\ref{fig:fan_markings} shows an example of fan markings for a Planet Four tile.
After having evaluated several different clustering algorithms, we have identified the Density-based Spatial Clustering of Applications with Noise (DBSCAN) clustering algorithm of~\cite{ester1996} as the most appropriate one for our application.
DBSCAN has the advantage of not requiring the number of expected clusters as input, instead it is controlled by two input parameters describing the minimum number of members of a cluster (\nolinkurl{min_samples}) and the maximum distance for a data point to be included into a cluster (\nolinkurl{epsilon}).
(Details on how we determine these parameters are described in Section~\ref{sec:cluster_parameters}.)
We set up our clustering pipeline using the DBSCAN implementation in the scikit-learn Python library \citep{pedregosa2011}.
All volunteer responses are treated the same with equal weight in the clustering algorithm.
Due to the differences in the classification interface for marking fans and ellipse-shaped blotches --- fans are drawn from a base point vs blotches drawn from the center --- the fans and blotch markings are clustered separately at this stage, and require their own set of clustering parameters.
\begin{figure}
  \centering
  \includegraphics[width=0.5\columnwidth]{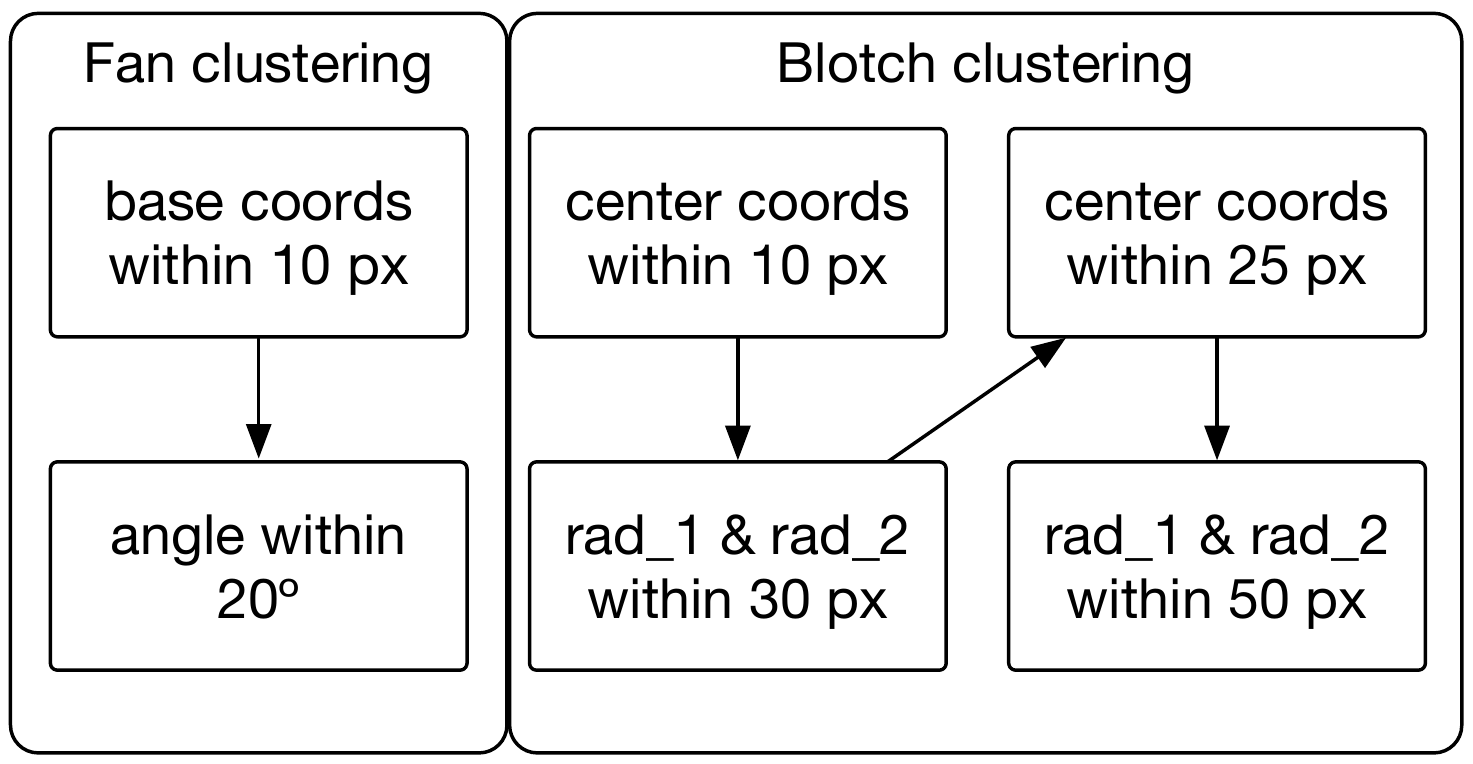}
  \caption{\label{fig:clustering_diagram}
  The sequence of clustering steps for both fan and blotch markings.
  It became apparent during our studies, that fan markings show less scatter, probably due to the tool having to be placed at a clearly identifiable base point.
  Blotches, however, do not show a clearly identifiable center, and their outline is often less sharply defined, creating a wider distribution of marking results, especially for larger blotches.
  This required a second run of clustering with more relaxed cluster parameters, as described in Section~\ref{sec:cluster_parameters} and in Table~\ref{tab:eps_values}.%
}
\end{figure}
\begin{figure}
\centering
\includegraphics[width=0.77\columnwidth]{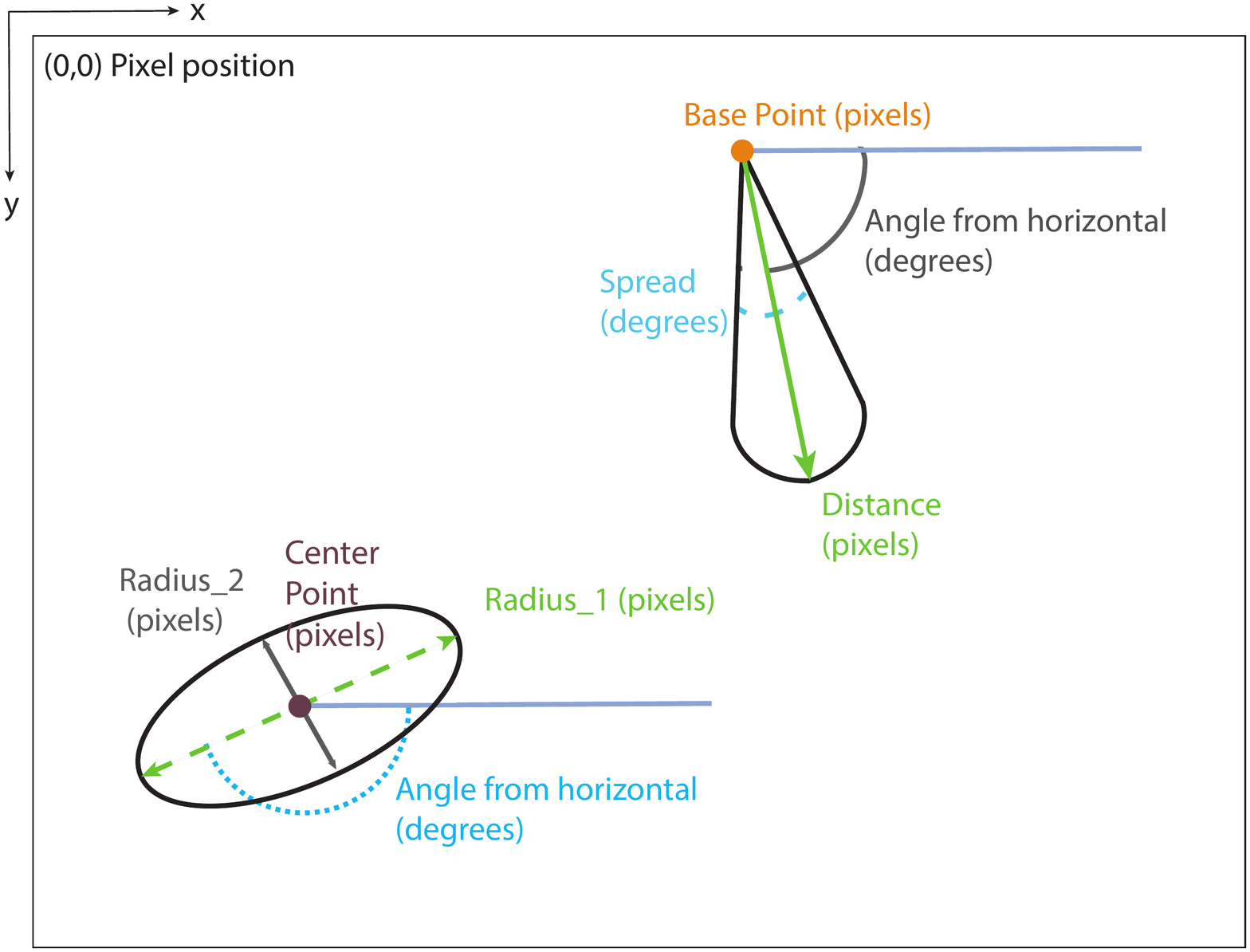}
\caption{\label{fig:marking_coords} The different coordinates available in the Planet Four marking catalog are described here.
Fans possess (x, y) base coordinates, an angle from horizontal for their pointing and a spread angle.
Blotches possess center (x, y) coordinates, semi-major and minor axis radii and also an angle indicating their alignment towards the horizontal. %
}
\end{figure}

\begin{figure}
\centering
\includegraphics[width=1.0\columnwidth]{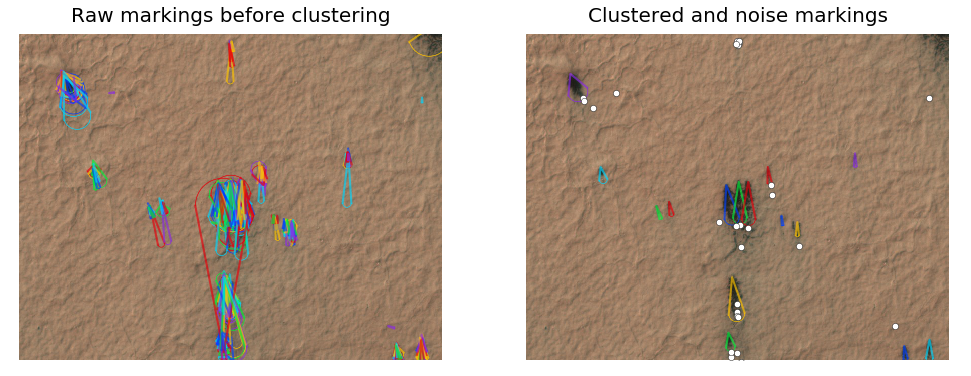}
\caption{\label{fig:fans_clustered} Fans from Figure~\ref{fig:fan_markings} for Planet Four subject ID \nolinkurl{APF00001cl} after applying our clustering pipeline.
\textbf{Left:} For direct comparison, this shows the same as Fig.~\ref{fig:fan_markings} on the right, on page~\pageref{fig:fan_markings}.
\textbf{Right:} Results after clustering, identification of noise markings, and averaging the cluster members' data into one object per cluster.
Markings that do not become member of a cluster are defined as \emph{noise} and will be discarded from further processing (shown as white dots).%
}
\end{figure}

In a first stage, we cluster the data for Planet Four tiles each on the (x,y)-pixel-coordinates of the base point of fans and of the center for blotches (see Fig.~\ref{fig:marking_coords} for a visual description of the available coordinates of the markings.).
Figure~\ref{fig:fans_clustered} shows the result of clustering in two dimensions of the x and y base coordinates of the fan markings, using a multi-step approach as shown in Fig.~\ref{fig:clustering_diagram}, as described below.
Once the clusters for a given set of parameters (see Section~\ref{sec:cluster_parameters} for details on the parameter tuning) have been defined, the original marking data for each cluster members are averaged to create one average marking object per cluster, including average directions for fan objects, e.g.\ in Fig.~\ref{fig:fans_clustered}.
The number of markings that went into the creation of the averaged object is stored for later.

\begin{figure}
  \centering
  \includegraphics[width=0.7\columnwidth]{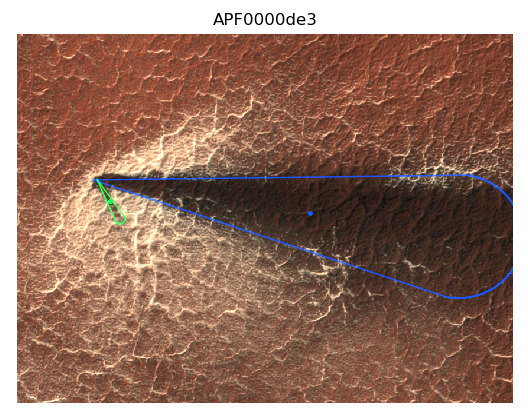}
  \caption{\label{fig:multiple_fans} Planet Four tile \nolinkurl{APF0000de3} from HiRISE image \nolinkurl{ESP_011961_0935}.
  It shows the prevalence and precise identification of \ce{CO2} jet deposits with multiple directions that start from the same base point, indicating multiple eruptions under different wind directions.
  The large fan is the second longest recorded in the catalog, with a length of approx.\ \SI{368}{m}.%
}
\end{figure}

After having clustered both fans and blotches on their base and center coordinates respectively, we apply a second stage of clustering on the markings.
For fan deposits, the major objective of this work is to determine the wind direction they indicate.
Due to this we want to be able to distinguish between different wind directions from the same source point, i.e.\ multiple subsequent eruptions, where later eruptions occurred with a different prevalent wind direction.
In the Planet Four help content we have emphasized that the volunteers should outline several fans if they appear to start from the same source point.
This is very relevant for data like that in Fig.~\ref{fig:multiple_fans}, to identify several wind directions indicated by the fans, from multiple subsequent jet eruptions.
By clustering not only on the base coordinates (x, y) but also on the recorded alignment angle of the fan markings, we are able to distinguish these subsequent fan deposits with different wind directions.

We have determined by reviewing the clustering results of a subset of the data that 20 degrees as a clustering value for angles enables this objective.
It means that fan markings that have an alignment angles further away from each other than 20 degrees are clustered into their own sub-cluster, even if they start at the same base point.
Blotches, on the other hand, are used for deposits that do not clearly indicate a direction, which is why we do not apply an angle clustering here.
However, blotches do not show a clearly identifiable center, and their outline is often less sharply defined, creating a wider distribution of marking results, especially for larger blotches.
Thus, we cluster also on the resulting ellipse radii for the blotches to ensure that we identify the statistically most common shape of the volunteer's blotch markings.

The values of the clustering parameters strongly influence the number of identified features.
We therefore studied extensively, how precisely they affect our results by reviewing random subsets of the data-set, which led to the empirical determination of the clustering parameter values that we eventually used for the catalog production.
These procedures will now be discussed in the following sections (see Fig.~\ref{fig:parameter_scan} for an example of reviewing parameter values).
The results of the clustering stage are then shown in the lower right (blotches) and lower middle (fans) parts of Figures~\ref{fig:p4_pipeline} and \ref{fig:pipeline_examples}.

\subsubsection{Cluster parameters}%
\label{sec:cluster_parameters}
\begin{table}
\centering
\begin{tabular}{llll}
\toprule
\textbf{Marking} & \textbf{Dimension} & \textbf{Small} & \textbf{Large} \\
\midrule
Fans      & xy (base)   & 10 px & NA \\
          & angle (deg) & 20    & NA  \\
Blotches  & xy (center) & 10 px & 25 px \\
          & radius (px) & 30 px & 50 px \\
\bottomrule
\end{tabular}
\caption{{\label{tab:eps_values}Empirically determined \nolinkurl{epsilon} values for the clustering pipeline.
NA: Fan markings did not require a second clustering run with relaxed precision on the distance, apparently the fact that a fan requires drawing from a distinguishable starting point helped the volunteers to keep the scatter small, both in base coordinates and angle precision.}
}
\end{table}
\paragraph{\nolinkurl{min_samples}}%
\label{sec:min_samples}
As described in Section~\ref{sec:classification_database}, Planet Four tiles have varying numbers of user classifications, thus the classifications for each Planet Four tile are clustered separately, with a variable requirement on the \nolinkurl{min_samples} clustering parameter.
More classifications for a Planet Four tile means that we have a higher ``sensitivity'' to smaller features (see for example Fig.~\ref{fig:fan_markings}, right), so to achieve a uniform detection efficiency, we implement a scaling factor on the required number of samples per cluster.
This results both in a higher sensitivity to have seasonal fans and blotches marked and higher precision averaged objects at the end of the clustering process.
In other words, the signal-to-noise ratio (SNR) is higher for a Planet Four tile that was classified by a larger number of volunteers and we adapted the clustering process to normalize for that fact.

To address the variable SNR in our data, we empirically determined a scaling factor \nolinkurl{min_samples_factor} (MSF) that, multiplied with the number of classifications that contain blotch or fan markings, results in the \nolinkurl{min_samples} value for the DBSCAN algorithm:
\[\texttt{min\_samples}=round\left(\texttt{min\_samples\_factor}\cdot n_{\texttt{markings}}\right),
\]
with \(n_{\texttt{markings}}\leq n_{\texttt{classifications}}\), the number of classifiers that have added either blotch or fan markings as classifications.

The best value for MSF was empirically found to be at \num{0.13}.
For example, when a Planet Four tile has \(n_{\texttt{class}}=30\) classifications (our current retirement value), \(n_{\texttt{class}}\) will be 4.
This value now provides the number of cluster members \nolinkurl{min_samples} that is required for a cluster to be created.
When a tile has 70 submissions, however, it would result in the requirement of having 9 cluster members to be deemed a real detection and to be entered into the next stage of the pipeline.
This way, we are exploiting the higher sensitivity from the larger number of submitted classifications.

\begin{landscape}
\begin{figure}[p]
\centering
\includegraphics[width=1\columnwidth]{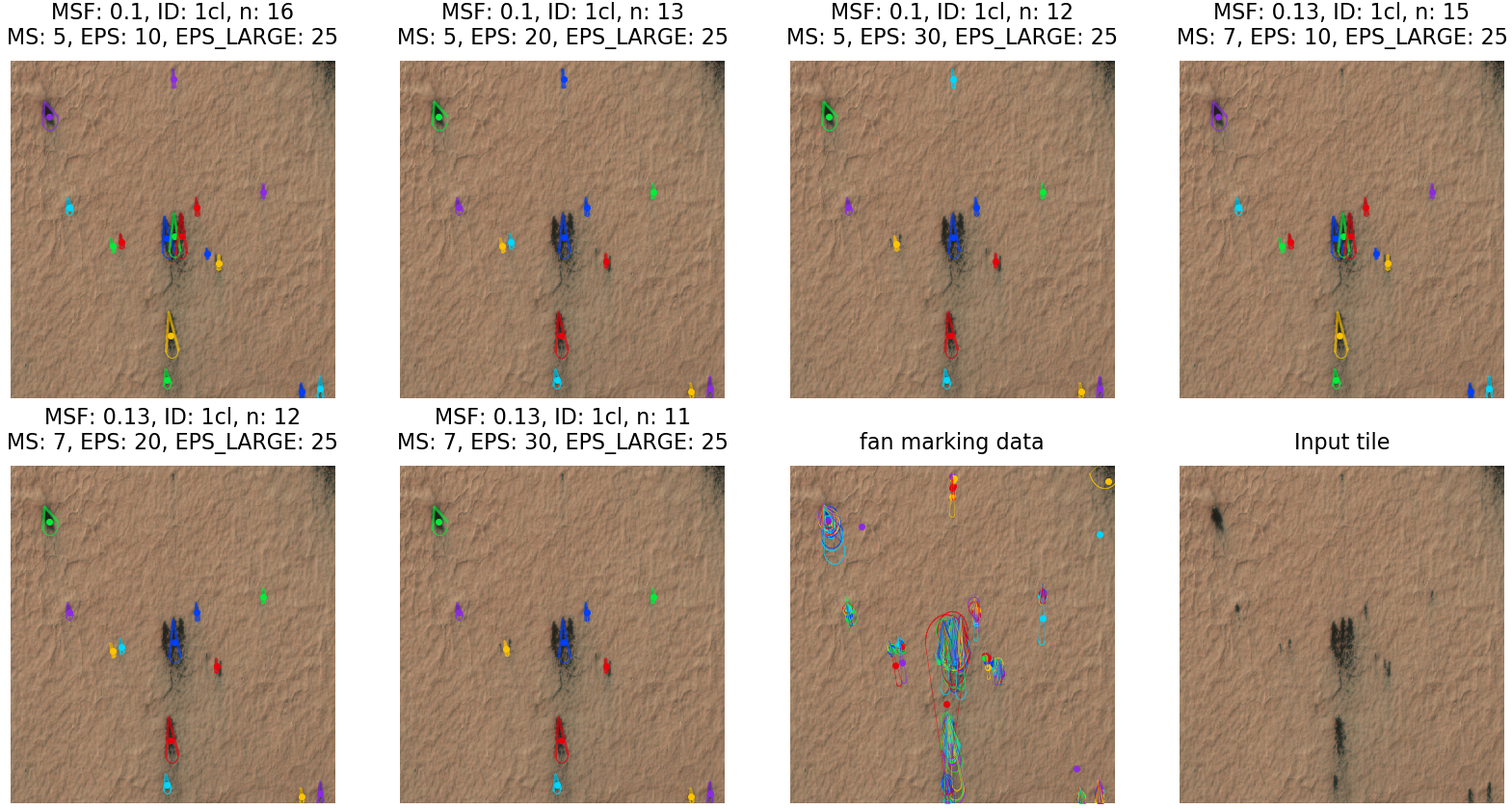}
\caption{\label{fig:parameter_scan} This figure shows our review plots for determining the best clustering parameters for Planet Four tile ID \emph{1cl}.
In this example, we review the fan clustering with a group of 2 different \nolinkurl{min_samples} values, controlled by using a \nolinkurl{min_samples_factor} of 0.1 and 0.13 respectively, leading to \nolinkurl{min_samples} values of 5 and 7.
Additionally, we are scanning the \nolinkurl{epsilon} (EPS) value for small deposits with the settings 10, 20, and 30 pixels, while the \nolinkurl{EPS_LARGE} value stays at 25 pixel for these runs (having no effect in this case due to the small size of markings).
The upper left 3 plots are for the setting of MSF=0.1 (resulting in a \nolinkurl{min_samples} value of 5), and EPS between the 10, 20, and 30 pixel values.
Then, the second group with an MSF of 0.13 (resulting in \nolinkurl{min_samples}=7), starts in the upper right with the fourth plot in the upper row, and continues in the lower left with the first two plots, again showing the tests for EPS values 10, 20, and 30 pixels respectively.
The last two plots in the lower row provide us with what the volunteers actually marked and what they received as input for the markings, the Planet Four tile, cut out from the larger HiRISE images.
The number of fans clustered varies significantly for different clustering parameter values, with \(n\) between 11 and 16.
We favor the setting in the upper right plot, for identifying correctly all small center fans, while not creating an object for the small black spot at the top of the image tile.
}
\end{figure}
\end{landscape}

\begin{landscape}
\begin{figure}[p]
\centering
\includegraphics[width=1\columnwidth]{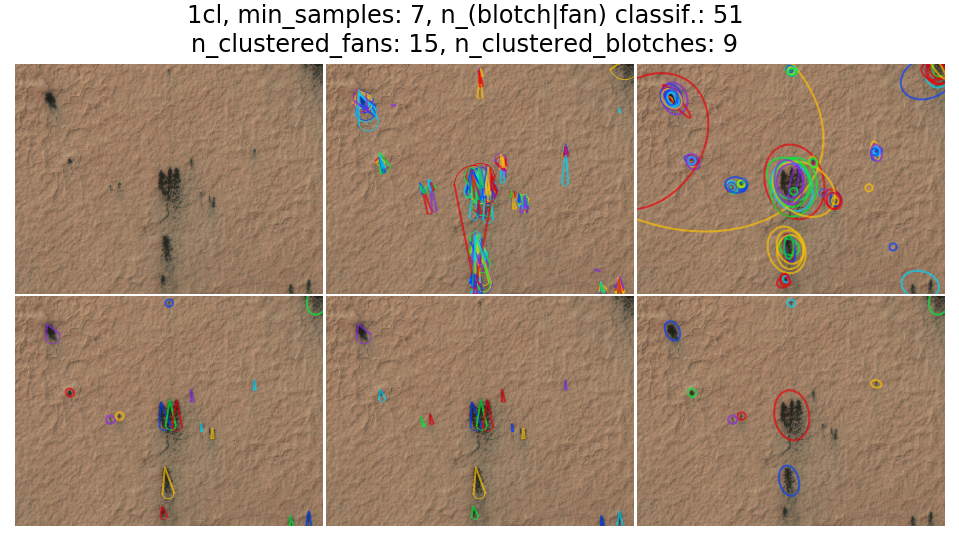}
\caption{\label{fig:p4_pipeline} This figure shows the final pipeline result of the tile from Fig.~\ref{fig:parameter_scan}.
\textbf{Upper Left}: The input tile; \textbf{Upper Middle}: Fan markings of the volunteers; \textbf{Upper Right}: Blotch markings of the volunteers; \textbf{Lower Right}: Blotch markings after clustering and averaging the cluster members;
\textbf{Lower Middle}: Fan markings after clustering and averaging the cluster members;
\textbf{Lower Left}: These are the final catalog entries.
To reach this, the results from Lower Middle and Lower Right are being compared, and the higher voted markings at comparable locations win.
How high that winning ratio must be to be entering the final catalog is determined by the threshold value (see Section~\ref{sec:thresholding}).
Note, how the center fans are cleanly identified and winning in the voting competition with the blotch at the same location.
The opposite is true for the the small object identified at the middle left, where a red blotch marking has won against the small cyan fan.
}
\end{figure}
\end{landscape}

\begin{figure}[p]
\centering
\vspace{-2cm}
\includegraphics[width=0.9\columnwidth]{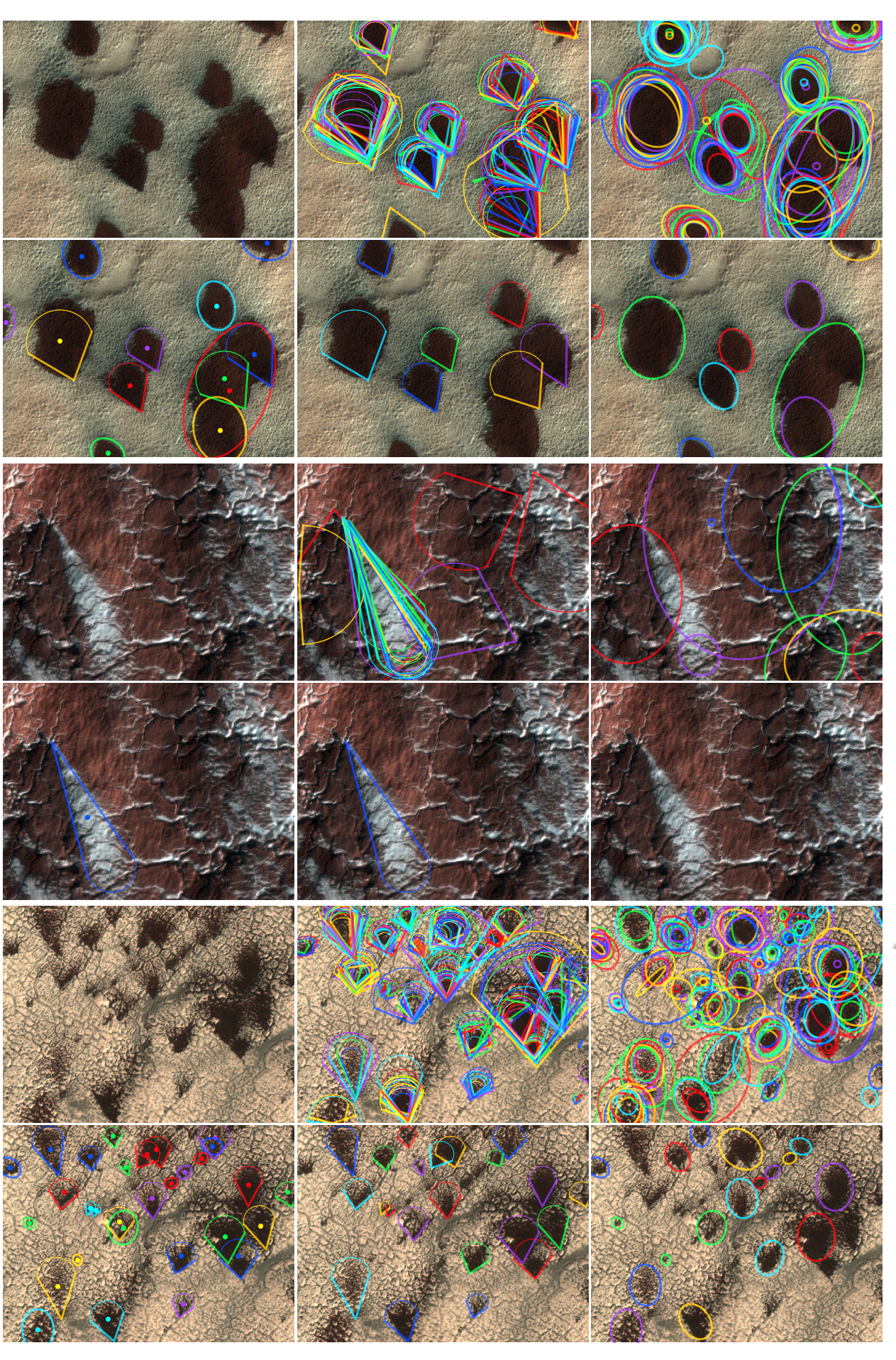}
\caption{\label{fig:pipeline_examples}
Three example Planet Four tile pipelines, for APF0000b0t, APF0000ops, and APF0000bk7.
See Fig.~\ref{fig:p4_pipeline} for a detailed description of the pipeline plotting sequence.}
\end{figure}

\paragraph{\nolinkurl{epsilon}}%
\label{sec:epsilon}
The second DBSCAN parameter, \texttt{epsilon}, describes the largest distance that two points are allowed to have, for them to be considered to be in the same cluster.
The dimension for this measurement depends on what mathematical feature is currently being clustered.
When we cluster on the base point coordinates of fans, the central point coordinates or semi-radii of blotches, the feature space is measured in pixels, while fan angles are clustered in degrees.
The size scale of the dark fans and blotches varies significantly between different regions of interest at the south pole of Mars.
Trying to cluster our data with only one value of \texttt{epsilon}, we realized that it was not possible to simultaneously resolve small markings on the order of 20 pixels properly that were precisely positioned by the volunteers, while also clustering successfully markings of much larger deposits that could stretch more than half of the Planet Four tile that was shown to the volunteers.
The spread in marking coordinates is smaller for smaller features --- we think because of an increased focus to detail for smaller features ---, and thus, to ensure identification of large features, we implemented a second stage of clustering with larger allowed values for \texttt{epsilon}.
The resulting values in Table~\ref{tab:eps_values} were selected empirically after review of a random subset of the pipeline output.
Fig.~\ref{fig:parameter_scan} shows an example parameter scan review graphic that the science team used to determine the parameter values that work best for our task.

\subsection{Combination}%
\label{sec:fnotching}

When the direction of fan deposits are not very pronounced, i.e.\ the prevalent winds were weak at the time of the jet eruption, there is ambiguity in identifying the deposit as a fan or a blotch.
This can result in a given ground source having both survived clusters of fan and blotch markings that need to be combined in a strategic way to create a final object category for the observed ground source that will be listed in the resulting object catalog.
We make use of the relative frequency of which marking tool was used to create both marking clusters to identify how fan-like a source is.
For example, if 5 people classified a marking as a fan, but 5 other people marked it as a blotch, we assign a fan probability P(fan) of 0.51 by applying
\[ P\left(\texttt{fan}\right)=\frac{n_{\texttt{fans}}+0.01}{n_{\texttt{fans}}+n_{\texttt{blotches}}}, \]
with \(n_{\texttt{fans}}\) and \(n_{\texttt{blotches}}\) the number of volunteers that marked either.
The fudge value 0.01 is required to be able to make an either-or decision for the object when \(n_{\texttt{fans}}=n_{\texttt{blotches}}\),  flipping the switch in this close call for fans instead of blotches, due to the usefulness of fans for further scientific analysis.

We determine to which markings this procedure is applied by calculating the pair-wise Euclidean distance for all clustered objects and check if clusters are within a chosen limit of 30 pixels with each other.
We chose this value for allowing slightly more imprecision in the markings' positioning as the clustering algorithm that went into creating these average, but without combining too many markings that really should be individual items.
We have reviewed several hundred subsets of data and determined 30 pixels to be a good compromise on these competing tasks.
If a distance pair meets the combination criterion, we use above formula to calculate P(fan) for this pair of markings.
This value goes from 0 to 1 with 0 being a definite blotch when \(n_{fans}=0\) and 1 indicating a definite fan when \(n_{blotch}=0\), in other words either none or all volunteers had drawn a fan or a blotch, respectively.
We then create a meta-object for this pair, storing P(fan) under the name `vote\_ratio' in the catalog files, together with all other data for both objects.
We do this to enable future users of the catalog to decide on their own how reliably a marking is required to be a fan before it shall be used as such, with its data entering a study.
In other words, a specific study might require to only use the most clear fan markings, maybe with a P(fan) of larger than 0.8.
Applying such a cut is called Thresholding in our pipeline, described in the next section.

\subsubsection{Thresholding}%
\label{sec:thresholding}
For concrete applications, e.g.\ for this publication, a scientist can now apply a cut on P(fan), that will write out the decision to a new catalog file with fans and blotches.
For example, a cut on P(fan) of 0.8 would mean that all meta-objects with a value of smaller than 0.8 will be written out as the underlying blotch, while for meta-objects with a value of larger than 0.8 the stored fan will be written out.
In both cases, the remaining data of the meta-object that was thresholded against will be dropped for the newly created catalog file, but it is still available for other thresholding operations as an intermediate data product.
An example use case would be that a scientist wants to study the sensitivity of their research on the applied cut, for example, if we want to provide wind direction data to a mesoscale climate simulation, we might want to make sure that only the most certain directions are being used and would apply a higher cut on the meta-object value.

For the catalog that we deliver with this work, we chose a simple majority threshold of 0.5, so that the catalog offers the broadest use case.
Choosing simple majority means that we take a marking as a fan from the moment that at least an equal amount of volunteers have classified an object as a fan and as a blotch.
Catalog files with this applied P(fan) threshold of 0.5, all intermediate data products, and instructions on how to apply a threshold for writing out new catalog files will be provided as supplementary products (see \ref{sec:pipeline_outputs} for more details).

\subsection{Ground Projection}%
\label{sec:ground_projection}
For each Planet Four tile, the clustering in volunteer-drawn markings to identify seasonal sources is performed using the pixel positions of Planet Four tiles.
Once the cluster dimensions and position has been identified, the source's true location on the South Pole must be calculated.
However, the  HiRISE team-generated non-map projected color mosaics the Planet Four tiles are  derived from do not contain the spacecraft information necessary to compute the latitude and longitude per pixel.
We partially reconstruct the mosaics from the raw HiRISE image products or Experiment Data Records (EDRs) building a red filter only composite image with the necessary spacecraft information required to perform coordinate transforms.
The HiRISE EDRs were obtained from the NASA's Planetary Data System (PDS) HiRISE PDS Data Node.
We developed a reduction pipeline in Python using the US Geological Survey's (USGS) Integrated Software for Imagers and Spectrometers (ISIS)\footnote{http://isis.astrogeology.usgs.gov/} \citep{anderson2004,becker2007a} and the ISIS-3 Python wrapper Pysis\footnote{https://github.com/wtolson/Pysis} for this purpose.

\begin{figure}
\centering
\includegraphics[width=0.7\columnwidth]{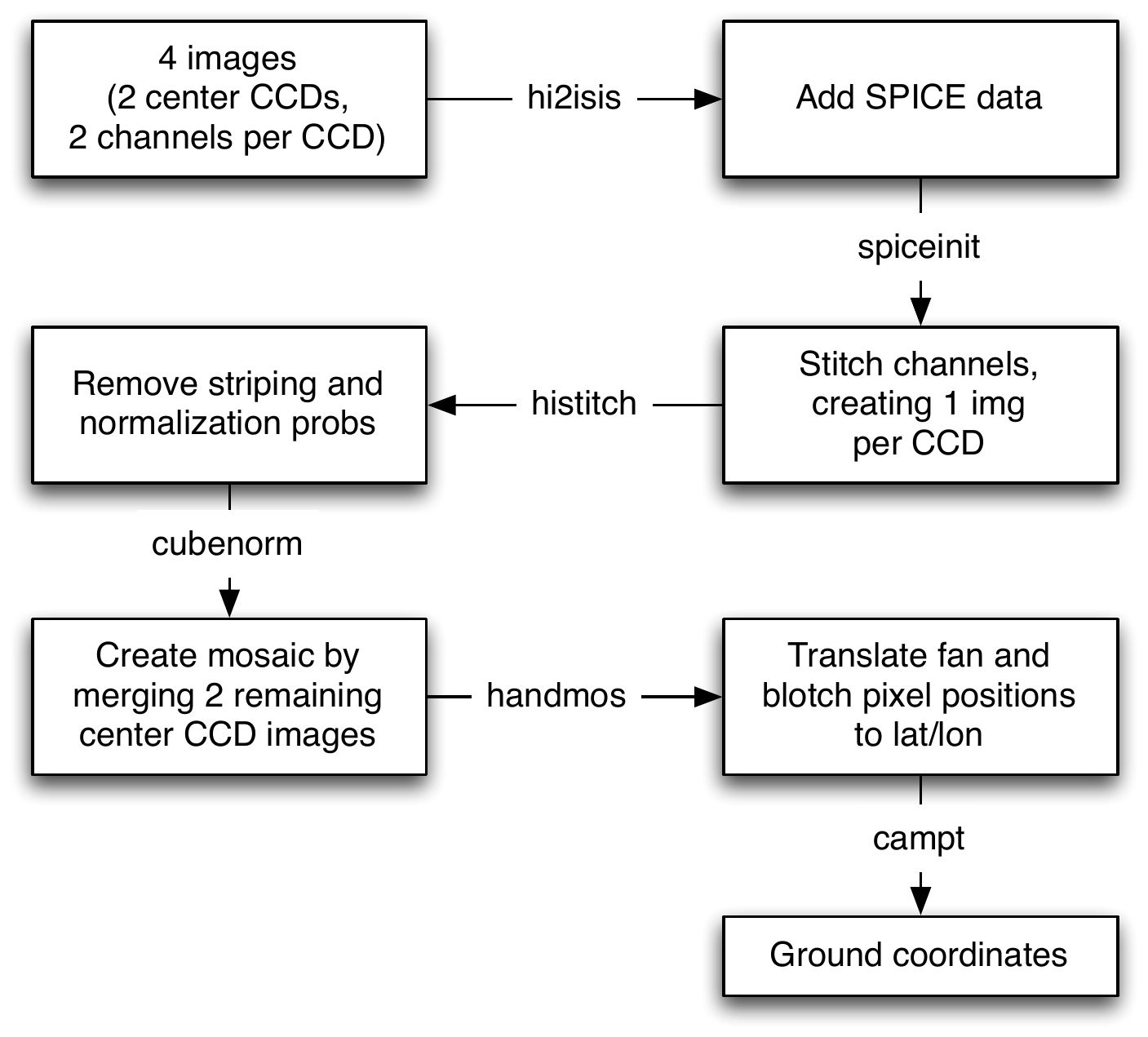}
\caption{\label{fig:isis_pipeline}
Process for creating single channel non-map projected mosaics with the required SPICE header information used to convert Planet Four feature pixel coordinates to geographical lat/lon coordinates.
The required ISIS-3 applications for each stage are listed in the arrows.%
}
\end{figure}

We briefly summarize the steps as shown in Fig.~\ref{fig:isis_pipeline} including the required ISIS-3 application names, to generate the red filter-only mosaic.
We start with the center two RED filter CCDs (RED 4 and 5), each with two readout channels.
All four EDR files (2 for each CCD) are read in and converted to ISIS-3 cube format, and the SPICE (Spacecraft \& Planetary ephemerides, Instrument C-matrix and Event kernels) information for MRO is added to the EDR headers.
For each CCD, we combine the two channel EDRs into a single image.
The combined image is then normalized to remove both the striping and left/right normalization effects.
This is not a necessary step for obtaining map project information but makes it easier to visually inspect the final combined mosaic.
Once both CCDs have been reduced they are combined in a final mosaic accounting for the 48 pixel (in 1\(\times{}\)1 binning) overlap.

Once the single filter red mosaic is made, we are able to translate any fan and blotch pixel position to latitude and longitude on the south pole using ISIS-3's \emph{campt} application.
The catalog tables \nolinkurl{P4_catalag_v1.0_L1C_cut_0.5_fan_meta_merged.csv} --- and \nolinkurl{_blotch_meta_merged.csv} respectively ---, provided as supplemental files include the cluster coordinates as latitude/longitude derived from this process, as well as a set of positional coordinates (X,Y,Z) in the body-fixed reference frame for Mars, measured in kilometers.

\subsection{Overlap regions}%
\label{sec:overlap}

\begin{figure}[p]
\centering
\vspace{-2cm}
\includegraphics[width=0.75\columnwidth]{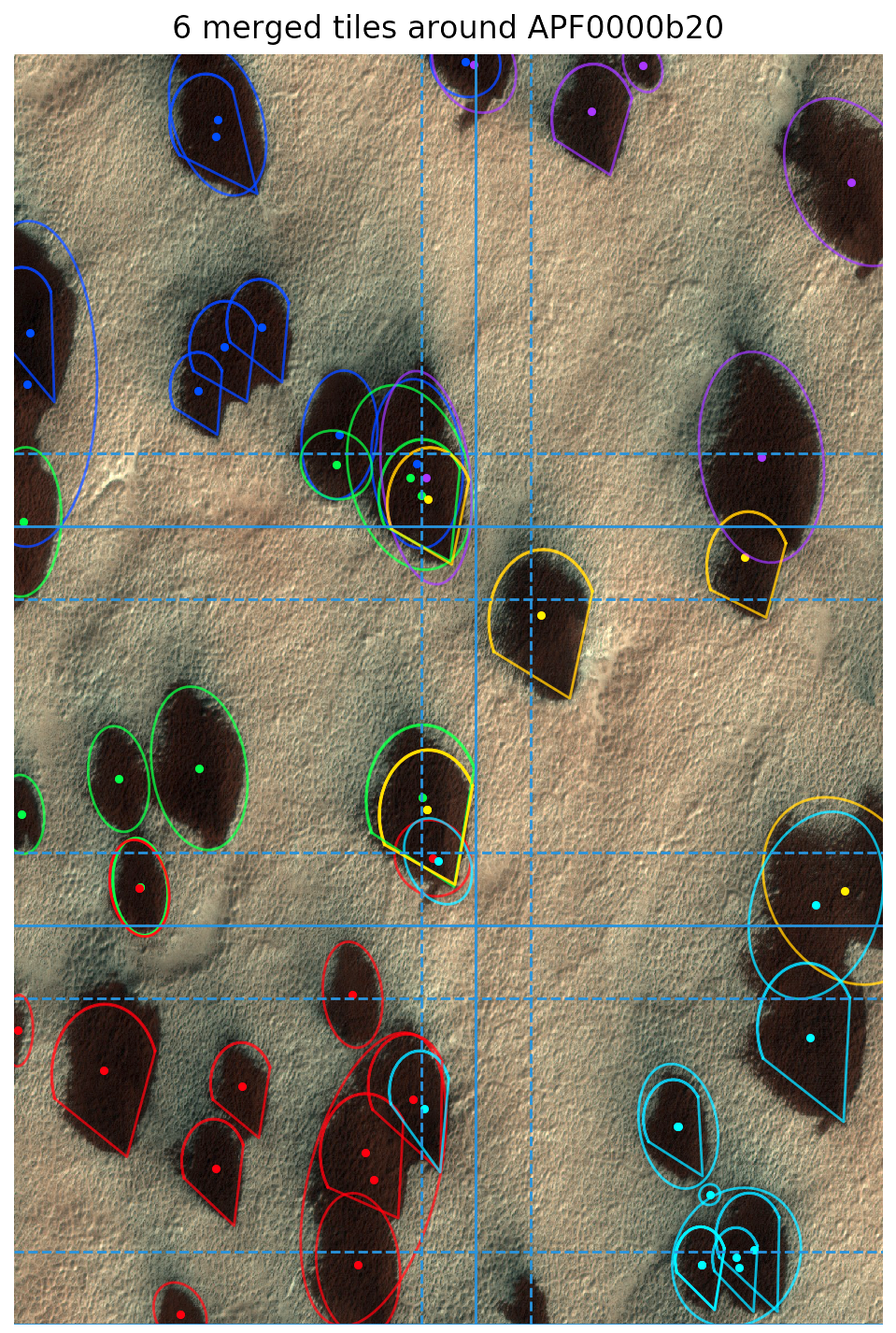}
\caption{\label{fig:overlap_merged}
Six neighboring Planet Four tiles of HiRISE image \nolinkurl{ESP_011931_0945} are merged in this plot.
The tiles have the following tile coordinates within the HiRISE image and Planet Four tile\_ids: Upper Left: (1, 33),b1j; Upper Right: (2, 33), b10; Middle Left: (1, 34), b0p; Middle Right: (2, 34), b20; Lower Left: (1, 35), b0t; Lower Right: (2, 35), b0a (all 3 letter tile\_ids need to prepend `APF0000' for the full ID).
The shape of the tiles are distorted compared to their displayed on-screen size for this plot.
Each tile was clustered individually, indicated by the different marking colors.
The solid lines indicate where an unshared division between the tiles would lie, the dashed lines show the overlap region that was added to each tile to maximize available information for the volunteers.
This plot is instructive in showing how the marked fans, specifically their directions match very well, despite the fact that sometimes only a very small part of the whole fan marking was visible to the classifying volunteer.
For increased precision in total marking counts and the area covered by markings we will design an object merging procedure on these overlap regions (next paper).%
}
\end{figure}

As previously mentioned in Section~\ref{sec:HiRISE}, to avoid edge effects, the cutting down of HiRISE images into screen-sized tiles is performed such that there is a 100-pixel overlap with the neighboring tiles.
This way, at least in one of the tiles of an area fans and blotches that cross the boundary between tiles will be visible completely.
However, from our own Planet Four marking efforts and from analyzing results from Planet Four volunteers, we have determined that the classification tools do provide such high level of precision in placement, that many volunteers position and push a fan or blotch marking out of bounds of the shown image area to make it fit a partially shown fan or blotch.
This results in several markings for the same object stemming from different Planet Four tiles, as shown in Fig.~\ref{fig:overlap_merged}.
It can be seen in this figure that the directions of fans are matching, despite the fact that some tiles only showed a small part of a fan in the overlap area.
We hence conclude that a wind direction analysis is not adversely affected by this analysis artefact.
For a future study focusing on area covered by markings and counts of fan and blotch activity, we will implement a merging procedure to remove multiple markings, similar to the \emph{Combination} step in our pipeline, as described in Section~\ref{sec:fnotching}.

\section{Data Validation}%
\label{sec:data_validation}

To date, there is no published catalog of the locations and numbers of seasonal defrosting features for any of the HiRISE images of the Martian south polar region to compare to the Planet Four results.
In order to assess the accuracy and recall rate of Planet Four and confirm the majority of fans and blotches present in the HiRISE observations are identified when combining multiple classifier markings, we have created a `gold standard' data-set based on expert assessment.
Using the same classification interface and markings tools on the Planet Four website as the citizen scientists used, the Planet Four Science team reviewed a subsample of the Seasons 2 and 3 tiles and produced a catalog of markings.
Similar validation processes have been applied in analyses of our previous Planet Four publication for the sister project Planet Four: Terrains \citep{schwamb2017a} and to crater counting crowd-sourced data for the Moon \citep{robbins2014,bugiolacchi2016}.

To generate the gold standard data-set, 960 Season 2 tiles and 767 Season 3 tiles were randomly selected and equally divided amongst the three of the primary Planet Four Science Team members (GP, KMA, MES) to review.
This corresponds to \SI{3}{\percent} of the tiles from each season classified on Planet Four.
Additionally another 192 tiles, both from Season 2 and 3, were randomly chosen and classified by all science team gold standard classifiers in order to compare the science team markings to each other.
This corresponds to approximately \SI{0.4}{\percent} of each season's tiles.
The Planet Four tile\_ids of the gold standard classifications and the user names of the science team members that did the analysis are provided in supplemental data files \nolinkurl{P4_catalog_v1.0_gold_standard_ids.zip}.

\begin{figure}[htbp]
\centering
\includegraphics[width=1\columnwidth]{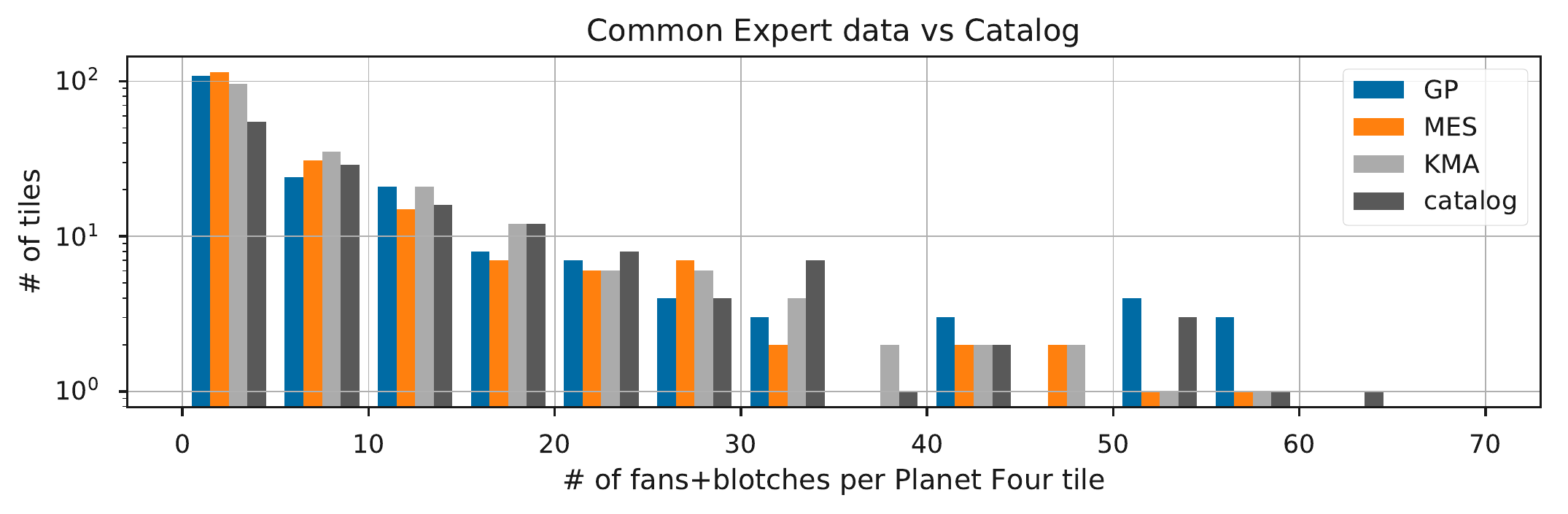}
\caption{\label{fig:gold_fans_blotches_common}
Comparing counts of identified objects (i.e.\ fans and blotches together) per Planet Four tile between experts and the catalog data; here, for the 192 common tile\_ids that were classified by all experts.
Bin size is 5, each bin is directly compared between the data from all experts GP (blue), MES (orange), KMA (grey) and the catalog results (brown).
Binning max was cut off at 75, omitting single entry bins above.
}
\end{figure}
\subsection{Counts of objects identified}
We use the expert classifications from the science team with our final catalog in order to explore how well fan and blotch features are identified and how accurately the shapes and dimensions are represented in the Planet Four catalog.
We show a tile-based comparison in Section~\ref{sec:gold_tile_comparison}, but first we examine the collective properties of the part of the Planet Four catalog that represents the gold standard tiles.
We compare and contrast these distributions to the expert classifications together and per expert reviewer.

\begin{figure}
\centering
\includegraphics[width=1\columnwidth]{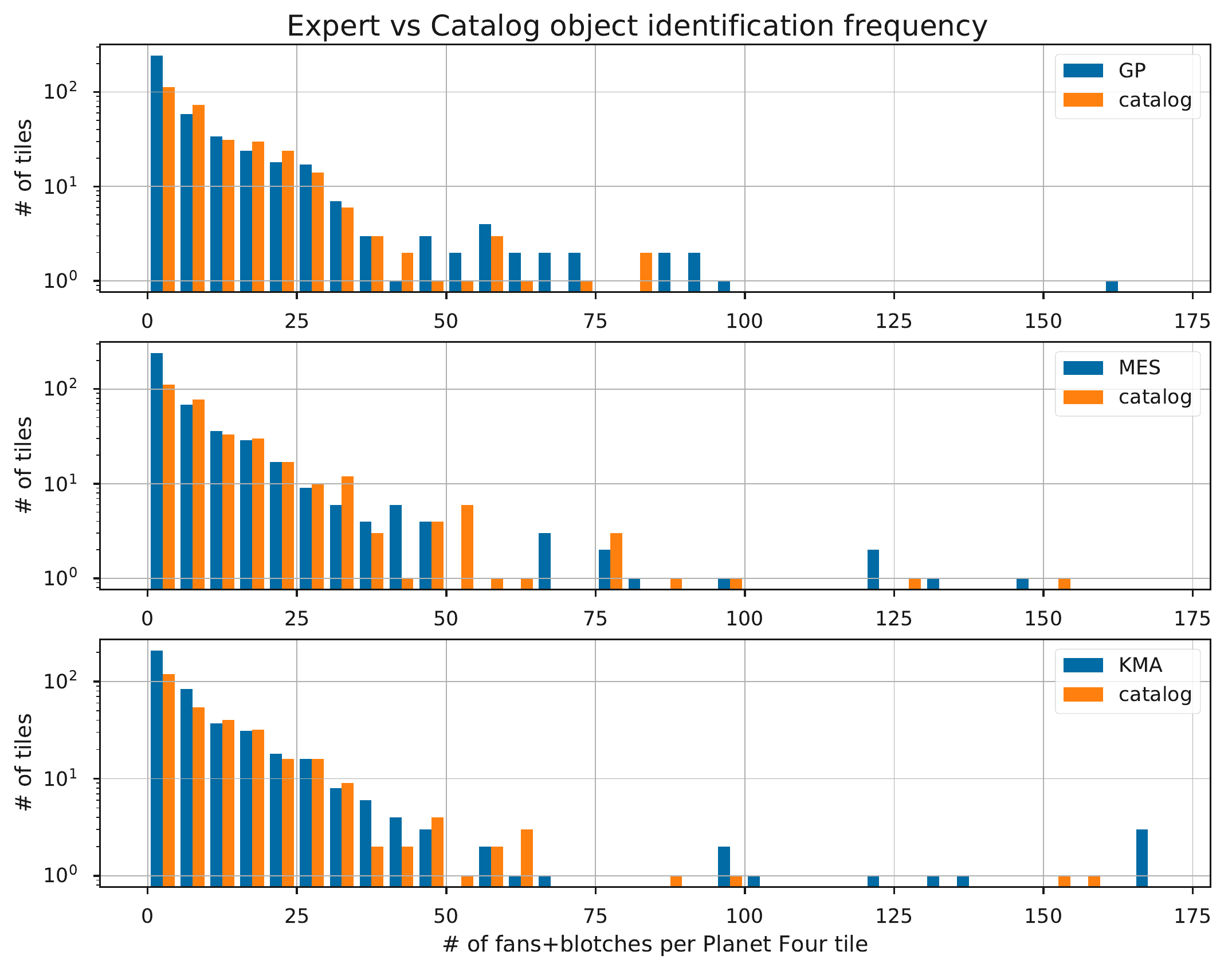}
\caption{\label{fig:gold_fans_blotches}
Comparing counts of identified objects (i.e.\ fans and blotches together) per Planet Four tile between experts and the catalog data.
Bin size is 5, each bin is directly compared between data from experts (in dark blue) and catalog data (in orange), with the experts GP, MES, and KMA respectively, from top to bottom.
Each histogram contains data for 432 tiles, with each expert classifying an independent data-set.
}
\end{figure}

Figure~\ref{fig:gold_fans_blotches_common} compares the number distribution of identified sources (i.e.\ fans + blotches) per Planet Four tile between experts and the catalog data for the 192 common tiles that were commonly classified by all three science team members (KMA, GP, MES).
Among the expert classifiers there are some visible differences especially where the interpretation of a single image or two dominates the value of the histogram bin.
The final catalog is within the variance of the individual expert assessments.
We can see this further in Figure~\ref{fig:gold_fans_blotches} which shows the number distribution of identified objects (i.e.\ fans and blotches together) per Planet Four tile when comparing the results for the tiles that were only classified by one of the science team members. We note that even tiles with 30 or 40 fans and/or blotches are still well represented in the catalog.

\subsection{Fan lengths and blotch areas}
\begin{figure}[htb]
\centering
\includegraphics[width=1\columnwidth]{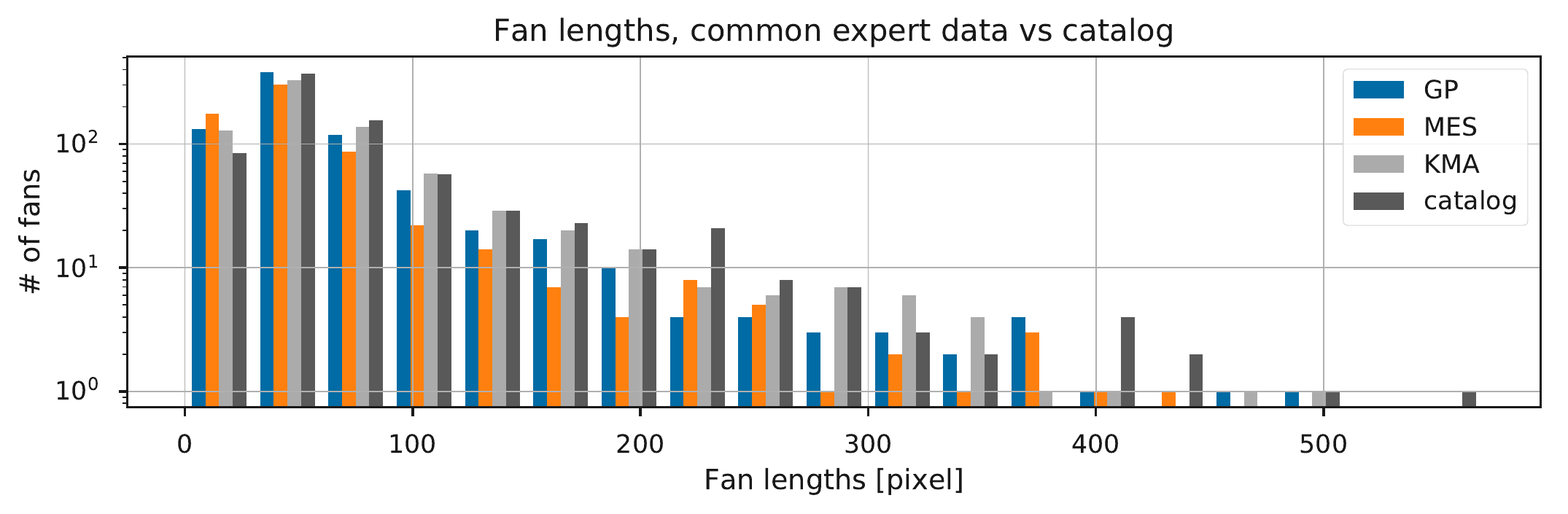}
\caption{\label{fig:gold_fan_lengths_common}
Comparing measured fan lengths between experts and the catalog data; here, for the 192 tile\_ids that were classified by all experts.
Bin size is 30, each bin is directly compared between the data from all experts GP (blue), MES (orange), KMA (grey) and the catalog results (brown).
Binning max was cut off at 600, omitting single entry bins above.
}
\end{figure}
\begin{figure}[htb]
\centering
\includegraphics[width=1\columnwidth]{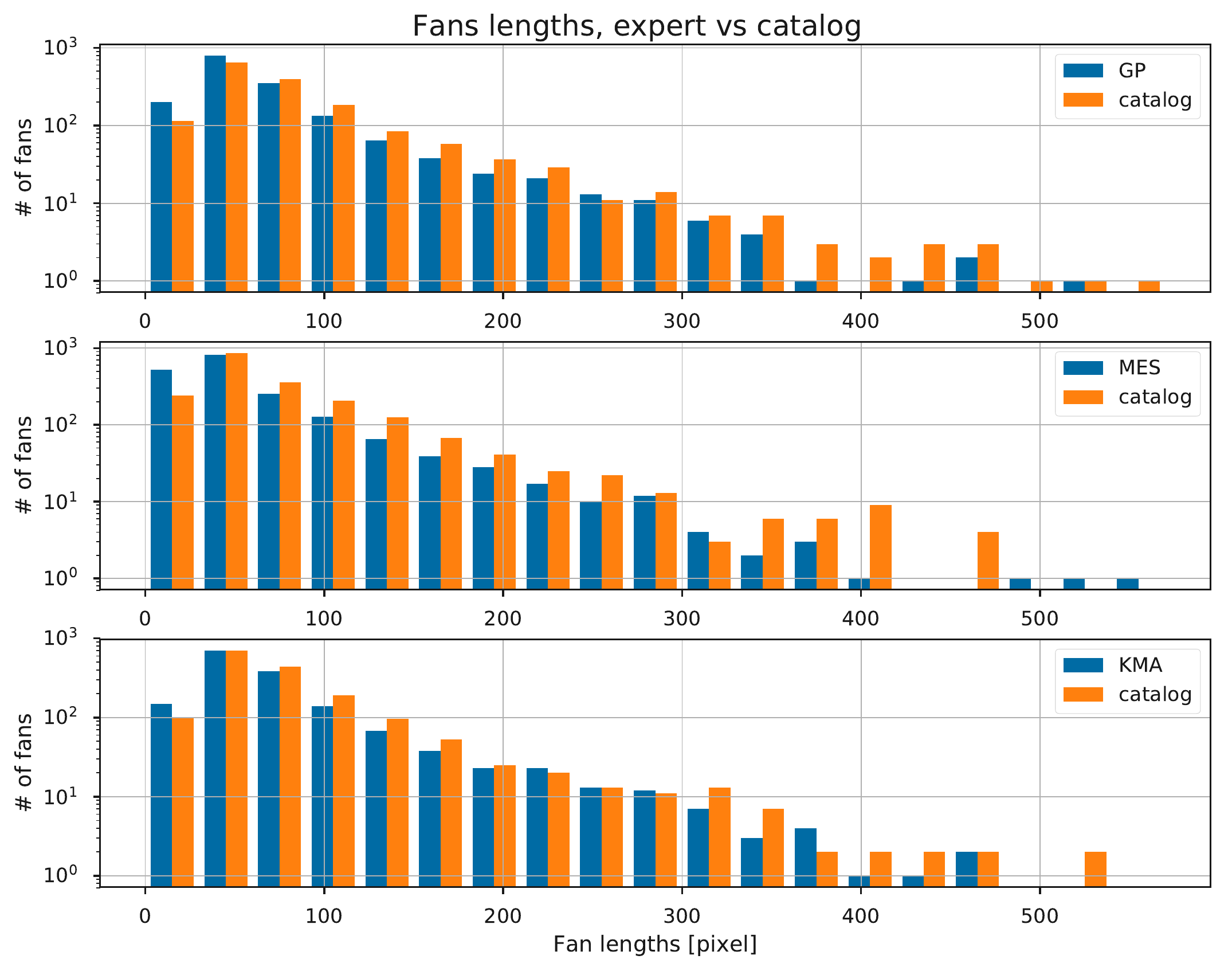}
\caption{\label{fig:gold_fan_lengths}
Comparing measured fan lengths between experts and the catalog data.
Bin size is 30, each bin is directly compared between the data from all experts GP (blue), MES (orange), KMA (grey) and the catalog results (brown).
Binning max was cut off at 600, omitting single entry bins above.
}
\end{figure}
\begin{figure}[p]
\centering
\vspace{-2cm}
\includegraphics[width=1\columnwidth]{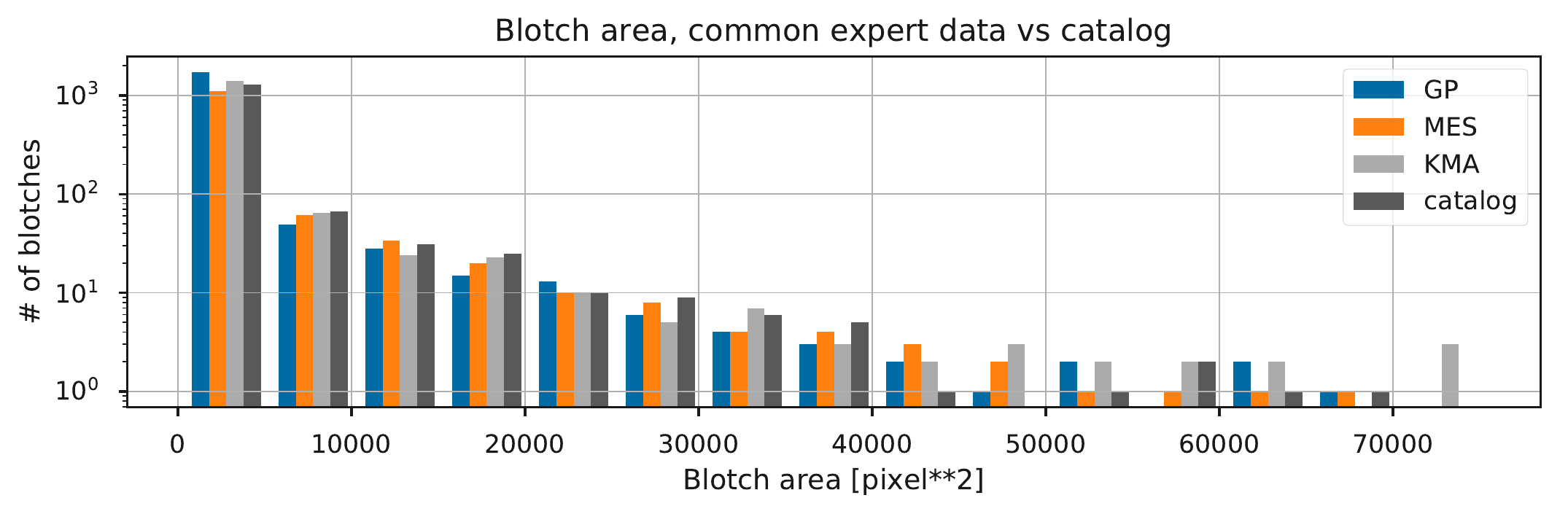}
\caption{\label{fig:gold_blotch_areas_common}
Comparing measured blotch areas between experts and the catalog data; here, for the 192 tile\_ids that were classified by all experts.
Bin size is 5000, each bin is directly compared between the data from all experts GP (blue), MES (orange), KMA (grey) and the catalog results (brown).
Binning max was cut off at \num{80000}, omitting single entry bins above.
}
\centering
\includegraphics[width=1\columnwidth]{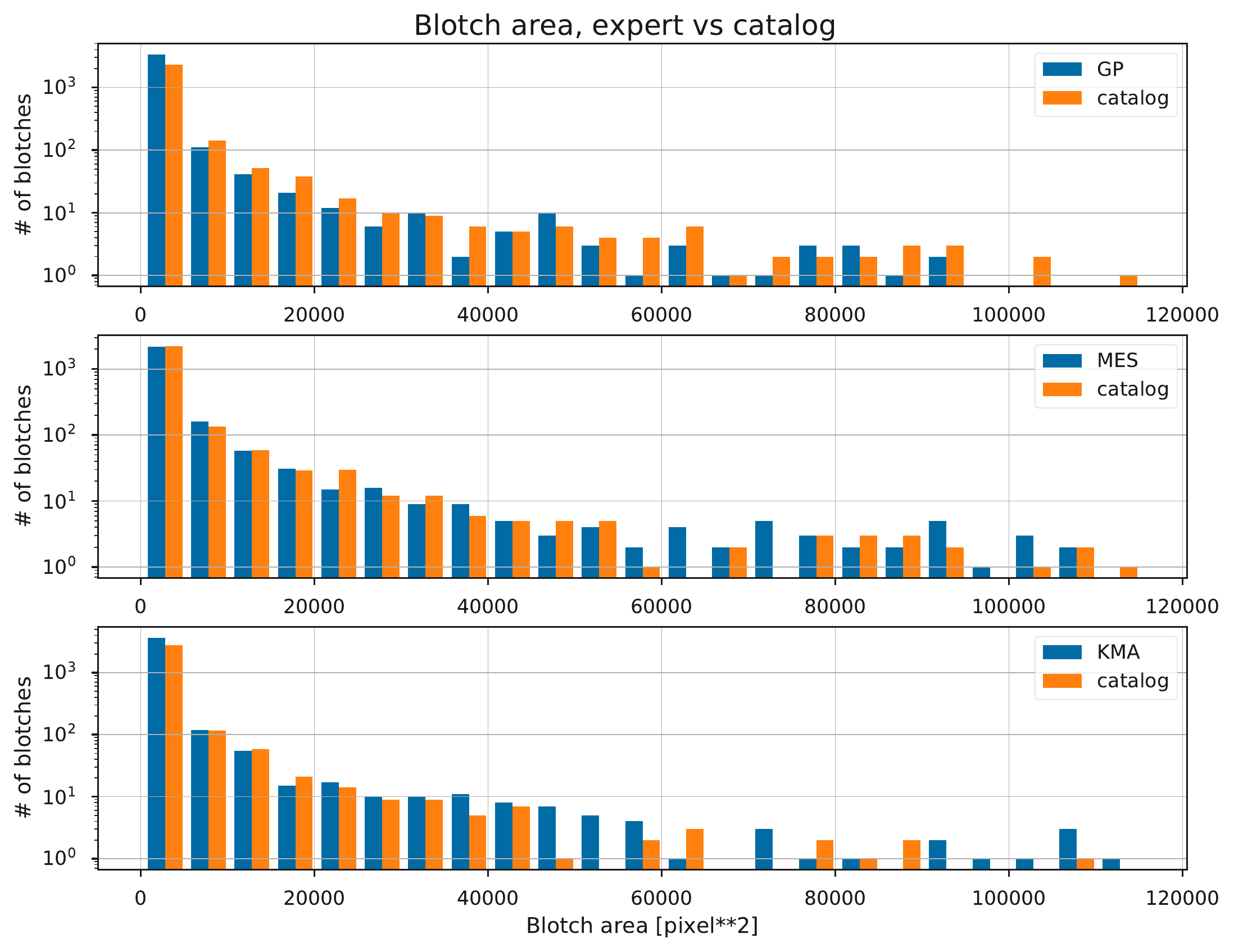}
\caption{\label{fig:gold_blotch_areas}
Comparing measured blotch areas between experts and the catalog data.
Bin size is 5000, each bin is directly compared between the data from all experts GP (blue), MES (orange), KMA (grey) and the catalog results (brown).
Binning max was cut off at \num{120000}, omitting single entry bins above.
}
\end{figure}

We also use our expert gold standard classifications to examine the physical sizes and areal coverage of the Planet Four catalog fans and blotches (see Figures \ref{fig:gold_fan_lengths_common} to \ref{fig:gold_blotch_areas}).
As in previous comparisons, there is good agreement.
The differences between the catalog is within the the variance seen between the individual expert classifiers.
Differences between the catalog and experts become more apparent when in small number regimes (when <10 sources  comprise the bin).
These differences between the distributions in these small sizes is consistent with small number Poisson uncertainty on the histogram values \citep{kraft1991}.
Thus, fan length and blotch areas are well reflected in the Planet Four catalog.


\subsection{Wind direction comparison}
\begin{figure}[htb]
\centering
\includegraphics[width=1\columnwidth]{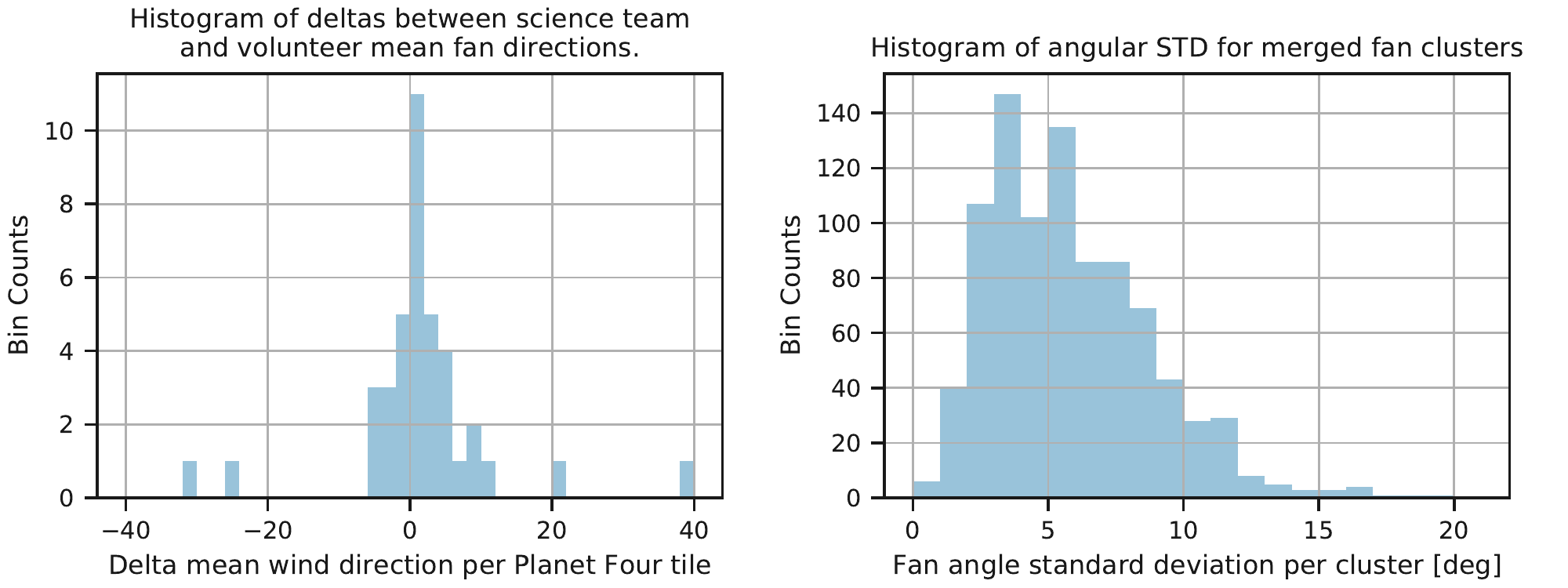}
\caption{\label{fig:gold_mean_fan_deltas_histo}
\textbf{Left:} From the 192 tiles that were analyzed by the science team, 82 resulted in fan catalog entries.
Of those, we used 39 that had more than 3 fans, for better statistics (the median number of fans per tile is 4, see Section~\ref{sec:results}).
In this histogram, we show the difference between the mean angle of the fans in these 39 Planet Four tiles between the science team and the volunteers.
Overall, we have a good agreement, with a few rare outliers, discussed in the text and in Figures~\ref{fig:gold_mean_diffs_lower} and \ref{fig:gold_mean_diffs_higher}. Bin size is 2.
\textbf{Right:} Standard deviations (STDs) of the directions of fan markings that went into each cluster, before they are merged into the average resulting catalog object.
This plot shows the distribution of these STDs for the set of 192 common gold tiles, which had a total amount of 904 fans. Bin size is 1.
}
\end{figure}
\begin{figure}[p]
\centering
\vspace{-2cm}
\includegraphics[width=0.85\columnwidth]{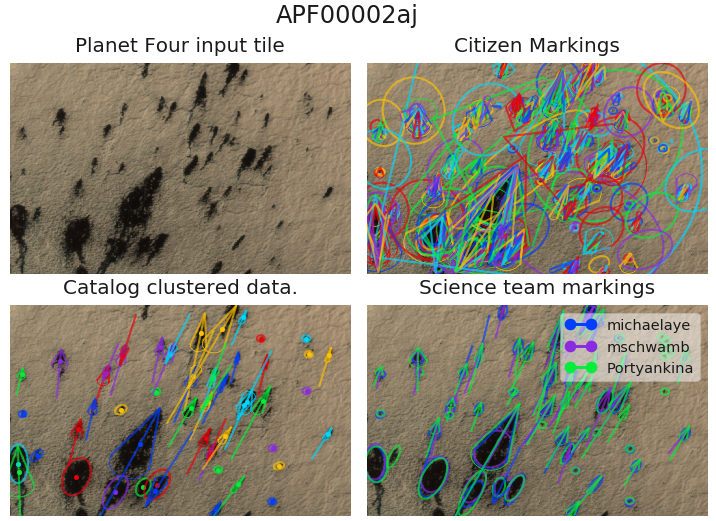}
\caption{\label{fig:gold_mean_diffs_lower} One of the outliers of Fig.~\ref{fig:gold_mean_fan_deltas_histo}, Planet Four tile ID \nolinkurl{APF00002aj} of HiRISE image \nolinkurl{ESP_012744_0985}.
The input image shows deposit shapes with less pronounced boundaries, leaking into the background.
There is also a visible gradient of directions across the tile (visible through the extended fan pointers). See the text for more interpretation.
}

\includegraphics[width=0.85\columnwidth]{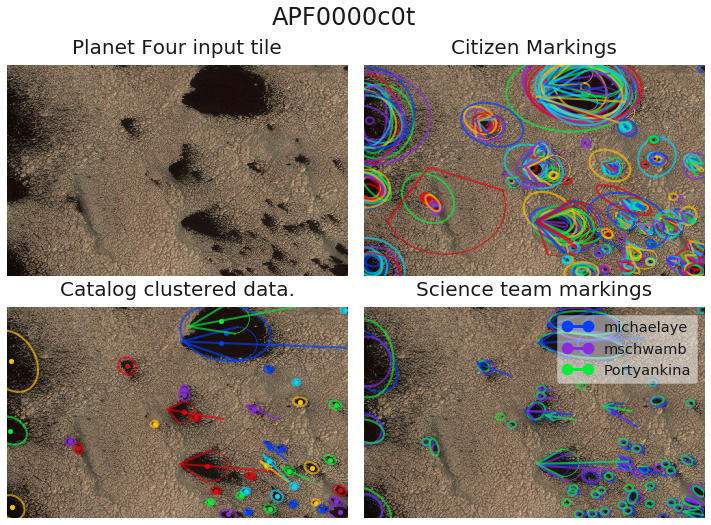}
\caption{\label{fig:gold_mean_diffs_higher}The highest outlier from Fig.~\ref{fig:gold_mean_fan_deltas_histo}, Planet Four tile ID \nolinkurl{APF0000c0t}, from HiRISE image \nolinkurl{ESP_012858_0855}.
While fans have been identified, their number is small, increasing the chance for variance between the experts and catalog data.
}
\end{figure}

Fig.~\ref{fig:gold_mean_fan_deltas_histo}, left, shows a histogram over the differences in the mean-over-tile fan directions between the catalog entries that are clustered from all the volunteers' markings and the average from the three science team members.
In general, the agreement is very good, with differences usually smaller than 10 degrees.
Another way to investigate our uncertainties is to calculate the angular standard deviation for each cluster member markings that are merged into the final catalog objects, independent on if the markings were done by an expert or a volunteer.
Fig.~\ref{fig:gold_mean_diffs_lower} discusses the lower outlier of, indicating that the respective Planet Four tile has a more difficult than usual scenario with a naturally occurring higher variance of the actual deposit directions on the ground.
Not only are the deposit shapes visible in the upper left more irregular than usual, there is a visible gradient of directions across this tile, as can be seen by the exaggerated fan pointers.
This gradient is probably caused by the basin shapes in the Inca City region that can create a topographical control of the alignment of fan deposits over the usual wind control.
However, our reduction pipeline is reliably reducing the markings for every deposit, but with higher than usual variance between orientation and size of the markings.
Having no single clear fan direction in the image tile, it is reasonable to expect a higher variance and hence, a higher delta when compared to the 3 science team members.

In a similar fashion, Fig.~\ref{fig:gold_mean_diffs_higher} discusses the high-side outlier of Fig.~\ref{fig:gold_mean_fan_deltas_histo}.
While fans have been identified, their counts is low, creating low statistics effects by letting small deviations having a larger effect on the comparison with the catalog data.
Additionally, the few fans that are visible appear to show different directions, leading to a less certain fan direction with a higher variance, which in turn can lead to larger differences when comparing their values, resulting from low statistics.

In Fig.~\ref{fig:gold_mean_fan_deltas_histo}, right, we plot the standard deviations for all 904 fan clusters for the 192 common tiles that were analyzed by all experts.
The right end of this histogram is cut off by our angular clustering parameter of \ang{20}, meaning larger angular differences are never clustered together.
However, the majority of standard deviations lie far below that safety cut-off value for the clustering.
We estimate an average uncertainty for our fan directions of about \SI[separate-uncertainty = true]{5(3)}{\degree}, using a half maximum width of this histogram.
The actual uncertainty highly depends on the quality of the data as given by the HiRISE binning mode and the local variability of winds, leading to increased diffusion of the deposits.
We believe these factors lead to the non-Gaussian skew of the histogram.

Additional validation results can be found in \ref{sec:extended_validation}.

\subsection{Summary}
In conclusion, our catalog has high completion in most cases. Outliers have been found to be caused by special circumstances with more challenging classification tasks, creating higher variance for all classifiers, including the experts.
The analysis of the gold standard sample demonstrates that the bulk composition of the Planet Four catalog represents a fairly complete picture of the seasonal fans and blotches captured in the HiRISE images.

\section{Results: Fan and Blotch Catalog }%
\begin{figure}[b!]
\centering
\includegraphics[width=0.8\columnwidth]{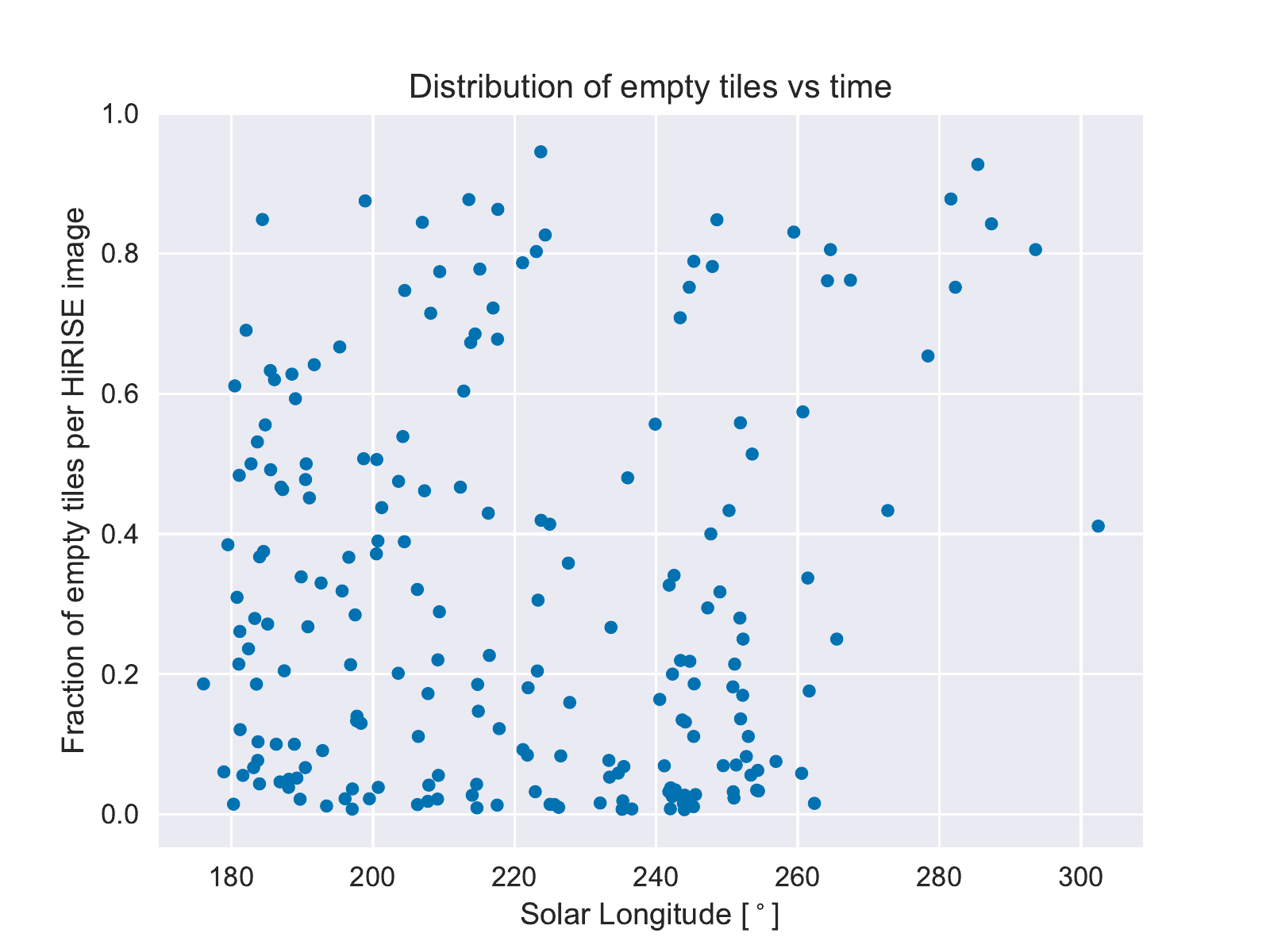}
\caption{\label{fig:empty_data} Distribution of empty tiles over time, measured in Mars Solar Longitude.
Until \Ls =\ang{260} the fraction of HiRISE images that can be empty varies randomly, reflecting the different ground surfaces imaged across all latitudes.
After \Ls =\ang{260} all \ce{CO2} is gone --- earlier at lower latitudes ---, and most of the HiRISE images appear empty in terms of identifiable blotches or fans, because any deposits blend with the ice-free background.}
\end{figure}

\label{sec:results}
From 221 HiRISE images from Mars years 29 and 30, cut up into \num{42904} Planet Four tiles, the Planet Four volunteers produced almost 2.8 million fan markings, that were clustered into \num{159558} fans in our MY29/MY30 catalog.
In Table~\ref{tab:fan_catalog_head} we show an example of fan catalog data.
For blotches, 3.46 million raw markings were combined into \num{250164} blotches.
\SI{29.6}{\%} of the image tiles (= \num{12693}) end up not having any clustered markings in our catalog.
Fig.~\ref{fig:empty_data} shows the distribution of the fraction of empty tiles per HiRISE image vs.\ solar longitude.
Visual checks of data with fractions above 0.8 confirmed that these HiRISE images are mostly free of \ce{CO2} jet deposits at spring times; in late summer, however, when the seasonal \ce{CO2} ice layer has fully sublimated,fan and blotch deposits are rendered mostly invisible, because they blend into the now ice-free background.
A notable exception to this general effect is the ROI Inca City where the summer data, after \Ls{} 250\degree--260\degree, regularly shows fan deposits still discernible.
This could point to an interesting difference in the ground soil compactification and its related observed texture.
New deposits from \ce{CO2} jet eruptions may be sufficiently different in texture from the background as a result from particle sorting and related phase function changes of the fresher surface.

\begin{landscape}
\begin{table}
\setlength\tabcolsep{2pt}
\footnotesize
\begin{tabular}{lrrlrrlrlrrrrr}
\toprule
{} &   angle &  distance &     tile\_id &  image\_x &   image\_y & marking\_id &  n\_votes &            obsid &  spread &  version &  vote\_ratio &       x &  x\_angle \\
\midrule
0 &  205.56 &    179.71 &  APF0000ci9 &  2270.76 &  24336.16 &    F000000 &       35 &  ESP\_012079\_0945 &   88.03 &        1 &        1.00 &  790.76 &    -0.90 \\
1 &  185.39 &    179.62 &  APF0000cia &  3391.21 &   5640.60 &    F000001 &       15 &  ESP\_012079\_0945 &   21.35 &        1 &        1.00 &  431.21 &    -1.00 \\
2 &  184.98 &    500.27 &  APF0000cia &  3509.96 &   5876.70 &    F000002 &       10 &  ESP\_012079\_0945 &   18.91 &        1 &        1.00 &  549.96 &    -1.00 \\
3 &  184.29 &    105.43 &  APF0000cia &  3716.27 &   5824.50 &    F000004 &        6 &  ESP\_012079\_0945 &   26.41 &        1 &        0.68 &  756.27 &    -1.00 \\
4 &  189.42 &    109.50 &  APF0000cia &  3452.17 &   6033.00 &    F000005 &        3 &  ESP\_012079\_0945 &   22.58 &        1 &        0.51 &  492.17 &    -0.99 \\
5 &  194.16 &    335.78 &  APF0000cib &  3565.47 &  15930.34 &    F000006 &       64 &  ESP\_012079\_0945 &   34.93 &        1 &        1.00 &  605.47 &    -0.97 \\
6 &  187.74 &    183.41 &  APF0000cib &  3143.15 &  15433.60 &    F000007 &       20 &  ESP\_012079\_0945 &   25.68 &        1 &        1.00 &  183.15 &    -0.99 \\
7 &  209.47 &    179.29 &  APF0000cid &   942.95 &  22257.99 &    F000008 &       58 &  ESP\_012079\_0945 &   49.11 &        1 &        1.00 &  202.95 &    -0.87 \\
8 &  199.91 &    220.64 &  APF0000cid &  1199.11 &  21994.01 &    F000009 &       54 &  ESP\_012079\_0945 &   35.37 &        1 &        1.00 &  459.11 &    -0.94 \\
9 &  218.88 &    118.16 &  APF0000cid &   815.95 &  22539.28 &    F00000a &       42 &  ESP\_012079\_0945 &   49.66 &        1 &        1.00 &   75.95 &    -0.77 \\
\bottomrule
\end{tabular}

\begin{tabular}{lrrrrrrrr}
\toprule
{} &       y &  y\_angle &      l\_s &  north\_azimuth &  map\_scale &  BodyFixedCoordinateX &  BodyFixedCoordinateY &  BodyFixedCoordinateZ \\
\midrule
0 &  224.16 &    -0.43 &  214.785 &     126.856883 &       0.25 &            -65.804336 &            261.407884 &          -3370.504345 \\
1 &  160.60 &    -0.09 &  214.785 &     126.856883 &       0.25 &            -67.219114 &            257.011589 &          -3370.631413 \\
2 &  396.70 &    -0.09 &  214.785 &     126.856883 &       0.25 &            -67.170611 &            257.055226 &          -3370.630794 \\
3 &  344.50 &    -0.07 &  214.785 &     126.856883 &       0.25 &            -67.127761 &            257.024926 &          -3370.635002 \\
4 &  553.00 &    -0.16 &  214.785 &     126.856883 &       0.25 &            -67.169940 &            257.096267 &          -3370.628302 \\
5 &  586.34 &    -0.24 &  214.785 &     126.856883 &       0.25 &            -66.258570 &            259.361039 &          -3370.571273 \\
6 &   89.60 &    -0.13 &  214.785 &     126.856883 &       0.25 &            -66.400170 &            259.284370 &          -3370.565666 \\
7 &  337.99 &    -0.49 &  214.785 &     126.856883 &       0.25 &            -66.296391 &            261.048812 &          -3370.492211 \\
8 &   74.01 &    -0.34 &  214.785 &     126.856883 &       0.25 &            -66.261274 &            260.965240 &          -3370.497183 \\
9 &  619.28 &    -0.62 &  214.785 &     126.856883 &       0.25 &            -66.300167 &            261.124709 &          -3370.487589 \\
\bottomrule
\end{tabular}

\begin{tabular}{lrrr}
\toprule
{} &  PlanetocentricLatitude &  PlanetographicLatitude &  PositiveEast360Longitude \\
\midrule
0 &              -85.427383 &              -85.480830 &                104.129523 \\
1 &              -85.493546 &              -85.546226 &                104.656897 \\
2 &              -85.493039 &              -85.545725 &                104.644396 \\
3 &              -85.493723 &              -85.546401 &                104.637107 \\
4 &              -85.492368 &              -85.545061 &                104.642019 \\
5 &              -85.459101 &              -85.512180 &                104.330752 \\
6 &              -85.459755 &              -85.512827 &                104.364183 \\
7 &              -85.431209 &              -85.484612 &                104.249678 \\
8 &              -85.432730 &              -85.486115 &                104.246813 \\
9 &              -85.429945 &              -85.483362 &                104.246483 \\
\bottomrule
\end{tabular}
\caption{\label{tab:fan_catalog_head}First ten lines of the fan catalog file \texttt{P4\_catalog\_v1.0\_L1C\_cut\_0.5\_fan\_meta\_merged.csv}, broken into three segments.}
\end{table}
\end{landscape}

\begin{figure}
\centering
\includegraphics[width=0.8\columnwidth]{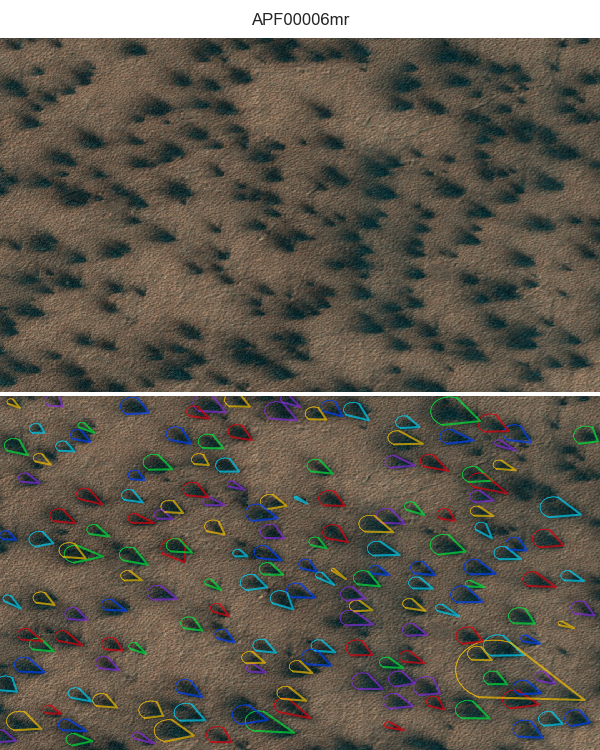}
\caption{\label{fig:most_fans} Planet Four tile \nolinkurl{APF00006mr} from HiRISE image \nolinkurl{ESP_011296_0975} has the highest number of resulting fan entries per tile. Top: Input tile as seen by volunteers; Bottom: Overlaid clustering results from the catalog.}
\end{figure}

\begin{figure}
\centering
\includegraphics[width=0.8\columnwidth]{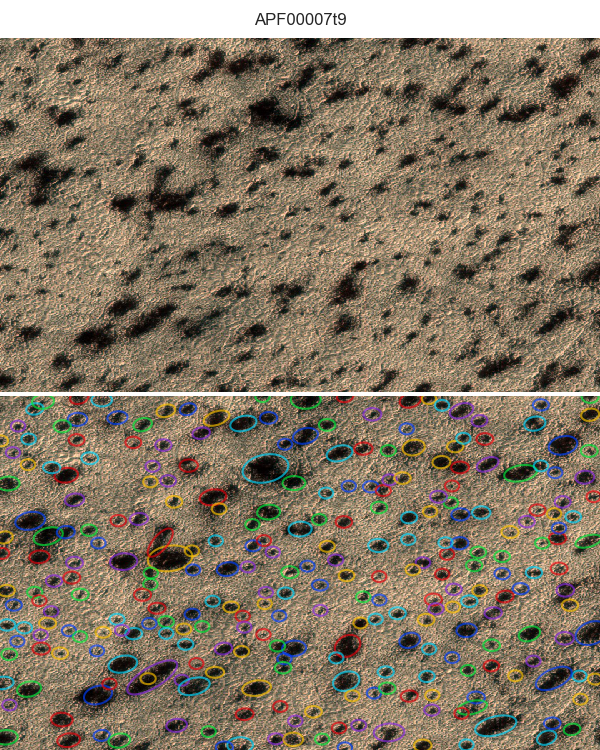}
\caption{\label{fig:most_blotches} Planet Four tile \nolinkurl{APF00007t9} from HiRISE image \nolinkurl{ESP_012604_0965} has the highest number of resulting blotch entries per tile. Top: Input tile as seen by volunteers; Bottom: Overlaid clustering results from the catalog.}
\end{figure}

\subsection{Catalog properties}
\subsubsection{Fan counts}
\begin{figure}
\centering
\includegraphics[width=1\columnwidth]{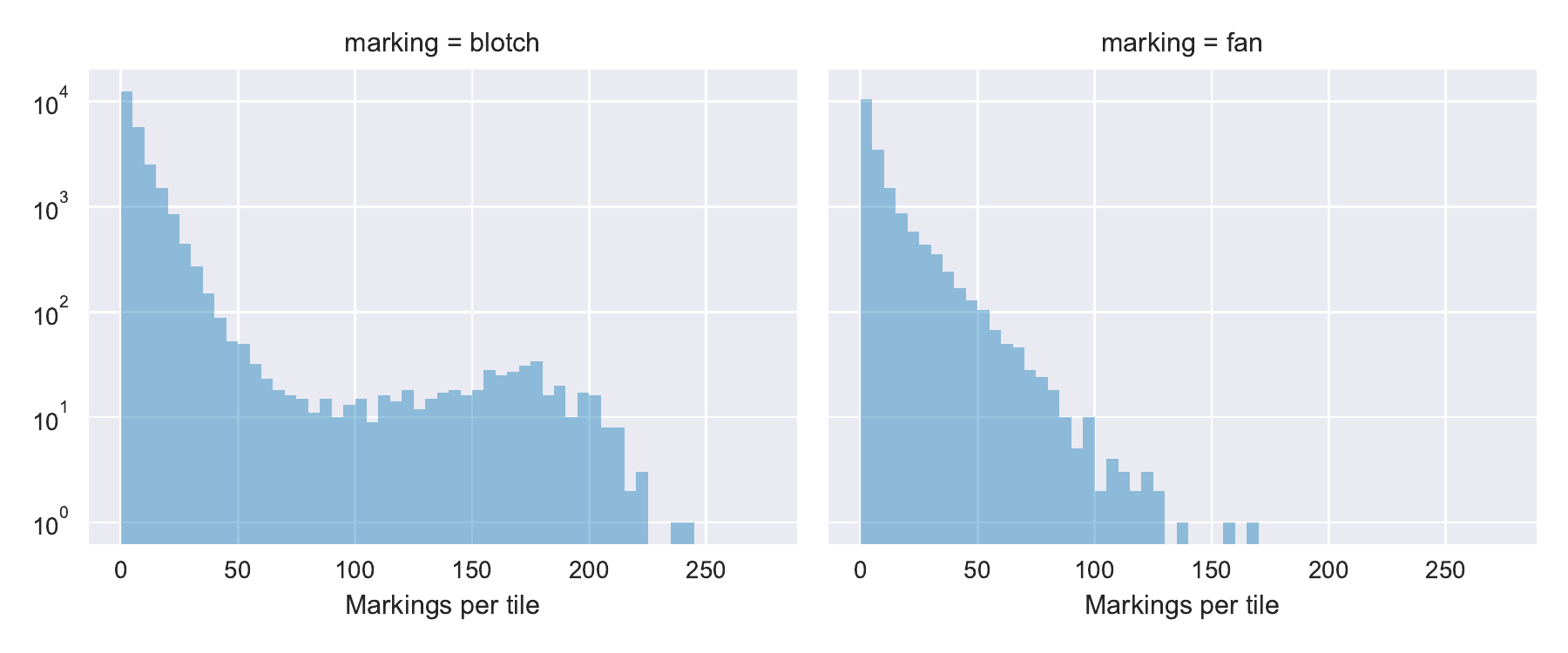}
\caption{\label{fig:number_distributions} Count distributions of catalog objects per tile, blotches on the left, fans on the right.
The bin size is 5 counts in both plots.}
\end{figure}

The highest counts of fans and blotches were 167 fans in the tile\_id \nolinkurl{APF00006mr} and 278 blotches in the tile\_id \nolinkurl{APF00007t9}, shown in Figures~\ref{fig:most_fans} and \ref{fig:most_blotches}.
These data serve as an indication of the dedication of the Planet Four volunteers producing results in such high spatial density.
The median count of fans and blotches per tile is 4.
The distribution of both numbers is shown in Fig.~\ref{fig:number_distributions}.

\subsubsection{Fan lengths}
\begin{figure}
\centering
\includegraphics[width=1\columnwidth]{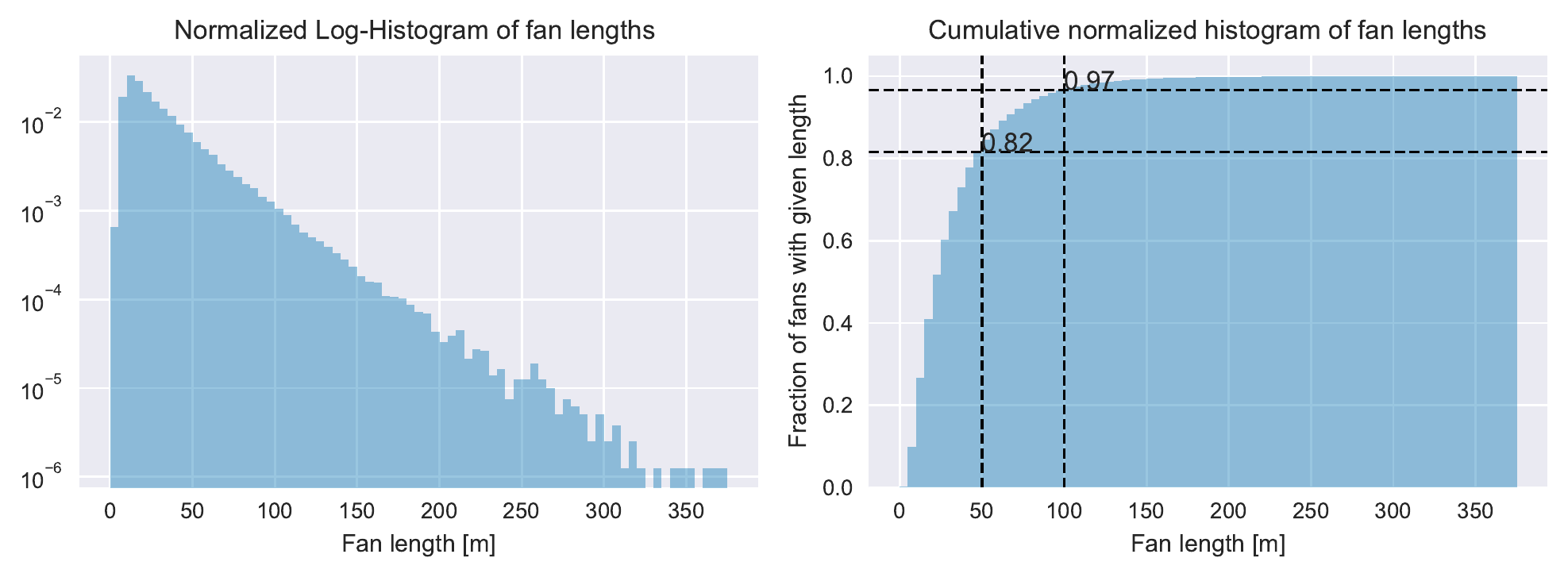}
\caption{\label{fig:fan_lengths} Normalized histograms of all fan lengths in the catalog. Left: Log-Histogram, Right: Cumulative Histogram.
The median (i.e.~fraction of 0.5) value is at \SI{24}{\meter}, with \SI{82}{\%} of the fans shorter than \SI{50} and \SI{97}{\%} shorter than \SI{100}{\m}, as indicated by the lines in the plot. %
}
\end{figure}

As an example of the possibilities of the produced catalog, we describe the measured fan lengths in the catalog.
The catalog column \emph{distance} requires scaling by the values in \emph{map\_scale}, to correct for the different HiRISE binning modes.
The distribution of these measurements are shown in Fig.~\ref{fig:fan_lengths}.
About \SI{97}{\%} of all fans are below \SI{100}{\meter} in length, with a median value of \SI{24}{\meter}.

The three largest fans measured are all from the same ROI called Manhattan Classic (Lat \SI{-86.39}{\degree}, Lon \SI{99}{\degree}), having lengths of \SIlist{373;368;361}{\meter} respectively.
They were identified in the HiRISE images \nolinkurl{ESP_013095_0935} (longest) and \nolinkurl{ESP_011961_0935} (second and third).
The two longest fan markings even identify the same fan, but at different times in the season, with the longest observed at \Ls =\ang{265}, and it's shorter self at \Ls =\ang{209}. Being only \SI{5}{m} different, we attribute the increased marking measure to both material being potentially moved around by winds during spring and a decrease of precision in identification after the \ce{CO2} has sublimed and the deposits start to fade into the background.
However, we interpret the fact to have identified the largest fan twice, as a further indication of the high reliability of our results, considering that the random image serving procedure of the Planet Four classification interface ensured that volunteers do not classify images in the order they have been taken, because that would have increased the chances of being biased by their previous classification.
In this case, where 119 volunteers classified \emph{APF0000dtk} with the longest fan, and 54 volunteers classified \emph{APF0000de3} with the second longest fan (shown in Fig.\ref{fig:multiple_fans}), only one volunteer was identified to be the same.

\begin{figure}
\centering
\includegraphics[width=0.85\columnwidth]{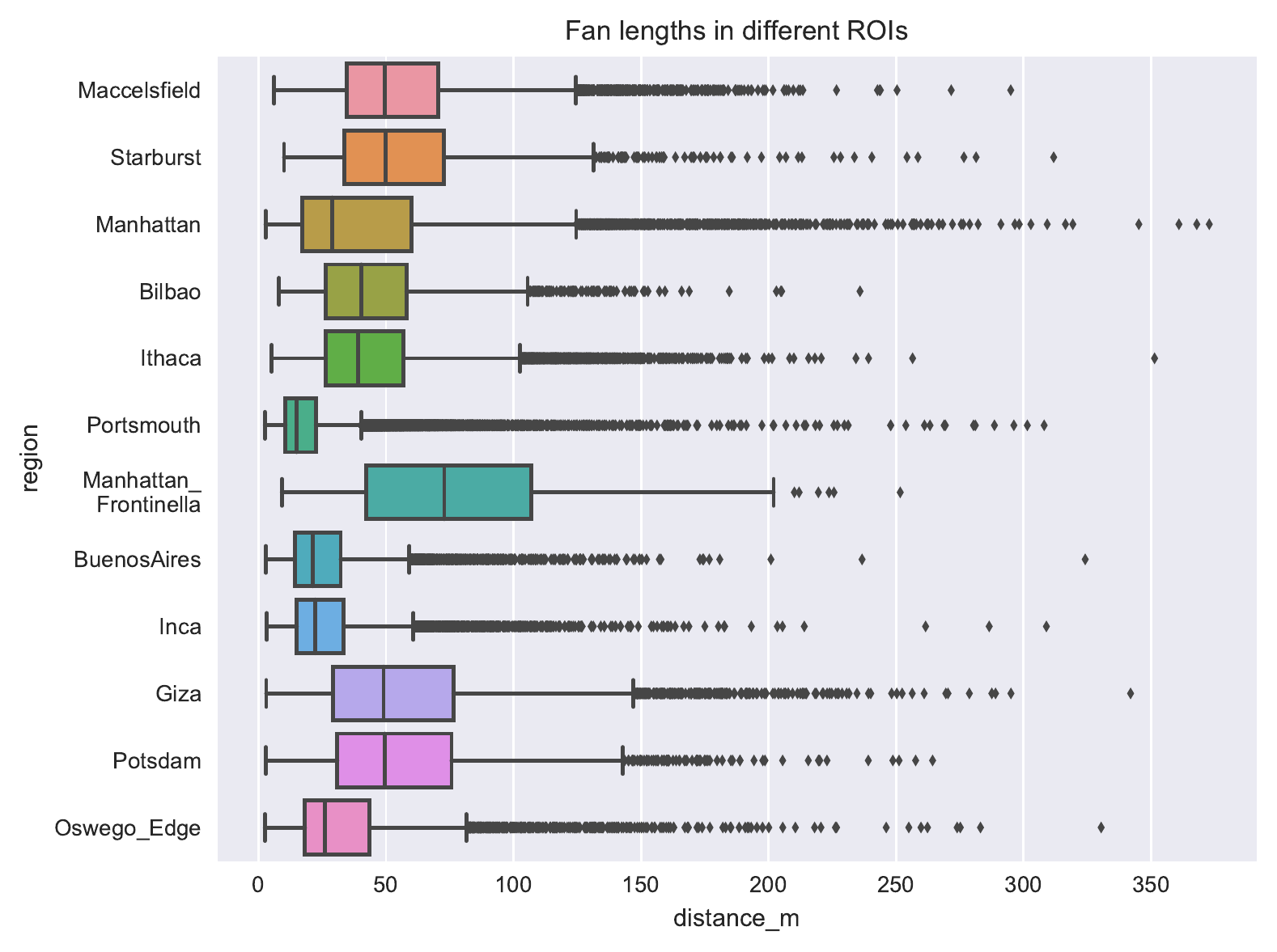}
\caption{\label{fig:fan_lengths_vs_regions}
Boxplot showing the distributions of fan lengths for our set of regions of interest at the Martian south pole, over both Martian years of data, MY 29 and 30 (Inca City and Inca City Ridges are combined here due to their proximity).
The boxplot setup uses the standard setup of interquartile range (IQR) for the box and its whiskers extending to 1.5xIQR, single dots for outliers.%
}
\end{figure}

An overview of the fan lengths distributions for all major ROIs over all 2 Martian years of data is shown in Fig.~\ref{fig:fan_lengths_vs_regions}.
When compared between Mars years 29 and 30, the total (over all ROIs) fan length statistics are very comparable, with a median of \SI{24.2}{\m} for MY29 and \SI{23.8}{\m} for MY30.
However, we identify specific ROIs that have different fan properties between MY29 and 30.
For example, the ROI \emph{Manhattan} has a median fan length of \SI{42}{\m} in MY29 and a decreased median of \SI{25}{\m} in MY30.
This is in contrast with the ROI \emph{Giza}, where the trend has the opposite direction, with a median fan length of \SI{44}{\m} in MY29, comparable with \emph{Manhattan's} median in the same year, but then increases to a median of \SI{59}{\m} for MY30.
Meanwhile, in ROI \emph{Ithaca}, both years are very similar, with median fan lengths of \SIlist[list-pair-separator={ and }]{39.5;38.6}{\m} respectively.

\section{Wind Direction Results from Four Sample Regions of Interest (ROIs)}%
\label{sec:regional}

Early in the mission, HiRISE has defined several regions of interest (ROIs) within the southern polar areas that have been extensively monitored for seasonal activity ever since (the list of original seasonal ROIs can be found in \citet{hansen2010}).
We have selected a sub-set of these ROIs to be analyzed by Planet Four, as shown in Table~\ref{tab:regions}).
The map of ROIs' distribution over the pole is shown in Fig.~\ref{fig:monitoring_overview}.

Below we will focus on 4 example ROIs to showcase the use of Planet Four data catalog and our ability to monitor wind directions using fan markings positions and locations.
We have picked these 4 ROIs (informally named Ithaca, Giza, Manhattan, and Inca City) for regional case studies of the seasonal winds because the temporal coverage over these locations is the highest.
We describe each ROIs' general settings and geomorphology based on observations of HiRISE and our previous works \citep{hansen2010, pommerol2011a}.
We then present the wind direction maps over spring season at each of these locations.
The wind rose diagrams for each HiRISE image separately are available in the supplementary files \nolinkurl{P4_catalog_v1.0_wind_rose_diagrams.pdf}

\subsection{Ithaca}

The Ithaca region is located at southern latitude 85.2\degree{}, eastern longitude 181.4\degree{}.
This location is away from the permanent polar cap, at the edge of the cryptic region and situated on a surface that is relatively smooth on a large scale: the digital terrain model produced by HiRISE (\nolinkurl{DTEPD_040189_0950_040216_0950_A01}) shows vertical elevation variations less than \SI{60}{\m} across the Ithaca region.
At the same time, on the meter scale the surface in Ithaca is rough, showing irregular and uneven bumps and pits.
No araneiforms (i.e.\ radially-organized channels) were detected Ithaca according to HiRISE imaging, while rare isolated troughs and patterned ground similar to araneiform troughs are present \citep{hansen2010}.

\begin{figure}
\includegraphics[width=0.98\columnwidth]{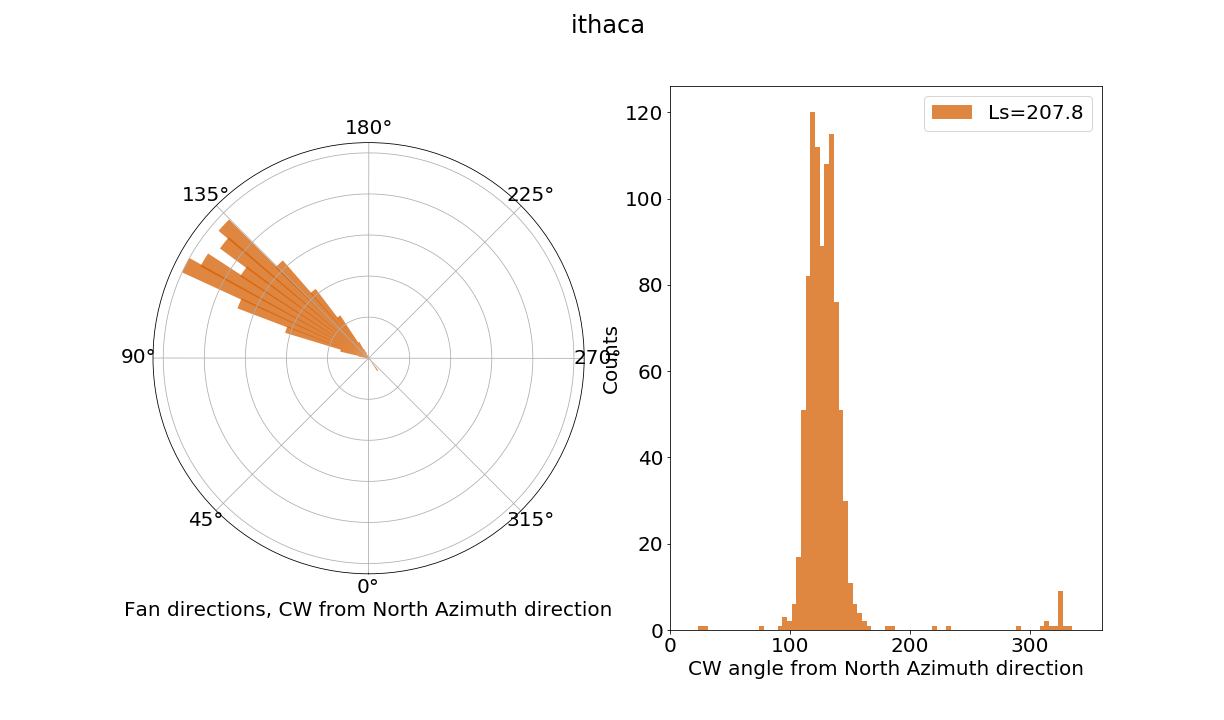}
\includegraphics[width=0.98\columnwidth]{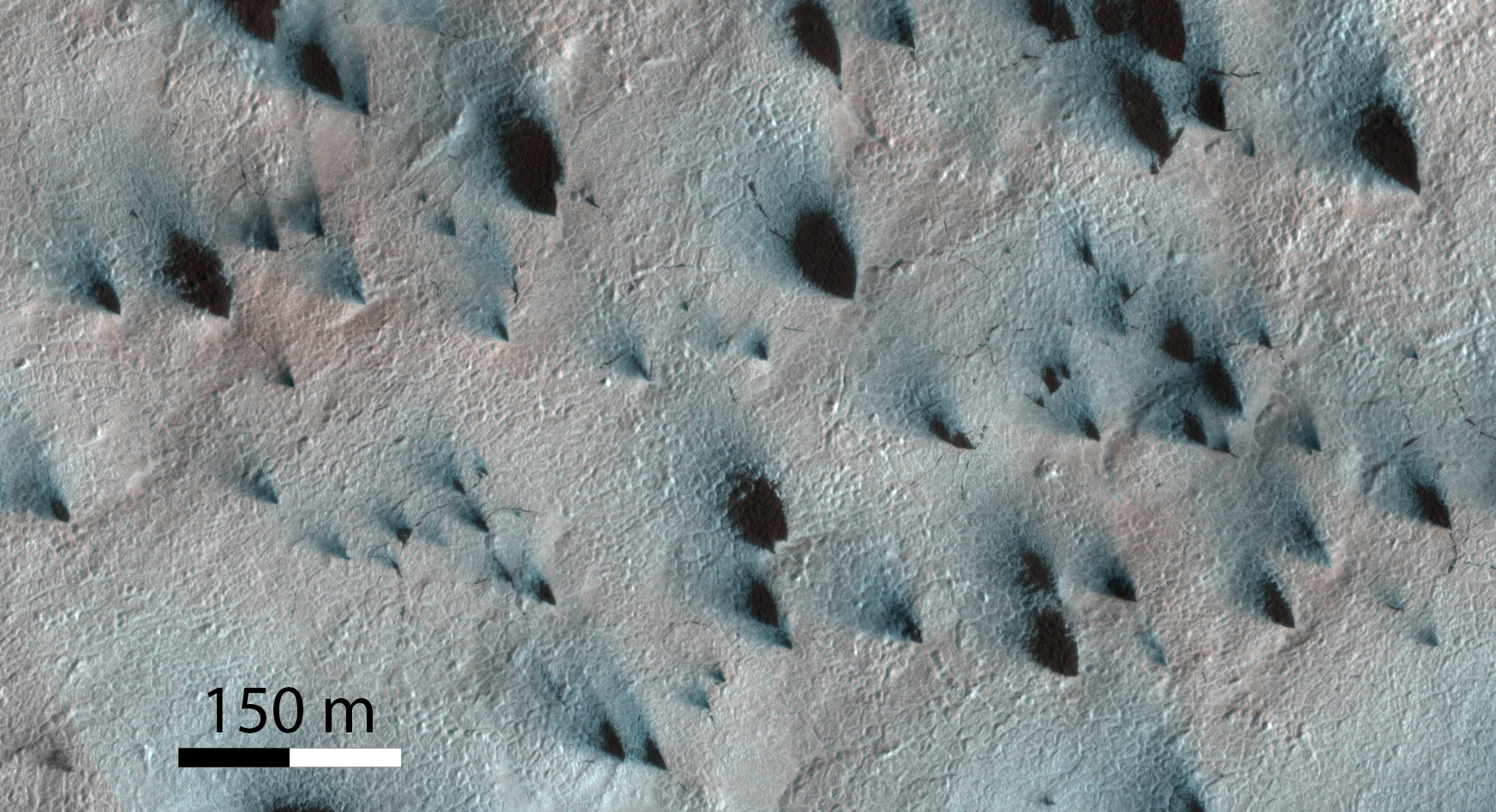}
\caption{\textbf{\label{fig:Ithaca_Ls181}}
Fan directions in Ithaca region at \Ls{}=207.8\degree{} (top) and a subframe of HiRISE image \texttt{ESP\_011931\_0945} that can be directly compared to the wind rose diagram in the top left panel.}
\end{figure}

During local spring, fan-shaped deposits densely cover the Ithaca region (see an example in Fig.~\ref{fig:Ithaca_Ls181}.
Opening angles and lengths of the fans were reported to evolve during spring while the nature of these changes was not quantified \citep{thomas2010}.
Multiple fans were observed to emerge from the common vents, at times merging together to create a wider singular fan.
The directions of the fan deposits were noted to be consistent from one Martian year to another with only little variation.

An interesting detail about Ithaca is very prominent bluish halos and fans that are repeatedly observed here \citep{thomas2010}.
In contrast to the more common dark fan-shaped deposits, these halos and fans have higher albedos, approaching the albedo of fresh ice deposits.
In Ithaca they are also distinctively bluer than the rest of the surface.
There are at least two types of such bright deposits.
One type resembles narrow fans that are located centrally over the older dark fans.
These appear early in spring, before \Ls{} = \ang{190}.
The other type resembles halos contouring the pre-existing dark fans.
They appear on average later than the narrow bright fans.

In summer (\Ls{} > \ang{270}), the seasonal deposits are mostly invisible in Ithaca.
Partially, this is because the low scale roughness creates a patchy-looking environment with pits being darker than  bumps either due to shadows or dust collecting in depressions.

Fig.~\ref{fig:Ithaca_Ls181} shows a typical plot that we will use to analyze derived wind directions in our ROIs.
This particular plot was created from Planet Four data for one HiRISE image (\nolinkurl{ESP_011931_0945}) taken in Ithaca at \Ls{} = \ang{207}.
To create this plot we took all the fan markings over the HiRISE image and plotted it as a histogram of their directions (top right panel of Fig.~\ref{fig:Ithaca_Ls181}).
Note that, in contrast to the standard wind rose diagrams showing the directions of the origin of winds, we use this diagram to show the measured deposition directions caused by the winds, i.e.\ the opposite from the wind origins.
We decided for this kind of display because it relates more to the actual measurements performed by the Planet Four project and does not imply any interpretation.
The fan direction is counted clock-wise (CW) from the North Azimuth (NA) direction, where \ang{0} always represents North, and \ang{270} West.
The histogram is not scaled, i.e.\ the y-axis shows the actual counts of the fan markings with the direction of each bin in the x-axis.
The maximum of the histogram is the most probable direction for the markings and the width indicates how variable the directions of the markings are for this particular \Ls{}.
The default size of each histogram bin is 3.6~\degree{}.
In exceptionally rare cases for a particular image the number of fan markings and thus number of wind measurements are low.
Such cases require special treatment and increase in bin size.
On the top left panel the same data are plotted in the wind rose diagram.
This time the histogram is normalized to highlight the difference in directions if several HiRISE images are plotted in the same frame.
Note that the position of zero (NA direction) depends on the location of ROI, i.e.\ the wind rose diagram is map-projected to the location of the data plotted.
Thus, the direction of the fans can be directly compared to the map-projected HiRISE image (bottom panel).
In this particular example one can see that the histogram has 2 peaks that indicate there are two distinct directions of the fans.\
This can either be
\begin{mylist}
    \item because of overlapping fan deposits from jets that erupted from the same vents at different times prior to \Ls{}=207.8\degree{} under different wind regimes; or
    \item because different areas of the ROI have distinctively different wind regimes.
\end{mylist}
In this example comparing the derived fan directions to the sub-frame of the HiRISE image indicates that the first case is more probable.

\begin{figure}
\includegraphics[width=0.98\columnwidth]{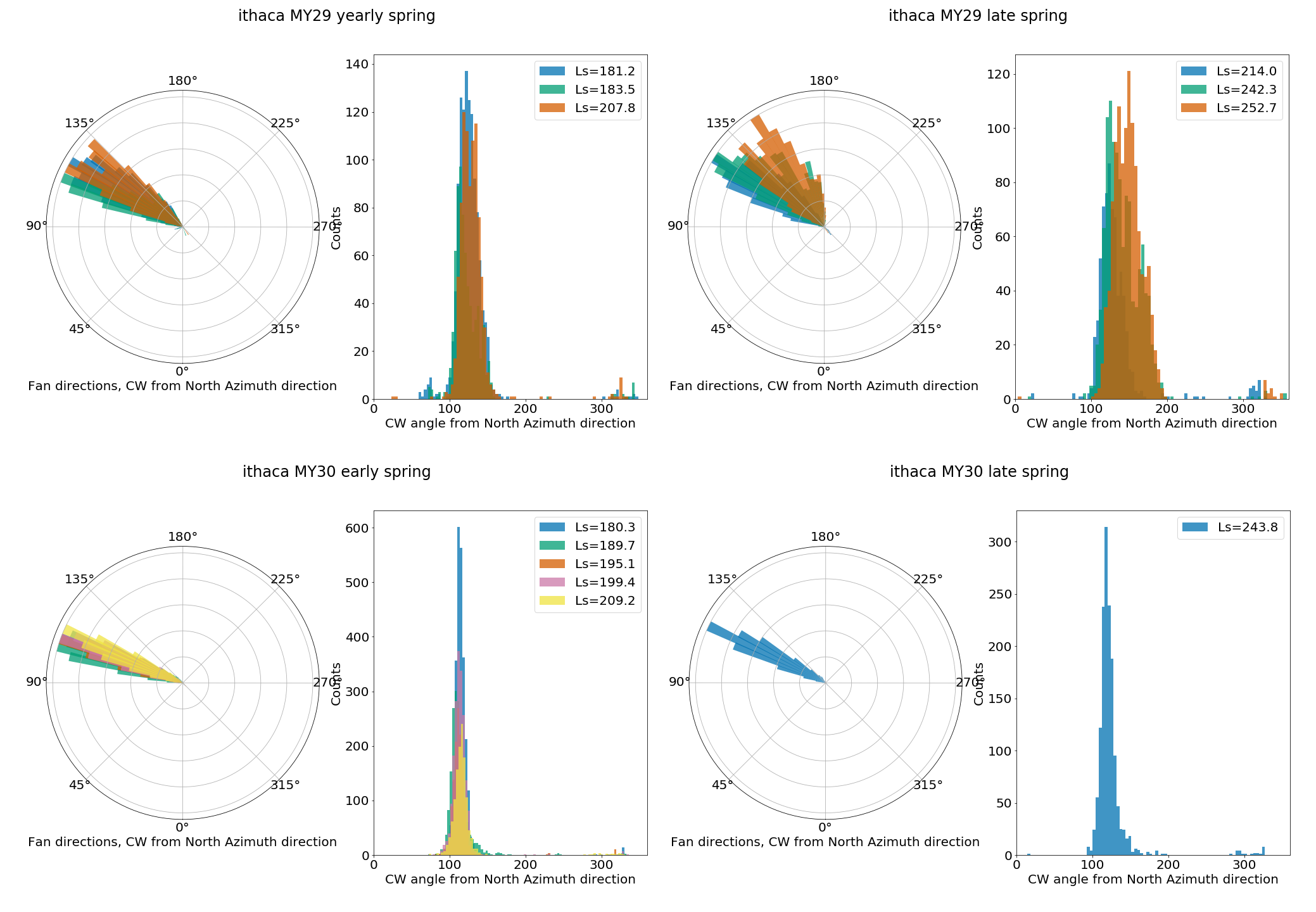}
\caption{\textbf{\label{fig:winds_Ithaca}}
Direction of fan markings in Ithaca region for early and late spring of MY29 and MY30.}
\end{figure}

Fig.~\ref{fig:winds_Ithaca} shows directions of the fan deposits in Ithaca as retrieved by the Planet Four project for two Martian years: MY29 and MY30.
We have separated the spring season into early spring, i.e.\ before \Ls{}=210\degree{}, and late spring, from  \Ls{}=210\degree{} to \Ls{}=270\degree{}.
The panels in this figure are organized in the way that columns show separation into early and late spring while rows show MY 29 and MY 30.

Ithaca fans sustain the same direction towards \SI{\approx{} 125}{\degree} through the whole spring in both years with only a little shift towards East (see also top left panel of Fig.~\ref{fig:winds_all_vs_Ls}).
In MY29 fan direction histograms are wider than in MY30.
A narrow histogram is an indication of small deviations of the governing winds at the times of jet eruptions.
The shift of the mean wind direction is less than 10\degree{} in the early spring and the maximum shift is 25\degree{} over the whole season in MY29.
Histograms widen with increase of \Ls{} and sometimes develop double maxima indicating more variability in the marked fan directions.
This is also reflected in the increase of the standard deviation towards the end of spring.
It can be attributed to larger wind variability later in spring or that winds become strong enough to lift the particles from the ground at times between jet eruptions.
Over-all MY30 show similar behavior to MY29.

\subsection{Giza}

The Giza region is at southern latitude 84.8\degree{}, eastern longitude 65.7\degree{}.
It is located closer to the edge of the permanent cap than Ithaca.
It is also near a trough with exposure of southern polar layered deposits while the area of Giza is flat on km-scale (see HiRISE DTM \nolinkurl{DTEPC_004736_0950_005119_0950_A01}).
On the smaller scales, as can be seen in multiple HiRISE images taken over this area (including those that were input for Planet Four) the region is covered in modulated bumps and small ripples.
One side of this ROI is covered in yardangs.

Very large and very intricate araneiform structures are located in this region.
Their troughs are narrow, long, with high degrees of branching.
These araneiforms are very active in spring: multiple long and narrow fans emerge from their troughs and cover an extended area.
HiRISE detected a dusty reddish haze over the araneiforms in Giza in several years indicating active loading of dust into the lower layer of atmosphere.
The directions of the fans in the late spring were previously noted to co-align with yardangs, suggesting that the wind regime in this area in summer stayed stable for an extended period of time \cite{hansen2010}.

\begin{figure}
\includegraphics[width=0.98\columnwidth]{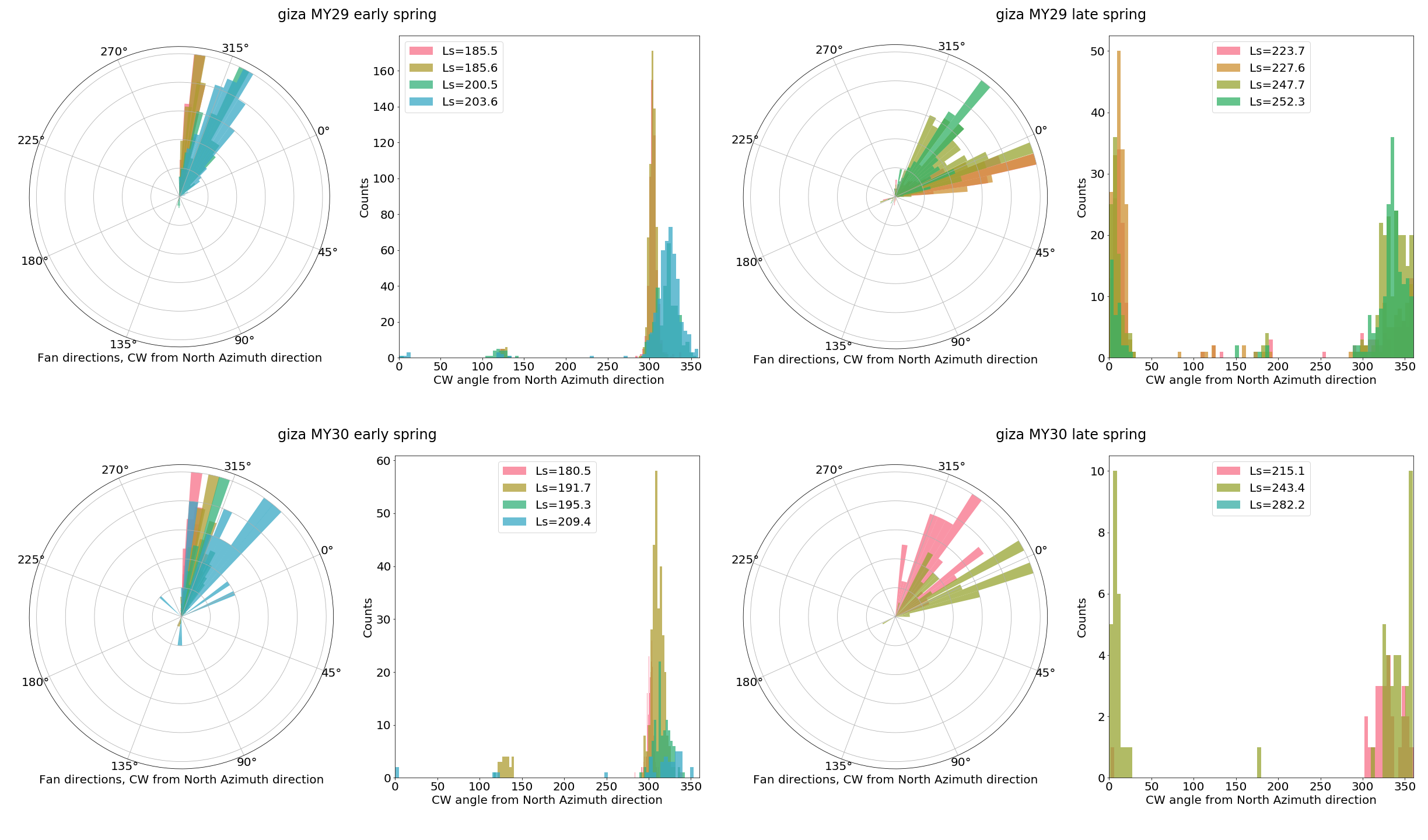}
\caption{\textbf{\label{fig:winds_Giza}}
Direction of fan markings in Giza region for early and late spring of MY29 and MY30. HiRISE images used:
\texttt{ESP\_011447\_0950}, \texttt{ESP\_011448\_0950}, \texttt{ESP\_011777\_0950},
\texttt{ESP\_011843\_0950}, \texttt{ESP\_012212\_0950},
\texttt{ESP\_012265\_0950}, \texttt{ESP\_012344\_0950}, \texttt{ESP\_012704\_0850},
\texttt{ESP\_012753\_0950}, \texttt{ESP\_012836\_0850},
\texttt{ESP\_012845\_0950}, \texttt{ESP\_020150\_0950}, \texttt{ESP\_020401\_0950},
\texttt{ESP\_020480\_0950}, \texttt{ESP\_020783\_0950},
\texttt{ESP\_020902\_0950}, \texttt{ESP\_021482\_0950}, \texttt{ESP\_022273\_0950}.
}
\end{figure}

Similar to Ithaca, in Giza we do not observe significant differences in fan directions between MY29 and MY30 (Fig.~\ref{fig:winds_all_vs_Ls} lower left panel and Fig.~\ref{fig:winds_Giza}).
Early images taken before \Ls{}=190\degree{} show very narrow histograms with a maximum between 300\degree{} and 310\degree{}.
The maximum, which marks the direction of most fans, slowly shifts towards 360\degree{}.
The shift rate is higher than in Ithaca (>~45\degree{} over the whole spring).
The number statistics of fan detection worsens in the late spring in both years, but it is particularly noticeable in late spring of MY30  (see histograms for the late spring of MY30).
This is explained by decreasing contrast between the fan deposits and undisturbed surface around fans in late spring images, i.e.\ the fans blending in with their environment.

\subsection{Manhattan}

The Manhattan region is in a very active area with at least 3 HiRISE ROIs that once were all considered under this same  name.
This area is around southern latitude 86\degree{}, eastern longitude 99\degree{}, as the two above, this is on the edge but still inside the cryptic region.
The ROI is located on the eastern side of a South Polar Layered Deposit (SPLD) trough that in spring is completely covered with seasonal activity.
The area is inclined towards the trough, i.e.\ in the north-west direction, however, rather insignificantly.
According to the HiRISE DTM (\nolinkurl{DTEPC_022259_0935_022339_0935_A01}), there is a \SI{270}{\m} elevation change over approximately \SI{8}{\km} (\SI{\approx{} 2}{\degree} slope).

Manhattan is covered in well developed interlaced araneiforms.
Similar to Giza, the araneiforms here have thin and long troughs and branch significantly.
Aside from araneiforms, the surface in Manhattan is smooth, even on tens to hundreds meters scales with just several exceptions of shallow irregular pits.

Seasonal activity is extensive in Manhattan, with dark fan deposits that at times develop bright halos.
Intriguingly, araneiforms' troughs become visibly brighter compared to the rest of surface around \Ls{} = 200\degree{} and stay bright almost until the region completely defrosts.

Fans in Manhattan are directed 230\degree{} from NA direction in the beginning of spring as shown by the first observations of both analyzed years.
This direction shifts during early spring and plateaus at 290\degree{} after \Ls{}=220\degree{}.

\begin{figure}
\includegraphics[width=0.98\columnwidth]{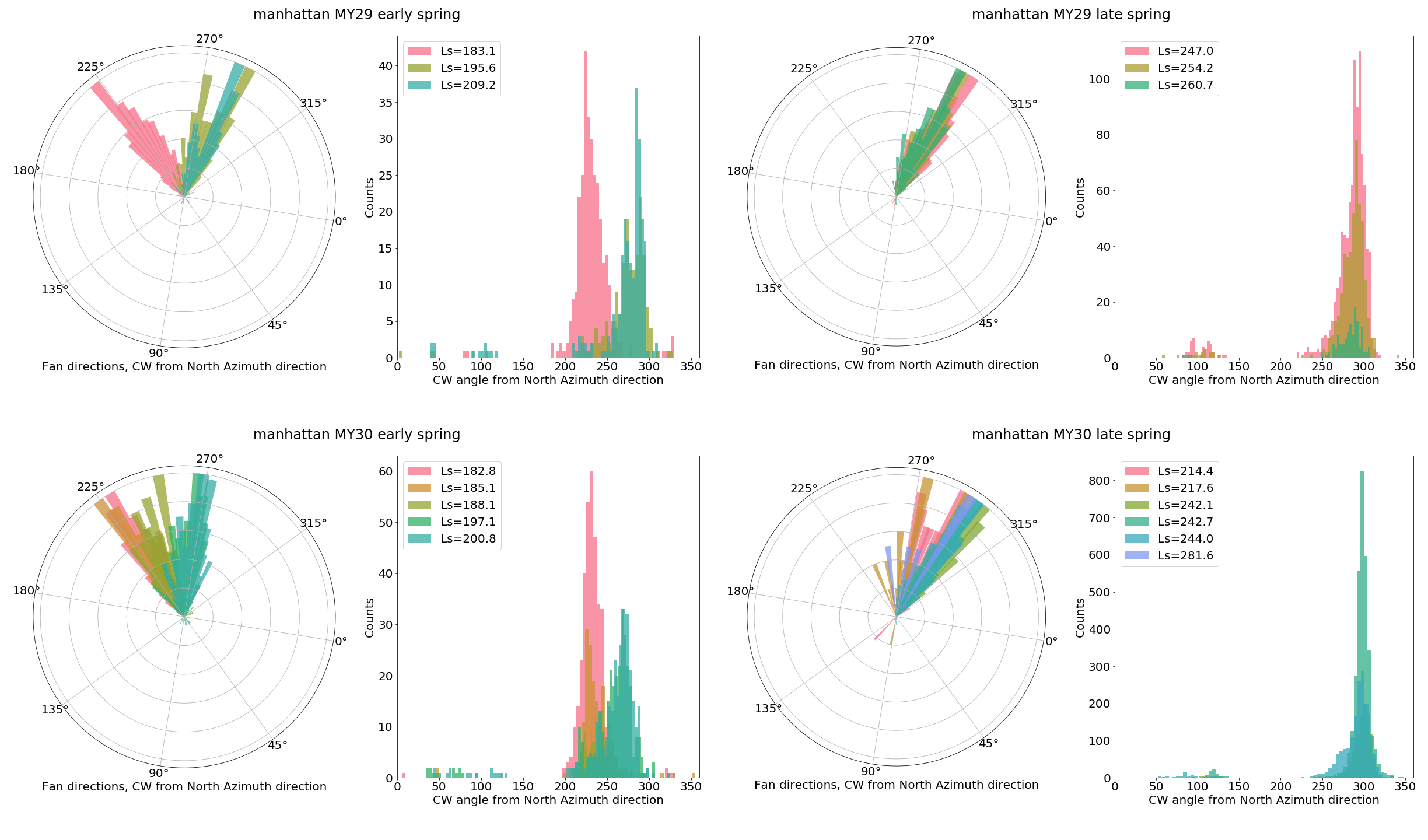}
\caption{\textbf{\label{fig:winds_MH}}
Direction of fan markings in Manhattan region for early and late spring of MY29 and MY30.}
\end{figure}

\subsection{Inca City}

Inca City is at latitude 81.3\degree{}, eastern longitude 295.7\degree{}; relative to the aforementioned ROIs it is on the opposite side of the permanent cap and the southern pole.
The topography of this location is the most complex in our list (HiRISE DTM \nolinkurl{DTEPC_022699_0985_022607_0985_A01}).
It is a system of over 300~m-high ridges that crisscross each other at almost right angles forming close-to-rectangular basins.
The slopes of the ridges sometimes exceed 13\degree{} providing a variety of insolation environments in a relatively small region.
The inner surface of the basins is flat and most of araneiforms of Inca City are carved in it.
The formation of the Inca City ridge system is debated but most commonly attributed to the interaction of irregularities of the local crust with an impact-induced compaction wave \citep{kerber2017}.

Araneiforms in Inca City are morphologically different from those in Giza and Manhattan.
They have a well-developed central depression with relatively short troughs extending outwards and are on average smaller.

Seasonal activity in Inca City starts at the slopes of the ridges \citep{thomas2010}.
Fan deposits extend downwards following gravity lines.
The fans are very narrow but do not have any features of the flows (dark flows come later in spring).
It is not fully clear if the fans are directed by the gravity or by downslope winds in this ROI.\@
The surface around and near araneiforms, in the basin floor, gets  covered mostly in blotches suggesting that no significant winds are active inside the basins.

\begin{figure}
\includegraphics[width=0.98\columnwidth]{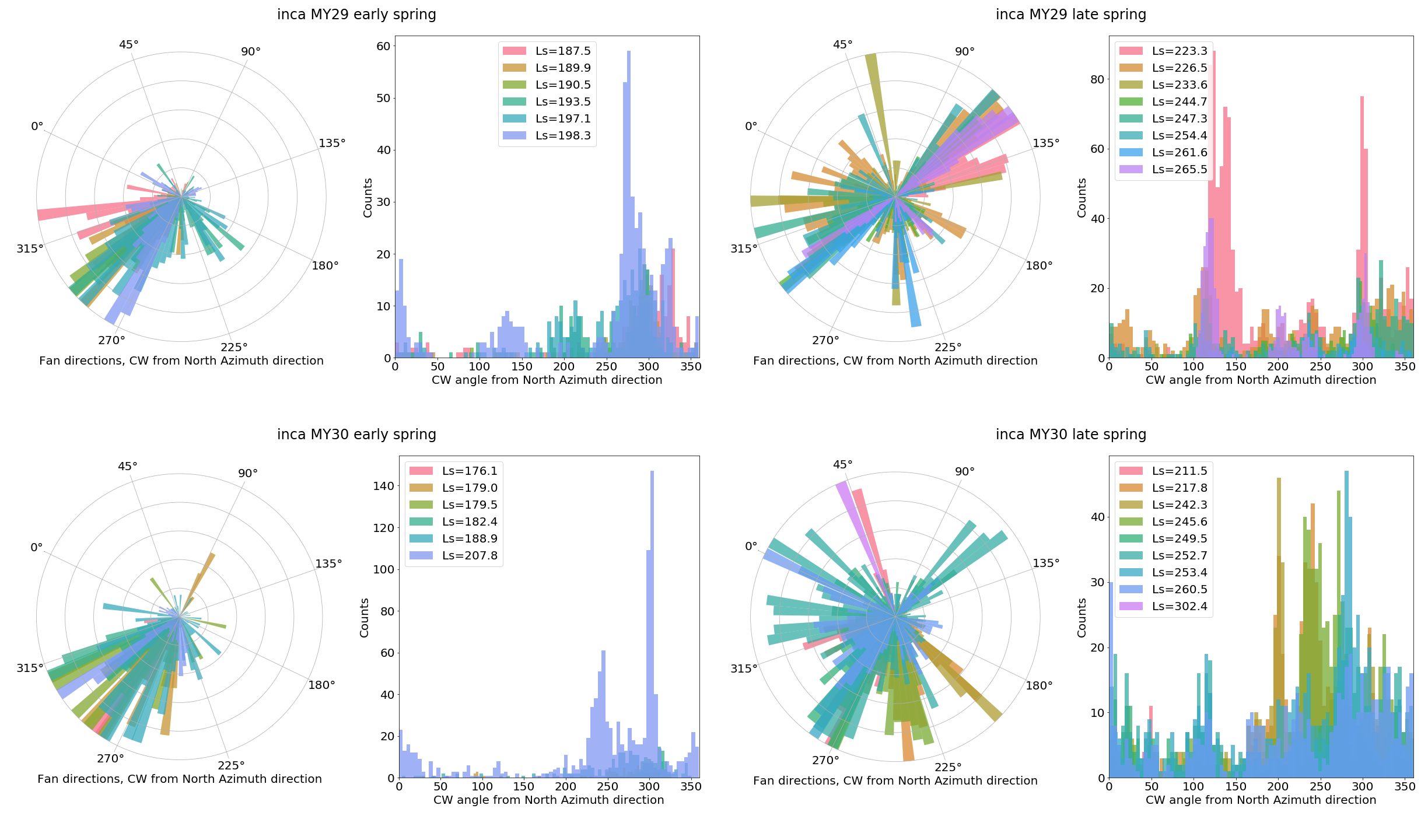}
\caption{\textbf{\label{fig:winds_Inca}}
Direction of fan markings in Inca City region for early and late spring of MY29 and MY30.}
\end{figure}

\begin{figure}
\includegraphics[width=0.98\columnwidth]{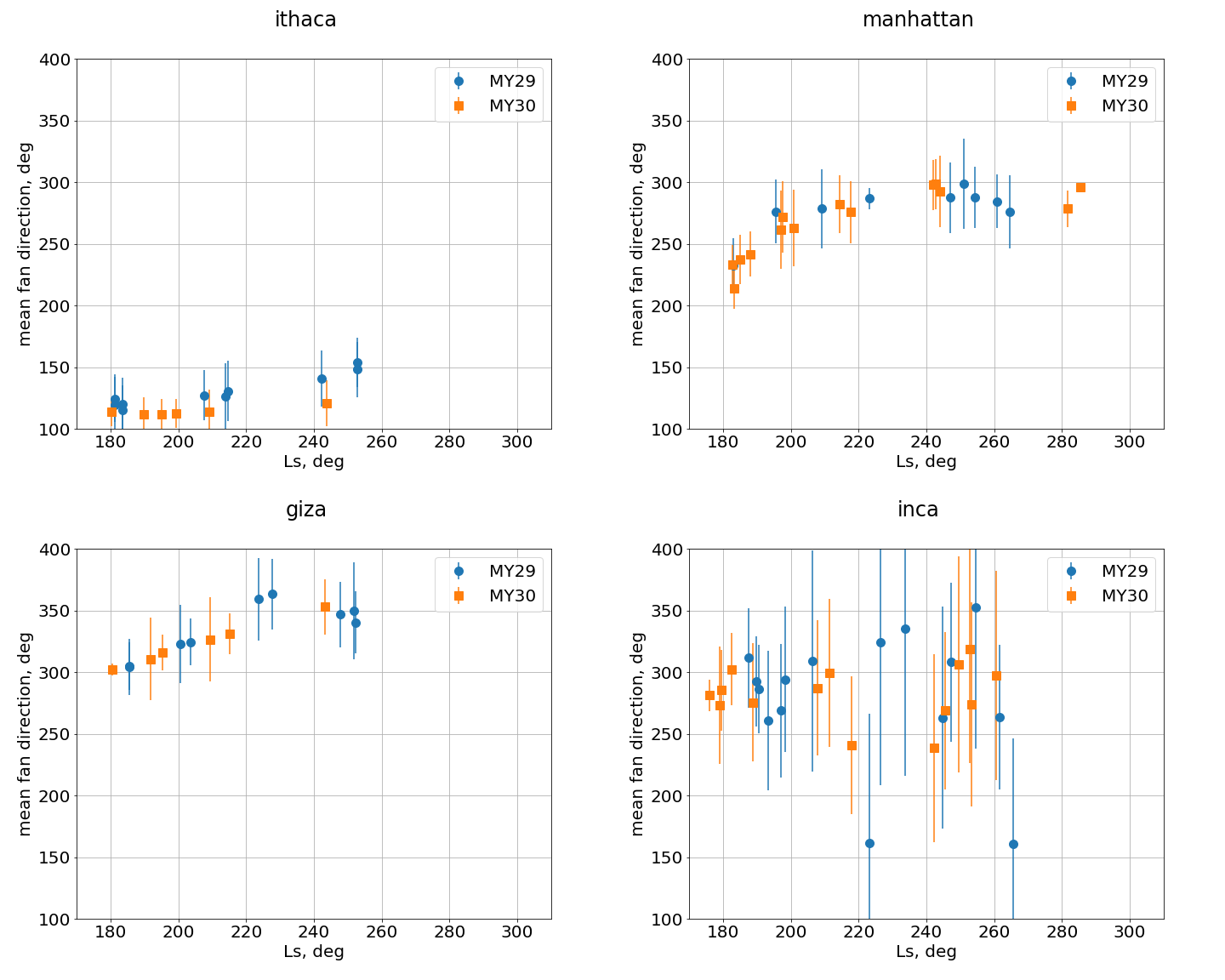}
\caption{\textbf{\label{fig:winds_all_vs_Ls}}
Direction of fan markings in 4 ROIs vs \Ls{} for MY29 and MY30.
Directions are plotted in degrees relative to NA direction.
Error bars represent the standard deviation of the data and not the error on the mean.
Prevailing winds control direction of fans in Ithaca, Manhattan, and Giza because the over-all topography in these ROIs is smooth and has no obstacles significantly modifying the winds.
In Inca City, however, the topography is more prominent, with 3~km-high ridges that break down the general winds and support creation of katabatic flows.
Thus, the fans here follow slopes of the ridges rather than wind direction, which is reflected in the large scatter of mean fan direction and large standard deviations on mean fan direction.}
\end{figure}

Directions of fans in Inca City are seemingly disordered, particularly in comparison to the 3 ROIs discussed above.
However, Inca City is special in this set because it has prominent topography that the other 3 ROIs lack.
Thus the analysis method that works well for our other ROIs might not be applicable to Inca City.
Inca City ridges affect the local deposition of solar energy and influence near-surface winds.
Directions of fans in Ithaca, Giza, and Manhattan are modified by near surface winds that normally pass undisturbed over the whole ROI.\@
In contrast, in Inca City fans are observed almost exclusively on the slopes of the ridges and are aligned with down-slope direction.
However, these fans appear on the slopes gradually through spring: the first fans according to our analysis are pointing to the south-west direction (270\degree{} from NA), i.e.\ located on south-west facing slopes.
Early observations have the smallest standard deviation indicating smallest variation in the fan directions (Fig.~\ref{fig:winds_Inca}).
However, even in the early histograms several local maxima may be detected.
The location of the secondary maxima are determined by the slopes that were covered by HiRISE image at each \Ls{}.
Later in spring the fans start to appear on the slopes with a different orientation than to the south-west.
This widens the histogram for each HiRISE image and makes the location of the histogram maximum a less and less relevant measure of the mean fan direction.
This results in the larger variation of the mean fan direction and large standard deviations (bottom right panel of Fig.~\ref{fig:winds_all_vs_Ls}).
Local maxima repeatedly occurring at the same directions from image to image in late spring and the whole scenario repeats in both years with only small variations.

\section{Conclusions}%
\label{sec:conclusions}

The Planet Four project has produced a catalog of \num{158476} fans and
\num{249801} blotches (ellipses), identifying locations of seasonal surface deposits
produced by the \ce{CO2} jet processes occurring during spring in the
Martian south polar region.
The catalog was generated by combining the
assessments made by Planet Four volunteers reviewing a set of \num{42904}
tiles derived from 221 HiRISE observations obtained over 2 Martian
Years, covering a set of 28 regions of interest (ROI) across the south pole.
To date, this catalog serves as the largest reporting of locations, sizes, and mapping of seasonal deposits on the Martian surface.
The Planet Four fan and blotch catalog constitutes a resource for studying polar winds, climate and polar processes.
Using south polar fans as regional wind markers, the Planet Four catalog can provide tests for and input to global and regional atmospheric circulation models.

Statistical comparisons between classifications produced by the science team and catalog results for the same image data (Section~\ref{sec:data_validation}) demonstrate that the bulk composition of the Planet Four catalog represents a fairly complete picture of the seasonal fans and blotches captured in the HiRISE images.
Trend consistency for fan directions between Mars Year 29 and 30, despite the fact that most data is being analyzed by different volunteers, further indicates reliability of the methods presented here (see summary Figure~\ref{fig:winds_all_vs_Ls}).
We have gone into considerable detail on the methodology behind the data in the catalog and are confident that its content can be productively used by our colleagues for their own research.

For 4 of the 28 ROIs we have presented mean fan directions.
In three of these, the fan deposits appear to be directly modified by near-surface winds at the time of jet eruption; the fourth ROI shows the strong influence of topography.
In ROIs Ithaca, Giza, and Manhattan:
The derived mean winds show no significant inter-annual variability between MY29 and MY30: their direction at the same \Ls{} are the same with less than \SI{10}{\degree} variations.
In Inca City: The mean direction of the fans coincides with the direction of slopes and changes over spring while more slopes become exposed to sunlight and cold jet eruptions happen.

Our analysis in this paper focused on HiRISE observations from seasons 2 (MY29) and 3 (MY30) of the HiRISE southern seasonal processes campaign, and research into inter-annual variability starts to be feasible.
However, the HiRISE campaign covers now 6 seasons of monitoring, and for a number of selected ROIs 5 of these have been or are being analyzed by the Planet Four project at the time of writing.
The results from the analysis of these longer timespans and additional areal coverage will be topics of future publications and data releases.

\section*{Acknowledgements}
The data presented in this paper are the result of the efforts of the Planet Four volunteers, generously donating their time to the Planet Four project, and without whom this work would not have been possible.
Their contributions are individually acknowledged at \href{http://www.planetfour.org/authors}{\nolinkurl{http://www.planetfour.org/authors}}.
Additionally we thank all those involved in BBC Stargazing Live 2013.
This publication uses data generated via the Zooniverse.org platform, development of which was supported by the Alfred P. Sloan Foundation.
The authors also thank Chris Lintott (University of Oxford), who had to decline authorship on this Paper. We thank him for his efforts contributing to the development of the Planet Four website and for his useful discussions.

MES is currently supported by Gemini Observatory, which is operated by the Association of Universities for Research in Astronomy, Inc., on behalf of the international Gemini partnership of Argentina, Brazil, Canada, Chile, and the United States of America.
MES was also supported in part by an Academia Sinica Postdoctoral Fellowship and by a National Science Foundation (NSF)  Astronomy and Astrophysics Postdoctoral Fellowship under award AST-1003258.

CM was supported by the 2014 Institute of  Astronomy and Astrophysics, Academia Sinica (ASIAA) Summer Student Program.
KMA and MES also thank the attendees of the Workshop on Citizen Science in Astronomy for the insightful conversations and acknowledge ASIAA and Taiwan's Ministry of Science and Technology (MOST) for supporting the workshop.
The authors also thank Greg Hines, Cliff Johnson, Margaret Kosmala, Chris Schaller, Brooke Simmons, and Ali Swanson for insightful discussions.

This work is also partially enabled by the National Aeronautics and Space Administration (NASA) support for the \emph{Mars Reconnaisance Orbiter} (MRO) High Resolution Imaging Science Experiment (HiRISE) team.
This paper includes data collected by the MRO spacecraft and the HiRISE camera, and we gratefully acknowledge the entire MRO mission and HiRISE teams' efforts in obtaining and providing the images used in this analysis.
The Mars Reconnaissance Orbiter mission is operated at the Jet Propulsion Laboratory, California Institute of Technology, under contracts with NASA.\@
The authors also thank Rod Heyd for guidance in extracting the geographic and location information for HiRISE non-map projected image.
This research has made use of the USGS Integrated Software for Imagers and Spectrometers (ISIS) and of NASA's Astrophysics Data System.

KMA and GP were supported for this work by NASA ROSES Solar System Workings grant NNX15AH36G.

All software created for the pipeline is based on the open source language Python, using the \emph{matplotlib} library \citep{hunter2007} for plotting, the \emph{pandas} library for data wrangling and analysis \citep{mckinney-proc-scipy-2010}, the \emph{scikit-learn} library \citep{pedregosa2011} for the clustering of Planet Four markings and other pre- and post-processing tasks, the \emph{IPython} and \emph{Jupyter} system for everday computing \citep{perez2007}, and the \emph{SciPy} tools on a daily basis \citep{jones2001}.

\clearpage
\appendix\section{The Zooniverse's Ouroboros Web Plateform}%
\label{sec:ouroboros}

In this Section, we briefly describe the Zooniverse's Ouroboros web platform and  describe how it interacts with the Planet Four classification interface.
The Planet Four website and the Ouroboros platform are both hosted on Amazon Web Services.
This enables the ability to rapidly scale up the number of servers based on the demand on the site, including handling the large number of classifiers during Stargazing Live 2013.
The Planet Four classification interface is a JavaScript and coffee script application that presents the classifier with the HiRISE tile and enables the volunteer to draw markers on the image and submit them for storage in the Planet Four classification database.
The Zooniverse's Ouroboros platform, written in Ruby on Rails, handles the back end storage of classifications in a Mongo database and determines the next tile that should be sent to a given Planet Four classifier for review.

Active tiles are shown to 30--100 classifiers before being retired from rotation.
Once a classification is complete, the Planet Four interface sends the information via the Ouroboros Application Programming Interface (API) to be stored in the database and to update the classification count for the respective tile.
If the activity on the website is low, this step is done immediately.
If site traffic is high, for example 70,000 people on the website at once (such as during launch of the project), Ouroboros is designed to queue the classifications and store them asynchronously to the database so as not to impair the speed and performance of the Planet Four website.
In this case the classification counts for the tiles and the list of tiles a registered or non-registered classifier has seen is not updated in live time

The Planet Four web interface queries the Ouroboros API to identify the next tile to present to a classifier.
Ouroboros checks the database and selects a random active tile that has not been previously reviewed by the Zooniverse registered user or non-logged-in session.
At any given time, Ouroboros readies a list of 5 tiles that the classifier has not seen.
When presented with a request to see another image, the next in this list is sent back by the API.\@
Typically this means the classifier rarely if ever is presented with a tile to review twice.
We note there was a bug in Ouroboros at launch that made repeats more prevalent.
In \ref{sec:dupes_and_spurious} we describe our methods to cleanse duplicate and spurious classifications from our final data reduction.

We note that in the Zooniverse's Ouroboros framework, refreshing the Planet Four interface in the browser will result in a new tile being selected and displayed without updating the classification database.
Refreshing the browser is just as easy as hitting the `Finished' and `Next' buttons to move on to a new image, so we do not believe this has any significant impact on classifier behavior.
We mention it for completeness only.
Also for the majority of the Season 2 and Season 3 classifications, a memory leak in the drawing library would cause the web browser to crash after a rather large number of fans were drawn in the image (approximately over 30--50 sources).
This impacted a very small fraction of tiles.

\section{Handling of Duplicate or Spurious Classifications}%
\label{sec:dupes_and_spurious}
With the Zooniverse Ouroboros queuing system (described in Appendix A), it is possible that a duplicate classification may occur, but these instances should be rare.
A software bug in the Ouroboros platform caused a number of classifiers to receive the same cutout they had previously classified before.
Duplicate classifications are only a small portion of the data-set, comprising \SI{1.9}{\%} percent of all classifications produced, and typically, a few classifications or less per Planet Four image tile were duplicates in those cases.

In order to treat each classification as an independent assessment, we removed all duplicate classifications, keeping only the first response for a given registered user/non-logged-in session for a given cutout.

We also found a concentration of markings positioned at the top left corner (x=0, y=0) of the marking interface, with nearly all having default values for the other recorded parameters.
Only \SI{0.12}{\%} of the \num{9631517} markings recorded for Seasons 2 and 3 are effected.
Further investigation shows that less than \SI{7}{\%} of fan and blotch markings with default parameters with x=0 or y=0 are not centered at the origin.
Thus, we believe these origin default-valued markings are due to a javascript error.
Therefore, we simply delete them from the database, but keep any other markings associated with those effected classifications.
Additionally 33 markings (\SI{\sim{}0.003}{\%} of all entries in the Planet Four classification database) do not have all of the required parameters that should have been recorded.
We believe this is to due a singularity in the drawing tool for that marker, and we remove that entry from the database.
There are also positions in the database recorded for a handful of fans and blotches significantly out of the bounds of the user interface.
A classifier can move a marker drawn outside the edge of the image, to better capture the center position of a feature, but these positions are well outside the image region.
This represents well less than \SI{1}{\%} of all classifications, and we have removed them from the analysis presented here.
All statistics and values reported in this Paper are after the filtering described above.

\section{Raw Classification Data}%
\label{sec:raw_classification}
Here we provide additional details about the raw classification data provided in the online supplementary data file\footnote{\nolinkurl{2018-02-11_planet_four_classifications_queryable_cleaned_seasons2and3.h5}}.
It is written in the binary HDF5 format, in the variant produced by the \emph{pandas} library (supported by the PyTables library\footnote{\href{http://pandas.pydata.org/pandas-docs/stable/io.html\#hdf5-pytables}{http://pandas.pydata.org/pandas-docs/stable/io.html\#hdf5-pytables}}).

The general structure is as follows:
Each classification submission by an individual volunteer creates a \emph{classification\_id}.
All objects created by this volunteer receives the same \emph{classification\_id}, with the marking data for each object being one entry in the classification database.
Each data row also has a \emph{marking} column that identifies if this data is for a fan, a blotch, an interesting feature that will have the string value ``interesting'' in the \emph{marking} column, or ``none'', when the volunteer did not create any marking object.
Below we describe the columns available in this database:

\begin{longtable}{p{0.25\linewidth}p{0.29\linewidth}p{0.38\linewidth}}
\toprule
\textbf{Column name} & \textbf{Example value} & \textbf{Description} \\
\midrule
classification\_id & 50ecaaf760d4050d21000414 & Unique ID for each classification by a Planet Four volunteer \\
created\_at & 2013-01-08 23:25:43 & time of submission \\
tile\_id & APF0000p9t & Planet Four tile identifier \\
image\_name & ESP\_021491\_0950 & HiRISE observation identifier \\
tile\_url & \url{http://www.planetfour.org/subjects/standard/50e741555e2ed211dc002346.jpg} & URL to image data for this Planet Four tile \\
user\_name &  abc & Originally, the Zooniverse username or non-logged-in session ID. For privacy concerns, we have converted these to anonymous IDs.\\
marking & blotch & identifier for what data in row is for: blotch, fan, interesting, none \\
x\_tile & 1 & x coordinate of tile inside larger HiRISE image frame. Starts at 1 in upper left of the HiRISE image, increases to the right. \\
y\_tile & 2 & y coordinate of tile inside larger HiRISE image frame. Starts at 1 in upper left of the HiRISE image and increase downwards. \\
acquisition\_date & 2011-01-01 00:00:00 & date only for HiRISE observation time (ignore hours) \\
local\_mars\_time & 5:43 PM & local mars time for given acquisition date \\
x             &         553.65 & x pixel coordinate of object in Planet Four tile. Starts at 0 in upper left, increases to the right. \\
y              &       355.817 & y pixel coordinate of object in Planet Four tile. Starts at 0 in upper left, increases downwards. \\
image\_x      &      2033.65 & x pixel coordinate of object in original HiRISE image. Starts at 0 in upper left, increases to the right. \\
image\_y       &     37071.8 & y pixel coordinate of object in original HiRISE image. Starts at 0 in upper left, increasing downwards. \\
radius\_1      &    295.195 & Semi-major axis of blotch object in pixels. NAN if not applicable (N/A) \\
radius\_2      &   294.715 & Semi-minor axis of blotch object in pixels. NAN if N/A \\
distance        &   NaN & Length of fan object in pixels. NAN if data row is for blotch or interesting \\
angle        &      27.4331 & Orientation of marking object with respect to tile image x-axis in degrees.
Positiv clock-wise, zero to image right (same definition as HiRISE) \\
spread       &                 NaN & Opening angle of fan objects in degrees. NAN if N/A \\
version        &                NaN & version of tool used to create fan. NAN if N/A \\
x\_angle        &          0.887549 & cartesian x coordinate of \emph{angle} column on unit circle\\
y\_angle        &         0.460713 & cartesian y coordinate of \emph{angle} column on unit circle\\
\bottomrule
\end{longtable}

The Planet Four classification interface recorded a different angle than the intended spread angle from the fan marking tool.
This was identified and subsequently fixed in the software.
The correct spread angle is recoverable from the values stored in the database.
We denote those markings generated before the patch with version flag set to 1.0 and those after with the version flag set to 2.0.
We provide the corrected spread angle for the fans affected, but leave that version flag in the final catalog, for reference.
To gather statistics on the understanding of the tutorial, the Planet Four classification database contains all the tutorial markings, indicated by a HiRISE image name of `tutorial'.
For the delivered raw classification database, the fan angles range has been converted from -180--180 to 0--360, while the range of the blotch angles have been converted to 0--180, due to their rotational symmetry.

\section{Pipeline outputs}%
\label{sec:pipeline_outputs}
The intermediate stages of the pipeline, as output by our clustering and combination pipeline are identified with different level identifiers 1A, 1B, and 1C, indicating different stages of the processing pipeline, where the processing is done on a per-tile-id level.
After this is done, the final step of  combines all the data from the ten-thousands of tile\_id folders into a set of summarizing CSV files.

\subsection{Directory file structure}
The directory file structure of the pipeline products are as follows (examples in parentheses):
\begin{itemize}
    \item HiRISE observation ID (\nolinkurl{ESP_011350_0945})
    \begin{itemize}
        \item Planet Four tile ID (\nolinkurl{APF0000any})
        \begin{itemize}
            \item Level 1A (\nolinkurl{L1A/APF0000any_L1A_fans.csv})
            \item Level 1B (\nolinkurl{L1B/APF0000any_L1B_fnotches.csv})
            \item Level 1C with cut value 0.5 in directory name (\nolinkurl{L1C_cut_0.5/APF0000any_L1C_cut_0.5_blotches.csv})
        \end{itemize}
    \end{itemize}
\end{itemize}

with the list of HiRISE observation IDs identifying the HiRISE observations that went into Planet Four for this database.

\subsection{Pipeline stage levels}
\subsubsection{Level 1A}
Level 1A is the data that is directly output from clustering and averaging the cluster members into average markings, as described in Section~\ref{sec:clustering}.
Here, the biggest reduction in terms of numbers of objects in the system occurs, as all the different volunteers data are being combined into one object when the clustering process has determined the markings to be part of one cluster.
All newly created average fans and blotches are summarized into one fan and blotch summary file respectively, which each line representing the mean object from averaging all cluster members.
As an example, the content of \nolinkurl{APF0000p3q_L1A_fans.csv} is shown below.
When the column name matches those given in Appendix~\ref{sec:raw_classification}, they have the same meaning.
The two new columns are \emph{n\_votes}, which records how many members the cluster had that was used to produce this averaged object, and \emph{marking\_id}, which have been created at this stage of the pipeline and serve as a tracer throughout the different pipeline outputs:

\vspace{0.5cm}
\setlength\tabcolsep{2pt}
\noindent\begin{tabular}{lrrrrrrrr}
\toprule
{} &  x\_tile &  y\_tile &           x &           y &     image\_x &       image\_y &  radius\_1 &  radius\_2 \\
\midrule
0 &     2.0 &    26.0 &  123.611111 &  455.666667 &  863.611111 &  14155.666667 &       NaN &       NaN \\
1 &     2.0 &    26.0 &  157.000000 &  391.800000 &  897.000000 &  14091.800000 &       NaN &       NaN \\
\bottomrule
\end{tabular}
\\
\begin{tabular}{lrrrrrrrl}
\toprule
{} &   distance &       angle &     spread &  version &   x\_angle &   y\_angle &  n\_votes &    image\_id \\
\midrule
0 &  81.884266 &  223.712817 &  71.559689 &      1.0 & -0.691035 & -0.660663 &        9 &  APF0000any \\
1 &  57.742472 &  248.754137 &  52.521798 &      1.0 & -0.360802 & -0.927999 &       10 &  APF0000any \\
\bottomrule
\end{tabular}
\\
\begin{tabular}{lll}
\toprule
{} &       image\_name & marking\_id \\
\midrule
0 &  ESP\_011350\_0945 &    F006de3 \\
1 &  ESP\_011350\_0945 &    F006de4 \\
\bottomrule
\end{tabular}

\vspace{0.5cm}
Additionally, each L1A folder contains a text file called \nolinkurl{clustering_setttings.yaml} that summarizes the clustering settings used for these data for reference.
\emph{epsilon} values are static and all the same, but the \emph{min\_samples} value is dynamically calculated, see Section~\ref{sec:cluster_parameters} for details.

\subsubsection{Level 1B}
At level 1B, the combination pipeline has determined with objects are so close to each other that they should be considered for merging (see Section~\ref{sec:fnotching}).
The outputs are between one and three files this time.
One only, in case all fans and blotches found were so close that they need to be evaluated by their classification votes.
Usually, though, there are two to three files, where one files stores the objects that need voting, and the other file(s) store the objects that don't have any close neighbors and will simply be copied over to the final level later. The fans and blotches in these latter files will receive the `vote\_ratio' value of 1.0, indicating that they had a ``perfect'' probability for being a fan, or blotch, respectively.
The third file that keeps the close objects for the later thresholding contains these temporary meta-objects in sets of 2 rows, one fan and one blotch, and has the term ``fnotch'' in its filename (fnotches: FaN--blOTCH).
This file contains all the clustering statistics data from L1A required to make a cut decision for L1C, with the data for each meta-object being sorted in alternating rows.
Here are the first four rows of the fnotch file \nolinkurl{APF0000any_L1B_fnotches.csv}:

\vspace{0.5cm}
\noindent
\begin{tabular}{lrrllrr}
\toprule
{} &       angle &   distance &    image\_id &       image\_name &     image\_x &       image\_y \\
\midrule
fan    &  223.712817 &  81.884266 &  APF0000any &  ESP\_011350\_0945 &  863.611111 &  14155.666667 \\
blotch &   67.261720 &        NaN &  APF0000any &  ESP\_011350\_0945 &  838.395834 &  14123.875000 \\
fan    &  247.146845 &  58.742330 &  APF0000any &  ESP\_011350\_0945 &  832.000000 &  14306.400000 \\
blotch &   70.684606 &        NaN &  APF0000any &  ESP\_011350\_0945 &  821.666667 &  14281.428571 \\
\bottomrule
\end{tabular}
\\
\begin{tabular}{llrrrrrrr}
\toprule
{} & marking\_id &  n\_votes &   radius\_1 &   radius\_2 &     spread &  version &           x &   x\_angle \\
\midrule
fan    &    F006de3 &        9 &        NaN &        NaN &  71.559689 &      1.0 &  123.611111 & -0.691035 \\
blotch &    B0071f2 &        8 &  49.309277 &  36.981958 &        NaN &      NaN &   98.395834 &  0.379131 \\
fan    &    F006de5 &        5 &        NaN &        NaN &  81.171448 &      1.0 &   92.000000 & -0.387419 \\
blotch &    B0071ed &        7 &  35.324591 &  26.493443 &        NaN &      NaN &   81.666667 &  0.217508 \\
\bottomrule
\end{tabular}
\\
\begin{tabular}{lrrrrr}
\toprule
{} &  x\_tile &           y &   y\_angle &  y\_tile &  vote\_ratio \\
\midrule
fan    &     2.0 &  455.666667 & -0.660663 &    26.0 &    0.539412 \\
blotch &     2.0 &  423.875000 &  0.907431 &    26.0 &    0.460588 \\
fan    &     2.0 &  606.400000 & -0.919245 &    26.0 &    0.426667 \\
blotch &     2.0 &  581.428571 &  0.852341 &    26.0 &    0.573333 \\
\bottomrule
\end{tabular}
\vspace{0.5cm}

This data stage L1B is what can be used to create a different significance threshold cut for the final data , by filtering on the data column \emph{vote\_ratio} in the fnotch file for the required threshold value.
For example, if a higher threshold on the probability for a fan is wanted, e.g.\ 0.8, one would filter out all rows that start with ``fan'' with a \emph{vote\_ratio} value below 0.8.
One then needs to decide if one wants to use this threshold as a general ``certainty'' filter and simply don't take any object with a \emph{vote\_ratio} < 0.8, or if one wants the blotch to appear instead of a fan.

\subsubsection{Level 1C}
This level contains the data of the final catalog files, but split-up into each Planet Four tiles.
At the end of the thresholding stage (Section~\ref{sec:fnotching}), appending the data for the rows that pass the threshold filters into the respective blotch and fan files and copying these completed files into the L1C directory completes that thresholding step and fills up the L1C folders.
A final tool walks through each folder and collects all the fan and blotch data into one summary file each, followed by merge operations with meta-data that is useful for future analysis.
These files are described in the next section, \ref{sec:catalog}.

\newpage
\section{Planet Four Catalog files description}%
\label{sec:catalog}

Our catalog product files consist of one CSV result file per fan and blotch markings, a Planet Four tile meta-data file, and a HiRISE observation meta-data file.
Below, each subsection describes the data columns for these files.

For convenience we provide both the planeto-centric and planeto-graphic latitudes for each fan's base and blotch's center point.
Longitudes are measured 0--360, increasing positive to the East.
Note that, because the HiRISE images were not co-registered, the conversion of pixel to geographical coordinates can be offset by up to 100 HiRISE pixels between data from different HiRISE images.

\subsection{Fan catalog}
\begin{longtable}{p{0.25\linewidth}p{0.15\linewidth}p{0.52\linewidth}}
\toprule
\textbf{Column name} & \textbf{Example value} & \textbf{Description} \\
\midrule
marking\_id & F00004ab & Consistent identifier for marking after clustering. Fxxx=Fan, Bxxx=Blotch \\
angle       & 185.4 & Alignment angle of marking measured from 3 o'clock direction, clockwise \\
distance    & 179.6 & Length of fan in pixels \\
tile\_id    & \nolinkurl{APF0000cia} & tile identifier in the Planet Four system \\
image\_x     & 3391.2 & Base X coordinate [px] in original HiRISE image \\
image\_y     & 5640.6 & Base Y coordinate [px] in original HiRISE image \\
n\_votes     & 15 & \# of markings that went into this average object. \\
obsid       & \nolinkurl{ESP_012079_0945} & HiRISE image observation id \\
spread      & 21.346 & Spreading angle of Fans \\
version     & 1 & Version number of Fan model used in Planet Four (see Appendix~\ref{sec:raw_classification})\\
vote\_ratio  & 1.0 & Ratio of votes from a potential combination step.
Value of 1.0 means only fan votes occurred. \\
x           & 431.206 & Base X pixel coordinate in the Planet Four tile \\
y           & 160.6 & Base Y pixel coordinate in the Planet Four tile \\
x\_angle     & -0.995088 & Polar X coordinate of alignment angle \\
y\_angle     & -0.0938355 & Polar Y coordinate of alignment angle \\
l\_s         & 214.785 & Solar longitude of HiRISE observation \\
map\_scale   & 0.25 & Factor for scaling distances to correct for HiRISE binning mode \\
north\_azimuth  & 126.857 & Direction of North in the original unprojected HiRISE input image \\
BodyFixedCoordinateX  & -67.2071 & Base X coord.\@ [km] in Mars-fixed ref.\@ frame \\
BodyFixedCoordinateY  & 257.05 & Base Y coord.\@ [km] in Mars-fixed ref.\@ frame \\
BodyFixedCoordinateZ  & -3370.63 & Base Z coord.\@ [km] in Mars-fixed ref.\@ frame \\
PlanetoCentricLatitude & -85.493 &  Latitude of catalog object (-centric) \\
PlanetoGraphicLatitude & -85.5457 & Latitude of catalog object (-graphic) \\
Longitude  & 104.652 & Longitude of catalog object \\
\bottomrule
\end{longtable}

\subsection{Blotch catalog}

\begin{longtable}{p{0.25\linewidth}p{0.15\linewidth}p{0.52\linewidth}}
\toprule
\textbf{Column name} & \textbf{Example value} & \textbf{Description} \\
\midrule
marking\_id & B00004ab & Consistent identifier for marking after clustering. Fxxx=Fan, Bxxx=Blotch \\
angle       & 185.4 & Alignment angle of marking measured from 3 o'clock direction, clockwise \\
tile\_id    & \nolinkurl{APF0000cia} & tile identifier in the Planet Four system \\
image\_x     & 3391.2 & Center X pixel coordinate in the original HiRISE image \\
image\_y     & 5640.6 & Center Y pixel coordinate in the original HiRISE image \\
n\_votes     & 15 & Number of markings used for the average object \\
obsid       & \nolinkurl{ESP_012079_0945} & HiRISE image observation id \\
radius\_1    & 10.4 & Semi-major axis of Blotch \\
radius\_2    & 15.2 & Semi-minor axis of Blotch \\
vote\_ratio  & 0.0 & Ratio of votes from a potential combination step.
Value of 0.0 means only blotch votes occurred. \\
x           & 431.206 & Center X pixel coordinate in the Planet Four tile \\
y           & 160.6 & Center Y pixel coordinate in the Planet Four tile \\
x\_angle     & -0.995088 & Polar X coordinate of alignment angle \\
y\_angle     & -0.0938355 & Polar Y coordinate of alignment angle \\
l\_s         & 214.785 & Solar longitude of HiRISE observation \\
map\_scale   & 0.25 & Factor for scaling distances to correct for HiRISE binning mode \\
north\_azimuth  & 126.857 & Direction of North in the original unprojected HiRISE input image \\
BodyFixedCoordinateX  & -67.2071 & Center X coord.\@ [km] in Mars-fixed ref.\@ frame \\
BodyFixedCoordinateY  & 257.05 & Center Y coord.\@ [km] in Mars-fixed ref.\@ frame \\
BodyFixedCoordinateZ  & -3370.63 & Center Z coord.\@ [km] in Mars-fixed ref.\@ frame \\
PlanetocentricLatitude & -85.493 &  Latitude of catalog object (-centric) \\
PlanetographicLatitude & -85.5457 & Latitude of catalog object (-graphic) \\
Longitude  & 104.652 & Longitude of catalog object (Positive East 360)\\
\bottomrule
\end{longtable}

\subsection{Planet Four tile catalog}
Here we provide the data required to position the Planet Four tiles both back into HiRISE images, if so required, or directly onto the Martian surface, by using the provided latitude/longitude values or their map-value equivalents in the BodyFixed-Mars frame in a rectangular coordinate system, measuring kilometers from the south pole.
The coordinate values come directly from the ISIS campt utility, while the x\_tile and y\_tile position indices of tiles inside the HiRISE image are the result of the splitting up routine that was developed by the Zooniverse team at the beginning of the project.
All coordinates were calculated at the tile center pixel coordinate of (420, 324).
The decimal digits precision was set to 7, guided by the Latitude/Longitude significant bits for a HiRISE pixel diameter on the ground for a 1x1 binning observation.

\vspace{0.5cm}
\noindent
\begin{tabular}{p{0.27\linewidth}p{0.15\linewidth}p{0.52\linewidth}}
\toprule
\textbf{Column name} & \textbf{Example value} & \textbf{Description} \\
\midrule
BodyFixedCoordinateX  & -67.2071 & Center X coord.\@ [km] in Mars-fixed ref.\@ frame \\
BodyFixedCoordinateY  & 257.05 & Center Y coord.\@ [km] in Mars-fixed ref.\@ frame \\
BodyFixedCoordinateZ  & -3370.63 & Center Z coord.\@ [km] in Mars-fixed ref.\@ frame \\
PlanetocentricLatitude & -85.493 &  Latitude of catalog object (-centric) \\
PlanetographicLatitude & -85.5457 & Latitude of catalog object (-graphic) \\
Longitude  & 104.652 & Longitude of catalog object (Positive East 360)\\
tile\_id   & \nolinkurl{APF0000cia} & tile identifier in the Planet Four system \\
obsid      & \nolinkurl{PSP_003092_0985} & HiRISE observation ID of the source image for this tile \\
x\_hirise  & 840 & X pixel coordinate of the tile center in the HiRISE image \\
x\_tile    & 5 & X index of the Planet Four tile inside the HiRISE image (1-based) \\
y\_hirise  & 648 & Y pixel coordinate of the tile center in the HiRISE image \\
y\_tile    & 11 & Y index of the Planet Four tile inside the HiRISE image (1-based) \\
\bottomrule
\end{tabular}

\subsection{HiRISE observations catalog}
This catalog provides the user with a list of HiRISE images and their meta-data that were used to create the Planet Four results presented here.
The columns with capital letters were directly taken from the published cumulative EDR index\footnote{\url{https://hirise-pds.lpl.arizona.edu/PDS/INDEX/EDRCUMINDEX.TAB}}.
The decimal digits precision was set to 7, guided by the Latitude/Longitude significant bits for a
HiRISE pixel diameter on the ground for a 1x1 binning observation.

\begin{longtable}{p{0.3\linewidth}p{0.22\linewidth}p{0.45\linewidth}}
\toprule
\textbf{Column name} & \textbf{Example value} & \textbf{Description} \\
\midrule
\nolinkurl{OBSERVATION_ID} & \nolinkurl{ESP_011296_0975} & HiRISE observation identifier \\
\nolinkurl{IMAGE_CENTER_LATITUDE} & -82.1965000 & Planetographic latitude of the HiRISE image center \\
\nolinkurl{IMAGE_CENTER_LONGITUDE} & 225.2530000 & Longitude of HiRISE image center (positive west 360) \\
\nolinkurl{SOLAR_LONGITUDE} & 178.8330000 & Solar longitude of HiRISE image. Equivalent to column \emph{l\_s} in the fan and blotch catalogs. \\
\nolinkurl{START_TIME} & 2008-12-23 16:15:26 & UTC time of observation start \\
\nolinkurl{map_scale} & 1.0000000 & Units: pixel/m. Calculated from EDRCUMINDEX by 0.25*BINNING \\
\nolinkurl{north_azimuth} & 110.6001067 & The median north azimuth value for the HiRISE image, recalculated with ISIS' campt, due to known errors in HiRISE EDR index file. \\
\texttt{\# of tiles} & 91 & the number of created Planet Four tiles per HiRISE observation. Depends on original image size. \\
\bottomrule
\end{longtable}

\section{Extended validation results}%
\begin{figure}[htb]
\centering
\includegraphics[width=1\columnwidth]{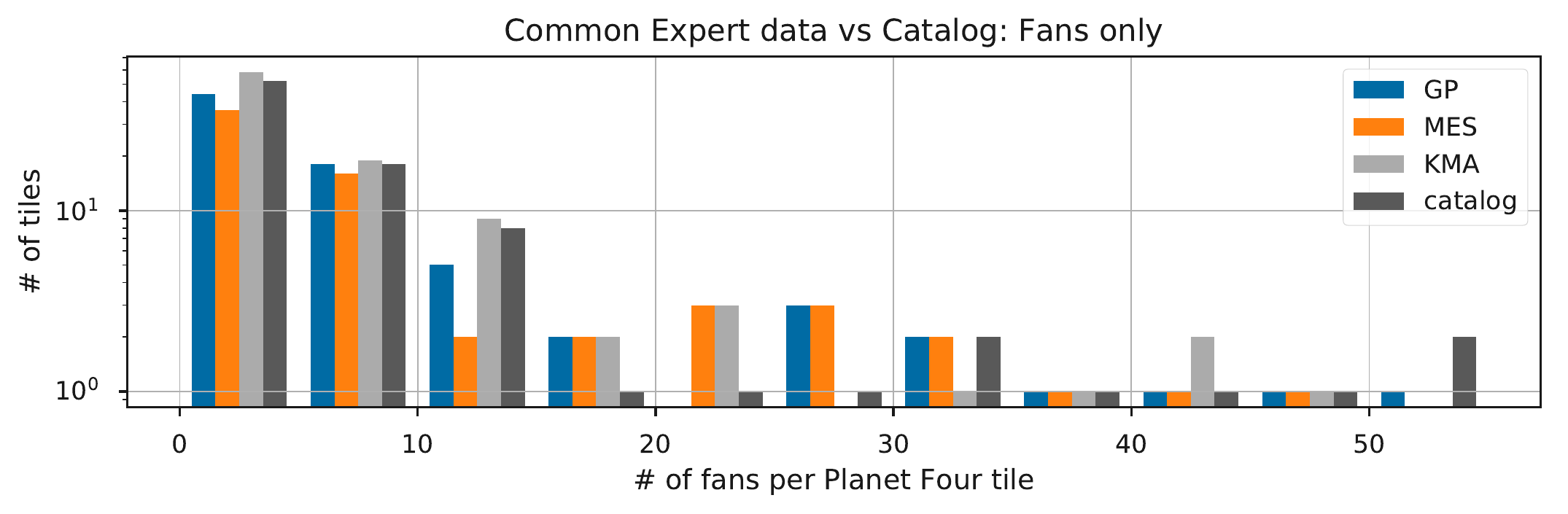}
\caption{\label{fig:gold_fans_common}
Comparing numbers of identified fans per Planet Four tile between experts and the catalog data; here, for the 192 tile\_ids that were classified by all experts.
Bin size is 5, each bin is directly compared between the data from all experts GP (blue), MES (orange), KMA (grey) and the catalog results (brown).
Binning max was cut off at 60, omitting single entry bins above.
}
\end{figure}

\begin{figure}[htb]
\centering
\includegraphics[width=1\columnwidth]{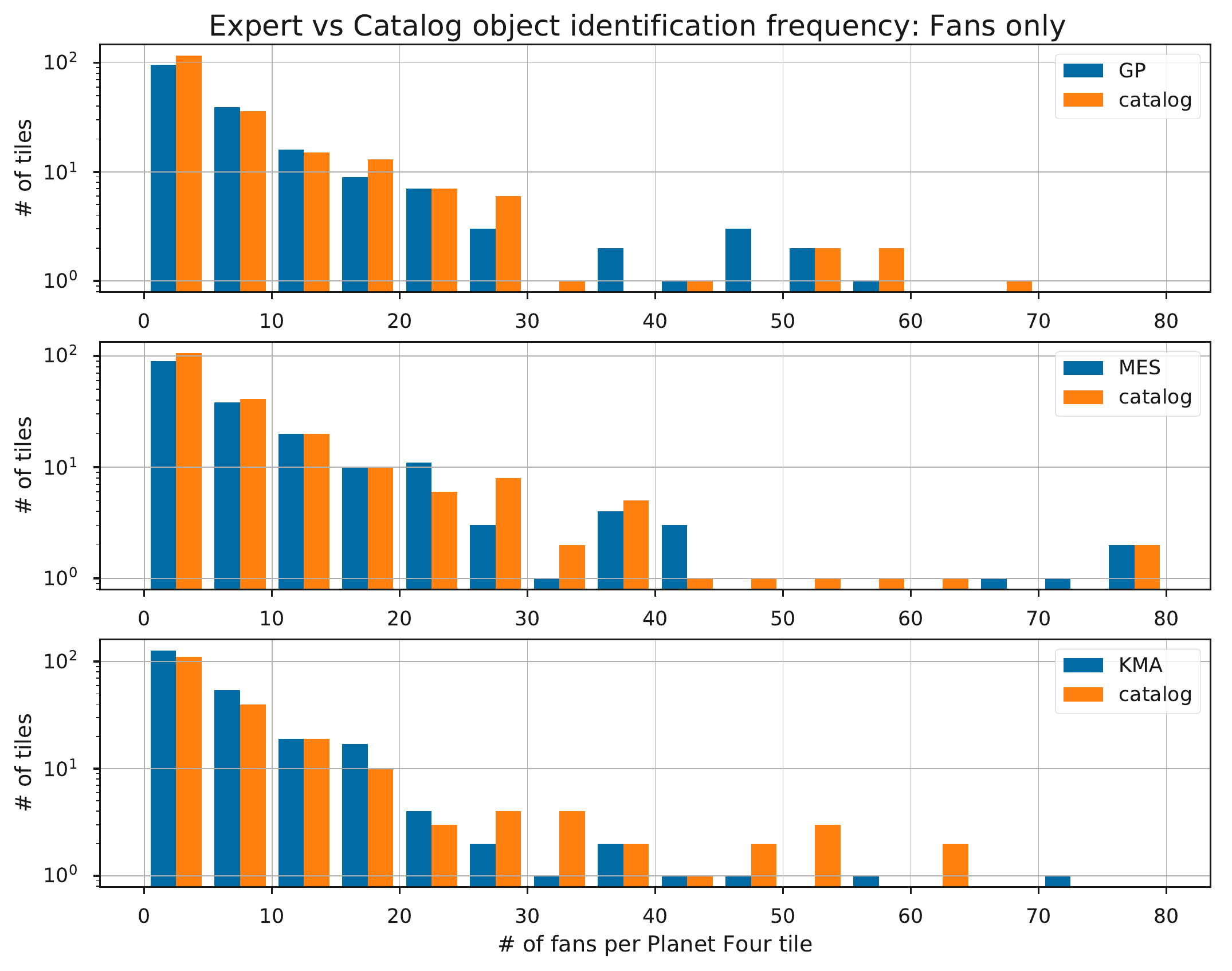}
\caption{\label{fig:gold_fans}
Comparing numbers of identified fans per Planet Four tile between experts and the catalog data.
Bin size is 5, each bin is directly compared between the data from all experts GP (blue), MES (orange), KMA (grey) and the catalog results (brown).
Binning max was cut off at 85, omitting single entry bins above.
}
\end{figure}

\begin{figure}[p]
\centering
\vspace{-2cm}
\includegraphics[width=1\columnwidth]{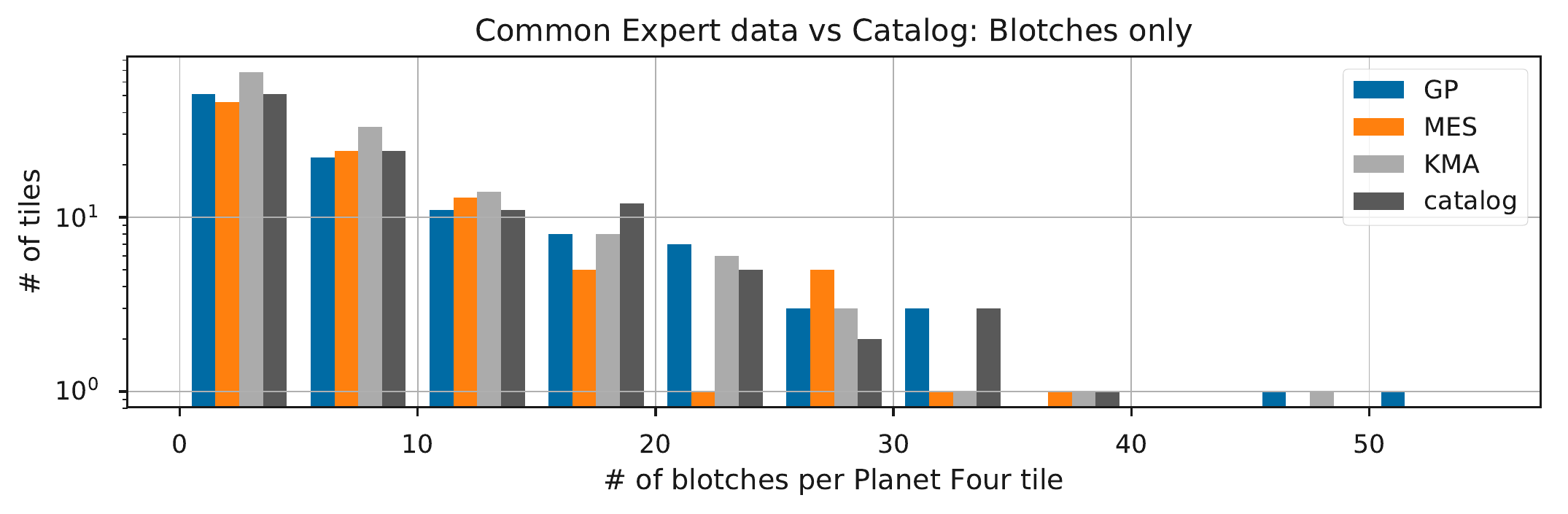}
\caption{\label{fig:gold_blotches_common}
Comparing numbers of identified blotches per Planet Four tile between experts and the catalog data; here, for the 192 tile\_ids that were classified by all experts.
Bin size is 5, each bin is directly compared between the data from all experts GP (blue), MES (orange), KMA (grey) and the catalog results (brown).
Binning max was cut off at 60, omitting single entry bins above.
}
\centering
\includegraphics[width=1\columnwidth]{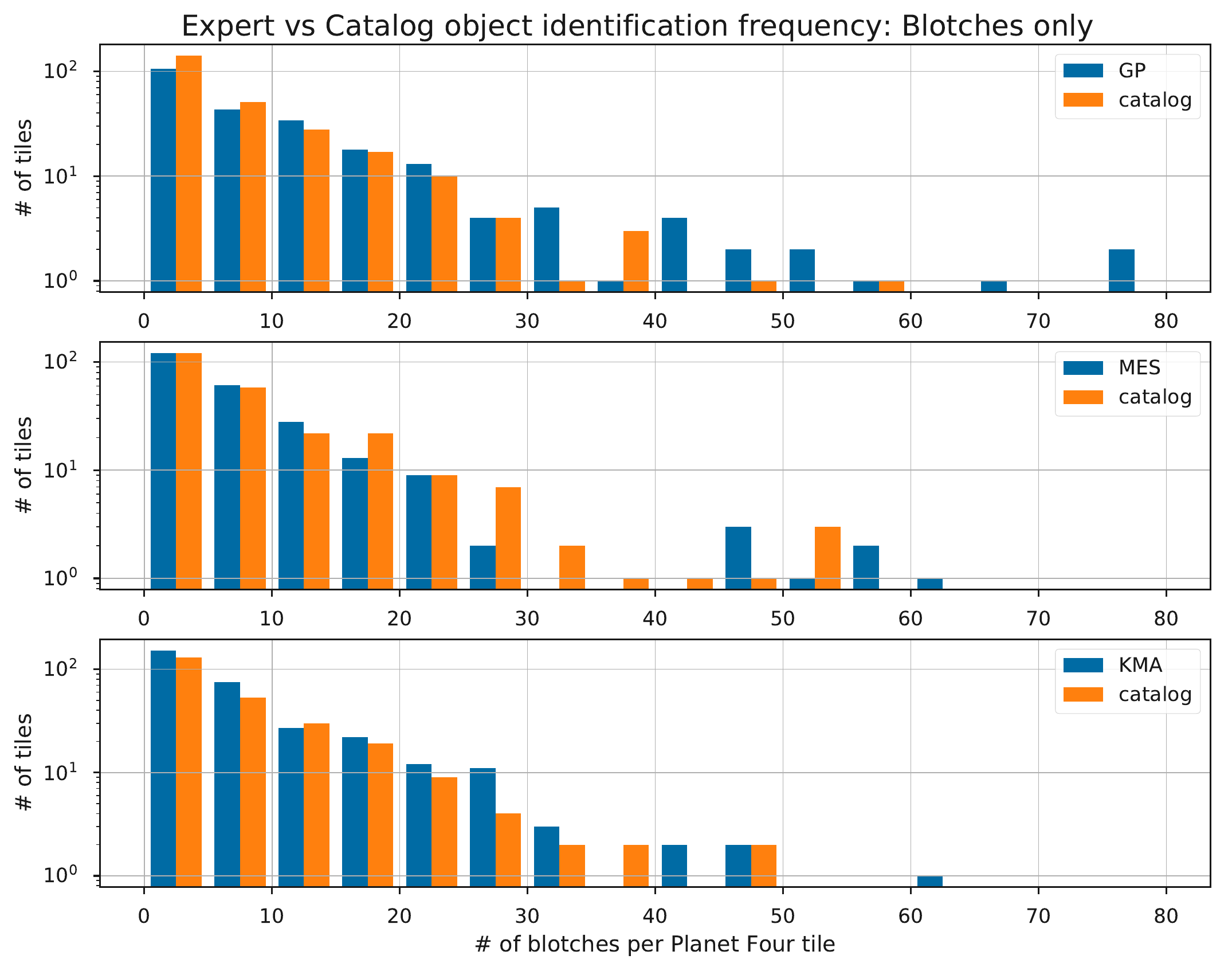}
\caption{\label{fig:gold_blotches}
Comparing numbers of identified blotches per Planet Four tile between experts and the catalog data.
Bin size is 5, each bin is directly compared between the data from all experts GP (blue), MES (orange), KMA (grey) and the catalog results (brown).
Binning max was cut off at 85, omitting single entry bins above.
}
\end{figure}

\label{sec:extended_validation}
In addition to the combined fan and blotch count we explored in Section~\ref{sec:data_validation}, we further explore here how well the Planet Four catalog identifies fans (those dark sources with a clear direction and starting point) versus blotches, separately.
We separate the catalog and gold standard classifications by marker type in Figures~\ref{fig:gold_fans_common} to \ref{fig:gold_blotches}.
The data processing pipeline plays a significant role in the completeness of the catalog.
At the Thresholding stage, our data processing algorithm determines which clusters will ultimately become fans with a value of P(fan) > 0.5.
Like for the total number of sources, the number distribution of fans and the number distribution of blotches matches the expert assessments and is within the 3-$\sigma$ uncertainty \citep{kraft1991}.
Thus, in most cases where the science team member marked a fan, the catalog also identifies this source as fan.
Based on these results, we have high confidence in our fan and blotches identifications within the Planet Four catalog.


\subsection{Example tile comparisons}
\label{sec:gold_tile_comparison}
\begin{figure}[htb]
\centering
\includegraphics[width=0.95\columnwidth]{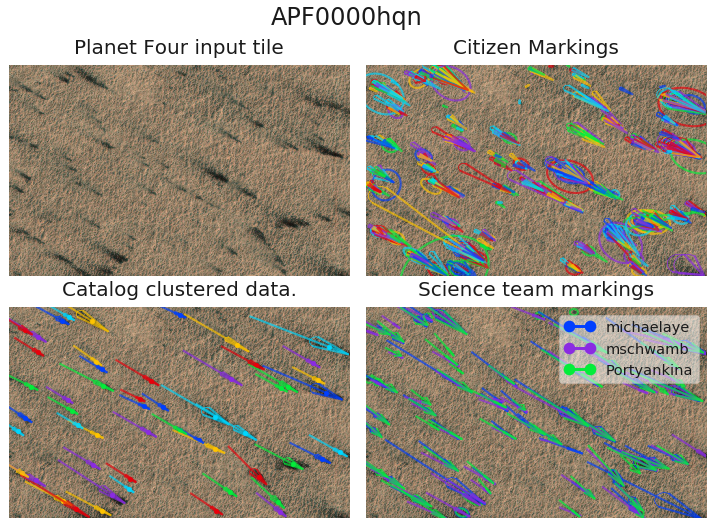}
\caption{\label{fig:gold_data62}Comparing volunteers' markings and the resulting clustering with the markings performed by science team members for Planet Four tile ID \nolinkurl{APF0000hqn} of HiRISE image \nolinkurl{ESP_012316_0925}.
The extended fan center lines are 3 times exaggerated fan lengths to indicate the general trend of fan directions for easy visual comparison.
The derived wind directions compare very well between the catalog and the science team data.%
 }
\end{figure}

\begin{figure}[htb]
\centering
\includegraphics[width=0.95\columnwidth]{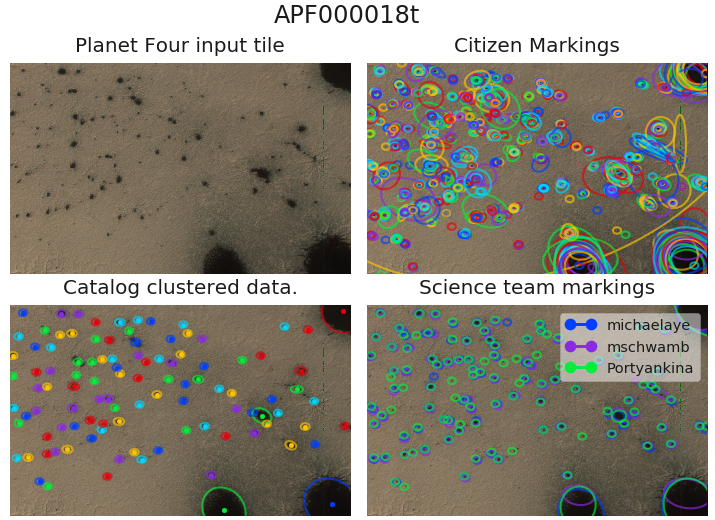}
\caption{\label{fig:gold_data03}Comparing volunteers' markings and the resulting clustering with the markings performed by science team members for Planet Four tile \nolinkurl{APF000018t} of HiRISE image \nolinkurl{ESP_012889_0985}.
The blotches are very well comparable between the science team and the volunteers, with slight disagreements between the science team members.%
}
\end{figure}

In Figures~\ref{fig:gold_data62} and \ref{fig:gold_data03} we show an example comparison of volunteer's markings with those performed by the science team.
The aforementiend slight deviations of the science team members with each other is visible, however, it is clear that the catalog wind directions in Fig.~\ref{fig:gold_data62} are well reproduced by both the specialists and the volunteers.
The results for blotches in Fig.~\ref{fig:gold_data03} are very comparable, with the added simplification that blotches have a much reduced directivity compared to fans.


\clearpage
\bibliographystyle{elsarticle-harv}
\bibliography{ZoteroLibrary}

\end{document}